\pdfoutput=1
\documentclass[acmsmall,screen]{acmart}
\usepackage{graphicx} % Required for inserting images
\usepackage[normalem]{ulem}
\usepackage[utf8]{inputenc}
\usepackage{amsmath}  
\usepackage{graphicx}
\usepackage{multirow}
\usepackage{enumitem}
\usepackage{array}
\usepackage{tcolorbox}
\usepackage{xcolor}
\usepackage{balance}
\usepackage{mathtools}
\usepackage{xspace}
\usepackage{url}
\usepackage{color}
\usepackage{float}
\usepackage{xcolor}
\usepackage{amsmath}
\usepackage{pifont}
\usepackage{bbding}
\usepackage{tcolorbox}
\usepackage{framed}

\usepackage{tcolorbox}

\definecolor{lightgray}{gray}{0.95}

\begin{document}
\title{LLM-Based Multi-Agent Systems for Software Engineering: \\ Literature Review, Vision and the Road Ahead}

\author{Junda He}
\email{jundahe@smu.edu.sg}
\affiliation{%
  \institution{Singapore Management University}
  \streetaddress{80 Stamford Rd.}
  \postcode{178902}
  \city{Singapore}
  \country{Singapore}
}

\author{Christoph Treude}
\email{ctreude@smu.edu.sg}
\affiliation{%
  \institution{Singapore Management University}
  \streetaddress{80 Stamford Rd.}
  \postcode{178902}
  \city{Singapore}
  \country{Singapore}
}

\author{David Lo}
\email{davidlo@smu.edu.sg}
\affiliation{%
  \institution{Singapore Management University}
  \streetaddress{80 Stamford Rd.}
  \postcode{178902}
  \city{Singapore}
  \country{Singapore}
}

 
\begin{abstract}
Integrating Large Language Models (LLMs) into autonomous agents marks a significant shift in the research landscape by offering cognitive abilities that are competitive with human planning and reasoning. 
This paper explores the transformative potential of integrating Large Language Models into Multi-Agent (LMA) systems for addressing complex challenges in software engineering (SE). 
By leveraging the collaborative and specialized abilities of multiple agents, LMA systems enable autonomous problem-solving, improve robustness, and provide scalable solutions for managing the complexity of real-world software projects. In this paper, we conduct a systematic review of recent primary studies to map the current landscape of LMA applications across various stages of the software development lifecycle (SDLC). To illustrate current capabilities and limitations, we perform two case studies to demonstrate the effectiveness of state-of-the-art LMA frameworks.
Additionally, we identify critical research gaps and propose a comprehensive research agenda focused on enhancing individual agent capabilities and optimizing agent synergy. 
Our work outlines a forward-looking vision for developing fully autonomous, scalable, and trustworthy LMA systems, laying the foundation for the evolution of Software Engineering 2.0.
\end{abstract}

\maketitle

\section{Introduction}
\label{sec:intro}

% introducing agent
Autonomous agents, defined as intelligent entities that autonomously perform specific tasks through environmental perception, strategic self-planning, and action execution~\cite{franklin1996agent, albrecht2018autonomous, mele2001autonomous}, have emerged as a rapidly expanding research field since the 1990s~\cite{maes1993modeling}. 
Despite initial advancements, these early iterations often lack the sophistication of human intelligence~\cite{unland2015software}. However, the recent advent of Large Language Models (LLMs)~\cite{kasneci2023chatgpt} has marked a turning point. This LLM breakthrough has demonstrated cognitive abilities nearing human levels in planning and reasoning~\cite{gpt4,kasneci2023chatgpt}, which aligns with the expectations for autonomous agents. As a result, there is an increased research interest in integrating LLMs at the core of autonomous agents~\cite{lo2023trustworthy, xi2023rise, wang2023survey} (for short, we refer to them as {\em LLM-based} agents in this paper).

Nevertheless, the application of singular LLM-based agents encounters limitations, since real-world problems often span multiple domains, requiring expertise from various fields.
In response to this challenge, developing {\em LLM-Based Multi-Agent (LMA)} systems represents a pivotal evolution, aiming to boost performance via synergistic collaboration.
An LMA system harnesses the strengths of multiple specialized agents, each with unique skills and responsibilities. These agents work in concert towards a common goal, engaging in collaborative activities like debate and discussion. 
These collaborative mechanisms have been proven to be instrumental in encouraging divergent thinking~\cite{liang2023encouraging}, enhancing factuality and reasoning~\cite{du2023improving}, and ensuring thorough validation~\cite{wu2023empirical}. As a result, LMA systems hold promise in addressing a wide range of complicated real-world scenarios across various sectors~\cite{horton2023large, wang2023voyager, wang2023describe}, such as software engineering~\cite{lo2023trustworthy, qian-etal-2024-chatdev, li2024camel, hong2023metagpt}.

The study of software engineering (SE) focuses on the entire lifecycle of software systems~\cite{kan2003metrics}, including stages like requirements elicitation~\cite{goguen1993techniques}, development~\cite{abrahamsson2017agile}, and quality assurance~\cite{tian2005software}, among others. 
This multifaceted discipline requires a broad spectrum of knowledge and skills to effectively tackle its inherent challenges in each stage. Integrating LMA systems into software engineering introduces numerous benefits:
\begin{enumerate}[leftmargin=*]
    \item \textbf{Autonomous Problem-Solving:} LMA systems can bring significant autonomy to SE tasks. It is an intuitive approach to divide high-level requirements into sub-tasks and detailed implementation, which mirrors agile and iterative methodologies~\cite{larman2004agile} where tasks are broken down and assigned to specialized teams or individuals. By automating this process, developers are freed to focus on strategic planning, design thinking, and innovation.
    \item \textbf{Robustness and Fault Tolerance: } LMA systems address robustness issues through cross-examination in decision-making, akin to code reviews and automated testing frameworks, thus detecting and correcting faults early in the development process. On their own, LLMs may produce unreliable outputs, known as {\em hallucination}~\cite{zhang2023siren, yang2023apidocbooster}, which can lead to bugs or system failure in software development. However, by employing methods like debating, examining, or validating responses from multiple agents, LMA systems ensure convergence on a single, more accurate, and robust solution. This enhances the system’s reliability and aligns with best practices in software quality assurance.
    \item \textbf{Scalability to Complex Systems:} 
    The growth in complexity of software systems, with increasing lines of code, frameworks, and interdependencies, demands scalable solutions in project management and development practices. LMA systems offer an effective scaling solution by incorporating additional agents for new technologies and reallocating tasks among agents based on evolving project needs. 
    LMA systems ensure that complex projects, which may be overwhelming for individual developers or traditional teams, can be managed effectively through distributed intelligence and collaborative agent frameworks.
\end{enumerate}

Existing research has illuminated the critical roles of these collaborative agents in advancing toward the era of Software Engineering 2.0~\cite{lo2023trustworthy}.
LMA systems are expected to significantly speed up software development, drive innovation, and transform the current software engineering practices. This article aims to delve deeper into the roles of LMA systems in shaping the future of software engineering. It spotlights the current progress, emerging challenges, and the road ahead.  
We provide a systematic review of LMA applications in SE, complemented by two case studies that assess current LMA systems' capabilities and limitations. 
From this analysis, we identify key research gaps and propose a comprehensive agenda structured in two phases: (1) enhancing individual agent capabilities and (2) optimizing agent collaboration and synergy.
This roadmap aims to guide the development of autonomous, scalable, and trustworthy LMA systems, paving the way for the next generation of software engineering

To summarize, this study makes the following key contributions:
\begin{itemize}[leftmargin=*]
    \item We conduct a systematic review of 71 recent primary studies on the application of LMA systems in software engineering.
    \item We perform two case studies to illustrate the current capabilities and limitations of LMA systems.
    \item We identify the key research gaps and propose a structured research agenda that outlines potential future directions and opportunities to advance LMA systems for software engineering tasks.
\end{itemize}

\section{Preliminary}
\label{sec:background}

\subsection{Autonomous Agent}
An \textit{autonomous agent} is a computational entity designed for independent and effective operation in dynamic environments~\cite{mele2001autonomous}. Its essential attributes are:

\begin{itemize}[leftmargin=*]
    \item \textbf{Autonomy:} Independently manages its actions and internal state without external controls.
    \item \textbf{Perception:} Detects the changes in the surrounding environment through sensory mechanisms.
    \item \textbf{Intelligence and Goal-Driven:} Aims for specific goals using domain-specific knowledge and problem-solving abilities.
    \item \textbf{Social Ability:} Can interact with humans or other agents, manages social relationships to achieve goals.
    \item \textbf{Learning Capabilities:} Continuously adapts, learns, and integrates new knowledge and experiences.
\end{itemize}

\subsection{LLM-based Autonomous Agent}
Formally speaking, an LLM-based agent can be described by the tuple \(\langle L, O, M, P, A, R\rangle\)~\cite{cheng2024exploring}, where:
\begin{itemize}[leftmargin=*]
    \item \(L\) symbolizes the \textbf{Large Language Model}, serving as the agent's cognitive core. It is equipped with extensive knowledge, potentially fine-tuned for specific domains, allowing it to make informed decisions based on observations, feedback, and rewards. Typically, an LLM suited for this role is trained on vast corpora of diverse textual data and comprises billions of parameters, such as models like ChatGPT\footnote{\url{https://openai.com/chatgpt/}}, Claude\footnote{\url{https://claude.ai/}}, and Gemini\footnote{\url{https://gemini.google.com/app}}. These models exhibit strong zero-shot and few-shot learning capabilities, meaning they can generalize well to new tasks with little to no additional training. Interaction with the LLM typically occurs through prompts, which guide its reasoning and responses.
    \item \(O\) stands for the \textbf{Objective}, the desired outcome or goal the agent aims to achieve. This defines the agent's focus, driving its strategic planning and task breakdown.
    \item \(M\) represents \textbf{Memory}, which holds information on both historical and current states, as well as feedback from external interactions.
    \item \(P\) represents \textbf{Perception}, which represents the agent’s ability to sense, interpret, and understand its surroundings and inputs. This perception can involve processing structured and unstructured data from various sources such as text, visual inputs, or sensor data. Perception allows the agent to interpret the environment, transforming raw information into meaningful insights that guide decision-making and actions.
    \item \(A\) signifies \textbf{Action}, encompassing the range of executions of the agent, from utilizing tools to communicating with other agents.
    \item \(R\) refers to \textbf{Rethink}, a post-action reflective thinking process that evaluates the results and feedback, along with stored memories. Guided by this insight, the LLM-based agent then takes subsequent actions.
\end{itemize}

\subsection{LLM-Based Multi-Agent Systems}
\label{subsec:LMA}
A multi-agent system is a computational framework composed of multiple interacting intelligent agents that interact and collaborate to solve complex problems or achieve goals beyond the capability of any single agent~\cite{wooldridge2009introduction}. These agents communicate, coordinate, and share knowledge, often bringing specialized expertise to address tasks across diverse domains.

With the integration of LLMs, \textit{LLM-Based Multi-Agent Systems} have emerged. In this paper, we define that an LMA system comprises two primary components: \textit{an orchestration platform} and \textit{LLM-based agents}.

\subsubsection{Orchestration Platform}
The orchestration platform serves as the core infrastructure that manages interactions and information flow among agents. It facilitates coordination, communication, planning, and learning, ensuring efficient and coherent operation. The orchestration platform defines various key characteristics:

\begin{enumerate}[leftmargin=*]
    \item  \textit{Coordination Models}: Defines how agents interact, such as cooperative (collaborating towards shared goals)~\cite{abdelnabi2023llm}, competitive (pursuing individual goals that may conflict)~\cite{wu2024shall}, hierarchical (organized with leader-follower relationships)~\cite{zhao2024hierarchical}, or mixed models.
    \item \textit{Communication Mechanisms}: Determines how the information flows between the agents: It defines the organization of communication channels, including centralized (a central agent facilitates communication~\cite{agashe2023llm}), decentralized (agents communicate directly~\cite{chen2023scalable}), or hierarchical (information flows through layers of authority~\cite{zhao2024hierarchical}). Moreover, it specifies the data exchanged among agents, often in text form. In software engineering contexts, this may include code snippets, commit messages~\cite{zhou2023ccbert}, forum posts~\cite{he2022ptm4tag, he2024ptm4tag+, he2024representation}, bug reports~\cite{bettenburg2008makes}, or vulnerability reports~\cite{imtiaz2021comparative}.
    \item \textit{Planning and Learning Styles}: The orchestration platform specifies how planning and learning are conducted within the multi-agent system. It determines how tasks are allocated and coordinated among agents. It includes strategies like \textit{Centralized Planning, Decentralized Execution (CPDE)} -- planning is conducted centrally, but agents execute tasks independently, or \textit{Decentralized Planning, Decentralized Execution (DPDE)} -- both planning and execution are distributed among agents.

\end{enumerate}

\subsubsection{LLM-Based Agents}
Each agent may have unique abilities and specialized roles, enhancing the system's ability to handle diverse tasks effectively. Agents can be:

\begin{enumerate}[leftmargin=*]
    \item \textbf{Predefined or Dynamically Generated}: Agent profiles can be explicitly predefined~\cite{hong2023metagpt} or dynamically generated by LLMs~\cite{wang2023survey}, allowing for flexibility and adaptability.
    \item \textbf{Homogeneous or Heterogeneous}: Agents may have identical functions (homogeneous) or diverse functions and expertise (heterogeneous).
\end{enumerate}

Each LLM-based agent can be represented as a node $v_i$ in a graph $G(V, E)$, where edges $e_{i,j} \in E$ represent interactions between agents $v_i$ and $v_j$. 

\section{Literature Review}
\label{sec:review}
In this section, we review recent studies on LMA systems in software engineering, organizing these applications across various stages of the software development lifecycle, including requirements engineering, code generation, quality assurance, and software maintenance. We also examine studies on LMA systems for end-to-end software development, covering multiple SDLC phases rather than isolated stages. 

\noindent \textbf{Search Strategy:} We conduct a keyword-based search on the DBLP publication database~\cite{dblp} to match paper titles. DBLP is a widely used resource in software engineering surveys~\cite{chen2020survey,zhang2018empirical,chen2024fairness}, which indexes over 7.5 million publications across 1,800 journals and 6,700 academic conferences in computer science. 

Our search included two sets of keywords: one set targeting LLM-based Multi-Agent Systems (called [agent words]) and the other focusing on specific software engineering activities (called [SE words]). 
Papers may use variations of the same keyword. For example, the term ``vulnerability'' may appear as ``vulnerable'' or ``vulnerabilities.'' To address this, we use truncated terms like ``vulnerab'' to capture all related forms. For LMA systems, we used keywords: \emph{``Agent'' OR ``LLM'' OR ``Large Language Model'' OR ``Collaborat''}. To ensure comprehensive coverage of SE activities, we incorporated phase-specific keywords for each stage of the SDLC into our search queries:

\begin{enumerate}
    \item \textbf{Requirements Engineering}: \emph{requirement, specification, stakeholder}
    \item \textbf{Code Generation}: \emph{software, code, coding, program}
    \item \textbf{Quality Assurance}: \emph{bug, fault, defect, fuzz, test, vulnerab, verificat, validat}
    \item \textbf{Maintenance}: \emph{debug, repair, review, refactor, patch, maintenanc}
\end{enumerate}

We focus on four key phases of the SDLC: requirements engineering, code generation, quality assurance, and software maintenance. For each phase, the relevant SE keywords are combined using the OR operator to capture all variations. The final search query for each SDLC phase follows the format: [agent words] AND [SE words]. 

Following the guide of previous work~\cite{meline2006selecting,zhou2024large,van2021automation}, we design the following inclusion and exclusion criteria. In the first phase, we filtered out short papers (exclusion criterion 1) and removed duplicates (exclusion criterion 2). In the second phase, we manually screened each paper’s venue, title, and abstract, excluding items such as books, keynote speeches, panel summaries, technical reports, theses, tool demonstrations, editorials, literature reviews, and surveys (exclusion criteria 3 and 4).
In the third phase, we conducted a full-text review to further refine relevant studies. Following Section \ref{subsec:LMA}, we exclude papers that do not describe LMA systems (exclusion criterion 5). Papers that rely solely on LLMs using non-agent-based methods or single-agent approaches are excluded. Further, we focused on LMA systems powered by LLMs with strong planning capabilities, such as ChatGPT and LLaMA, excluding models like CodeBERT and GraphCodeBERT. Since the release of ChatGPT is in November 2022, we limited our review to papers published after this date (exclusion criterion 6). Furthermore, we excluded papers unrelated to software engineering (exclusion criterion 7) and those that mention LMA systems only in discussions or as future work, without presenting experimental results (exclusion criterion 8).
After the third phase, we identified 41 primary studies directly relevant to our research focus. The search process is conducted on November 14th, 2024.

\begin{itemize}[leftmargin=1.5em]
\item[{\textcolor[RGB]{0,128,0}{\Checkmark}}] \emph{The paper must be written in English.}
\item[{\textcolor[RGB]{0,128,0}{\Checkmark}}] \emph{The paper must have an accessible full text.}

\item[{\textcolor[RGB]{0,128,0}{\Checkmark}}] \emph{The paper must adopt LMA techniques to solve software engineering-related tasks.}
\item[{\textcolor[RGB]{209,26,66}{\XSolidBrush}}] \emph{The paper has less than 5 pages.}
\item[{\textcolor[RGB]{209,26,66}{\XSolidBrush}}] \emph{Duplicate papers or similar studies authored by the same authors.}
\item[{\textcolor[RGB]{209,26,66}{\XSolidBrush}}] \emph{Books, keynote records,  panel summaries, technical reports, theses, tool demos papers, editorials}
\item[{\textcolor[RGB]{209,26,66}{\XSolidBrush}}] \emph{The paper is a literature review or survey.}
\item[{\textcolor[RGB]{209,26,66}{\XSolidBrush}}] \emph{The paper does not utilize LMA systems, e.g., using a single LLM agent.}
\item[{\textcolor[RGB]{209,26,66}{\XSolidBrush}}] \emph{The paper is published before November 2022 (the release date of ChatGPT).}
\item[{\textcolor[RGB]{209,26,66}{\XSolidBrush}}] \emph{The paper does not involve software engineering related tasks.}
\item[{\textcolor[RGB]{209,26,66}{\XSolidBrush}}] \emph{The paper lacks experimental results and mentions LMA systems only in future work or discussions.}
\end{itemize}

\noindent \textbf{Snowballing Search}
To expand our review, we conducted both backward and forward snowballing~\cite{wohlin2014guidelines} on the relevant papers identified in previous steps. This process involved examining the references cited by the relevant studies as well as publications that have cited these studies. We repeated the snowballing process until reaching a transitive closure fixed point, where no new relevant papers were found, resulting in an additional 30 papers identified.

\subsection{Requirements Engineering}
Requirements Engineering~\cite{van2000requirements, lo2024requirements} focuses on defining and managing software system requirements. This discipline is divided into several key stages to ensure requirements meet quality standards and align with stakeholder needs. These stages include elicitation, modeling, specification, analysis, and validation~\cite{hickey2004unified,christel1992issues}.

Elicitron~\cite{ataei2024elicitron} is an LMA framework that focuses specifically on the elicitation stage. It utilizes LLM-based agents to represent a diverse array of simulated users.
These agents engage in simulated product interactions, providing insights into user needs by articulating their actions, observations, and challenges.
MARE~\cite{jin2024mare} is an LMA framework that covers multiple phases of requirements engineering, including elicitation, modeling, verification, and specification. It employs five distinct agents, i.e., stakeholder, collector, modeler, checker, and documenter, performing nine actions to help generate high-quality requirements models and specifications.
Sami et al.~\cite{sami2024ai} propose another LMA framework to generate, evaluate, and prioritize user stories through a collaborative process involving four agents: product owner, developer, quality assurance (QA), and manager. The produce owner generates user stories and initiates prioritization. The QA agent assesses story quality and identifies risks, while the developer prioritizes based on technical feasibility. Finally, the manager synthesizes these inputs and finalizes prioritization after discussions with all agents

\subsection{Code Generation}

Code generation~\cite{herrington2003code,budinsky1996automatic} has consistently been a longstanding focus of software engineering research, aiming to automate coding tasks to boost productivity and minimize human error.

A prominent multi-agent setup for code generation typically on role specialization and iterative feedback loops to optimize collaboration among agents. We summarize the common roles identified in the literature, including the \textit{Orchestrator, Programmer, Reviewer, Tester, and Information Retriever}.

The \textit{Orchestrator} acts as the central coordinator, managing high-level planning and ensuring smooth task execution across all agents. Its responsibilities include defining high-level strategic goals, breaking them into actionable sub-tasks, delegating these tasks to the appropriate agents, monitoring progress, and ensuring that workflows align with overall project objectives~\cite{li2024codetree, zhang2024pair, ishibashi2024self, zan2024codes, cai2023low, phan2024hyperagent,li2024camel,josifoski2023flows}.
For instance, PairCoder~\cite{zhang2024pair} features a Navigator agent that interprets natural language descriptions to create high-level plans outlining solutions and key implementation steps. The Driver agent then follows these plans to handle code generation and refinement.
The Self-Organized Agents (SoA) framework~\cite{ishibashi2024self} employs a hierarchical design, with Mother agents managing high-level abstractions and delegating subtasks to specialized Child agents.
In CODES~\cite{zan2024codes},  the Orchestrator role is performed by the RepoSketcher, which converts high-level natural language requirements into a repository sketch. This sketch outlines the project structure, including directories, files, and inter-file dependencies. The RepoSketcher then delegates tasks to the FileSketcher and SketchFiller, ensuring the efficient and seamless creation of a complete, functional code repository.

During the implementation phase, the process typically begins with the Programmer, who is responsible for writing the initial version of the code. Once the initial code is produced, roles like the Reviewer and Tester step in to evaluate it, providing constructive feedback on quality, functionality, and adherence to requirements. This feedback initiates an iterative cycle, where the Programmer refines the code or the Debugger resolves identified issues, ensuring that the final code meets the desired standards and performs as expected~\cite{mathews2024test, wang2024intervenor, olausson2023self, chen2023agentverse, le2024indict, liu2023dynamic,lei2024autocoder,lin2024llm,dong2023self}.
For example, INTERVENOR~\cite{wang2024intervenor} pairs a Code Learner with a Code Teacher. The Code Learner generates the initial code and then compiles it to evaluate its correctness. If issues are identified, the Code Teacher analyzes the bug reports and the buggy code, subsequently providing repair instructions to address the errors.
Self-repair~\cite{olausson2023self} and TGen~\cite{mathews2024test} refine code by utilizing feedback obtained from running pre-defined test cases.

When predefined test cases are unavailable, the Tester can generate a variety of test cases, ranging from common scenarios to edge cases. These tests help uncover subtle issues that might otherwise go unnoticed and provide actionable feedback to guide subsequent refinement iterations~\cite{huang2023agentcoder, shinn2024reflexion, ishibashi2024self, hu2024self}.

Some frameworks employ the Information Retriever to gather relevant information to assist code generation. For instance, Agent4PLC~\cite{liu2024agents4plc} and MapCoder~\cite{islam-etal-2024-mapcoder} incorporate a Retrieval Agent tasked with sourcing examples of similar problems and extracting related knowledge. This agent provides essential contextual information and references tailored to the user's input, ensuring that solutions are well-informed and adhere to domain-specific best practices. Similarly, CodexGraph~\cite{liu2024codexgraph} employs a translation agent to facilitate interaction with graph databases, which are built using static analysis to extract code symbols and their relationships. By converting user queries into graph query language, this agent enables precise and structured information retrieval, enhancing the capability of LLM-based agents to navigate and utilize code repositories effectively.

Agent Forest~\cite{li2024more} adopts a different paradigm instead of role specialization. Instead, it utilizes a sampling-and-voting framework, where multiple agents independently generate candidate outputs. Each output is then evaluated based on its similarity to the others, with a cumulative similarity score calculated for each. The output with the highest score—indicating the greatest consensus among the agents—is selected as the final solution.

\subsection{Software Quality Assurance}

In this subsection, we review related work on testing, vulnerability detection, bug detection, and fault localization, with a focus on how LMA systems are being employed to enhance software quality assurance processes.

\noindent \textbf{Testing.} Fuzz4All~\cite{xia2024fuzz4all} generates testing input for software systems across multiple programming languages. In this framework, a distillation agent reduces user input while a generation agent creates and mutates inputs.
AXNav~\cite{taeb2024axnav} is designed to automate accessibility testing. It interprets natural language test instructions and executes accessibility tests, such as VoiceOver, on iOS devices. AXNav includes a planner agent, an action agent, and an evaluation agent.
WhiteFox~\cite{yang2023white} is a fuzzing framework that tests compiler optimizations. It uses two LLM-based agents: one extracts requirements from source code, and the other generates test programs.
Additionally, LMA systems are employed for tasks such as penetration testing~\cite{deng2023pentestgpt}, user acceptance testing~\cite{wang2024xuat}, and GUI testing~\cite{yoon2024intent}.

\noindent \textbf{Vulnerability Detection.}
GPTLens~\cite{hu2023large} is an LMA framework for detecting vulnerabilities in smart contracts. The system includes LLM-based agents acting as auditors, each independently identifying vulnerabilities. A critic agent then reviews and ranks these vulnerabilities, filtering out false positives and prioritizing the most critical ones.
MuCoLD~\cite{mao2024multi} assigns roles like tester and developer to evaluate code. Through discussions and iterative assessments, the agents reach a consensus on vulnerability classification.
Widyasari et al.~\cite{widyasari2024beyond} introduces a cross-validation technique, where multiple LLM's answer is validated against each other.

\noindent \textbf{Bug Detection.}
Intelligent Code Analysis Agent (ICAA)~\cite{fan2023static} is used for bug detection in static code analysis. The agents have access to tools like web search, static analysis, and code retrieval tools. A Report Agent generates bug reports, while a False Positive Pruner Agent refines these reports to reduce false positives.
Additionally, ICAA includes Code-Intention Consistency Checking, which ensures the code aligns with the developer's intended functionality by analyzing code comments, documentation, and variable names.

\noindent \textbf{Fault Localiztion.}
RCAgent~\cite{wang2023rcagent} performs root cause analysis in cloud environments by using LLM-based agents to collect system data, analyze logs, and diagnose issues. AgentFL~\cite{qin2024agentfl} breaks down fault localization into three phases. The Comprehension Agent identifies potential fault areas, the Navigation Agent narrows down the codebase search, and the Confirmation Agent uses debugging tools to validate the faults.

\subsection{Software Maintenance}
In this subsection, we explore related work on debugging and code review, highlighting how LMA systems contribute to automating and improving software maintenance processes.

\noindent \textbf{Debugging.} Debugging involves identifying, locating, and resolving software bugs. Several frameworks, including MASAI~\cite{arora2024masai}, MarsCode~\cite{liu2024marscode}, AutoSD~\cite{kang2023explainable}, and others~\cite{tao2024magis, ma2024understand, chen2024coder,lei2024infant}, follow a structured process consisting of stages like bug reproduction, fault localization, patch generation, and validation. Specialized agents are typically responsible for each stage.
FixAgent~\cite{lee2024unified} includes a debugging agent and a program repair agent that work together to iteratively fix code by analyzing both errors and repairs. The system refines fault localization by incorporating repair feedback. The agents also articulate their thought processes, improving context-aware debugging.
The MASTER framework~\cite{yang2024enhancing} employs three specialized agents. The Code Quizzer generates quiz-like questions from buggy code, the Learner proposes solutions, and the Teacher reviews and refines the Learner's responses.
AutoCodeOver~\cite{zhang2024autocoderover} uses an agent for fault localization via spectrum-based methods, collaborating with others to refine patches using program representations like abstract syntax trees.
SpecRover~\cite{ruan2024specrover} extends AutoCodeOver by improving program fixes through iterative searches and specification analysis based on inferred code intent.
ACFIX~\cite{zhang2024acfix} targets access control vulnerabilities in smart contracts, focusing on Role-Based Access Control. It mines common RBAC patterns from over 344,000 contracts to guide agents in generating patches.
DEI~\cite{zhang2024diversity} resolves GitHub issues by using a meta-policy to select the best solution, integrating and re-ranking patches generated by different agents for improved issue resolution.
SWE-Search~\cite{antoniades2024swe} consists of three agents: the SWE-Agent for adaptive exploration, the Value Agent paired with a Monte Carlo tree search module for iterative feedback and utility estimation, and the Discriminator Agent for collaborative decision-making through debate. 
RepoUnderstander~\cite{ma2024understand} constructs a knowledge graph for a full software repository and also uses Monte Carlo tree search to assist in understanding complex dependencies. 

\noindent \textbf{Code Review.} Rasheed et al.~\cite{rasheed2024ai} developed an automated code review system that identifies bugs, detects code smells, and provides optimization suggestions to improve code quality and support developer education. This system uses four specialized agents focused on code review, bug detection, code smells, and optimization. Similarly, CodeAgent~\cite{tang2024collaborative} performs code reviews with sub-tasks such as vulnerability detection, consistency checking, and format verification. A supervisory agent, QA-Checker, ensures the relevance and coherence of interactions between agents during the review process.

\noindent \textbf{Test Case Maintenance. }Lemner et al.~\cite{lemner2024exploring} propose two multi-agent architectures to predict which test cases need maintenance after source code changes. These agents perform tasks including summarizing code changes, identifying maintenance triggers, and localizing relevant test cases.

\subsection{End-to-end Software Development}
End-to-end software development encompasses the entire process of creating a software product. While conventional code generation is often limited to producing isolated components such as functions, classes, or modules, end-to-end development starts from high-level software requirements and progresses through design, implementation, testing, and ultimately delivering a fully functional and ready-to-use product.

In practice, developers and stakeholders typically adopt established software process models to guide collaboration, such as Agile~\cite{cohen2004introduction} and Waterfall~\cite{petersen2009waterfall}. Similarly, the design of LMA systems for end-to-end software development draws inspiration from these software process models.
The development process is organized into distinct phases, such as requirements gathering, software design, implementation, and testing. Each phase is managed by specialized agents with domain expertise.

\begin{figure}[t] 
    \centering       
    \includegraphics[width=\textwidth]{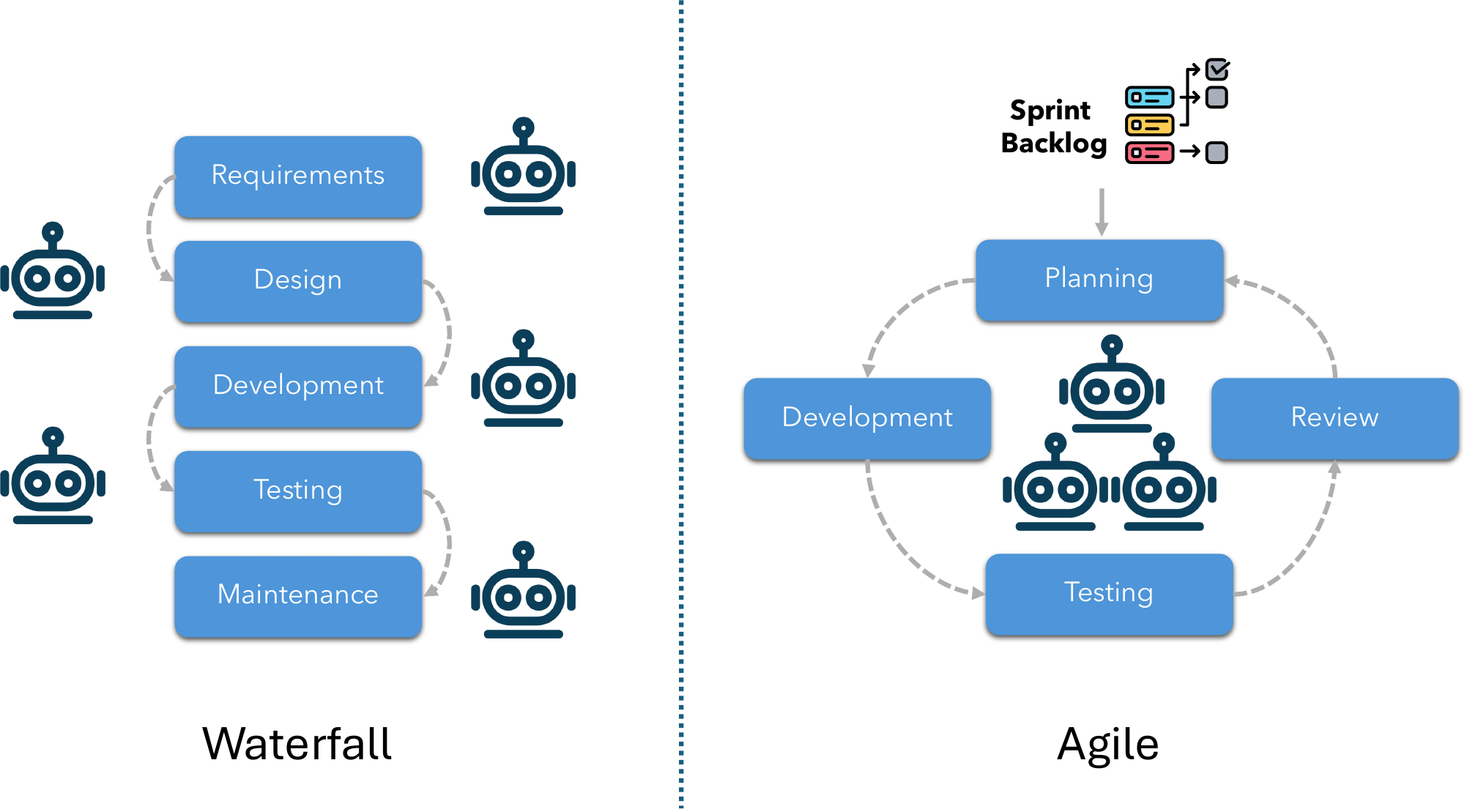} % Adjust width as needed
    \caption{Multi-Agent Systems in Software Development: Waterfall vs. Agile Models} 
    \label{fig:process_model}           
\end{figure}

It is important to note that works such as FlowGen~\cite{lin2024llm} and Self-Collaboration~\cite{dong2023self} emulate various software process models. However, their experiments focus on generating code segments rather than delivering fully developed software products. As a result, in this paper, these approaches are not considered to be designed for true end-to-end software development.

Several works~\cite{qian-etal-2024-chatdev, hong2023metagpt, zhang2024experimenting, du2024multi, zan2024codes, sami2024experimenting, rasheed2024codepori, holt2023l2mac} adopt the Waterfall model to automate software development.
The Waterfall model used in these multi-agent methods organizes the software development process into distinct, sequential phases, where each stage must be completed before proceeding to the next. The primary phases typically include Requirement Analysis, Architecture Design, Code Development, Testing, and Maintenance.
For instance, in MetaGPT~\cite{hong2023metagpt}, the Product Manager agent thoroughly analyzes user requirements. The Architect agent then transforms these requirements into detailed system design components. Subsequently, the Engineer implements the specified classes and functions as outlined in the design. Finally, the Quality Assurance Engineer creates and executes test cases to ensure rigorous code quality standards are met.
These approaches emphasize a linear and sequential design process, ensuring structured progression and clear accountability at each stage.

AgileCoder~\cite{nguyen2024agilecoder} and AgileGen~\cite{zhang2024empowering} adopt Agile process models for software development, emphasizing iterative development by breaking complex tasks into small, manageable increments.
AgileCoder~\cite{nguyen2024agilecoder} assigns Agile roles such as Product Manager and Scrum Master to facilitate sprint-based collaboration and development cycles.
AgileGen enhances Agile practices with human-AI collaboration, integrating close user involvement to ensure alignment between requirements and generated code. A notable feature of AgileGen is its use of the Gherkin language to create testable requirements, bridging the gap between user needs and code implementation.

While most methods rely on predefined roles and fixed workflows for software development, a few work~\cite{wang2024megaagent,li2023metaagents, lin2024think} investing in dynamic process models. 
Think-on-Process (ToP)~\cite{lin2024think} introduces a dynamic process generation framework. Since software development processes can vary significantly depending on project requirements, ToP moves beyond the limitations of static, one-size-fits-all workflows to enable more flexible and efficient development practices.
Given a software requirement, this framework leverages LLMs to create tailored process instances based on their knowledge of software development. These instances act as blueprints to guide the architecture of the LMA system, adapting to the specific and diverse needs of different projects.
Similarly, in MegaAgent~\cite{wang2024megaagent}, agent roles and tasks are not predefined but are generated and planned dynamically based on project requirements. 
Both ToP and MegaAgent highlight the shift from rigid, static workflows to dynamic, adaptive systems. These frameworks promise more efficient, flexible, and context-aware software development practices, aligning processes with project-specific requirements and complexities.

Additionally, instead of focusing on the process model, several works~\cite{qian-etal-2024-experiential,qian2024iterative} explore leveraging experiences from past software projects to enhance new software development efforts. Co-Learning~\cite{qian-etal-2024-experiential} enhances agents' software development abilities by utilizing insights gathered from historical communications. This framework fosters cooperative learning between two agent roles—instructor and assistant—by extracting and applying heuristics from their task execution histories.
Building on this, Qian et al.~\cite{qian2024iterative} propose an iterative experience refinement (IER) framework that enables agents to continuously adapt by acquiring, utilizing, and selectively refining experiences from previous tasks, improving agents' effectiveness and collaboration in dynamic software development scenarios.
\section{Case Study}
\label{sec:exp}

To demonstrate the practical effectiveness of LMA systems, we conduct two case studies. Specifically, we utilize the state-of-the-art LMA framework, ChatDev~\cite{qian-etal-2024-chatdev}, to autonomously develop two classic games: Snake and Tetris. ChatDev structures the software development process into three phases: designing, coding, and testing. ChatDev employs specialized roles, including CEO, CTO, programmer, reviewer, and tester. ChatDev's agents are powered by GPT-3.5-turbo\footnote{\url{https://platform.openai.com/docs/models/gp\#gpt-3-5-turbo}}. 
The temperature setting controls the randomness and creativity of the GPT-3.5's responses. Following the original ChatDev setting, we set the temperature of GPT-3.5-turbo as 0.2.

\subsection{Snake Game}
For the Snake game, we provide the following prompt to ChatDev to generate the game: 

\begin{tcolorbox}[colback=blue!5!white,colframe=blue!75!black,title=Snake Game Prompt]
\textit{``Design and implement a grid-based snake game displayed on the screen. Initialize the snake with a defined starting position, length, and direction. Enable continuous movement controlled by arrow keys. Introduce food that spawns randomly on the grid, ensuring it does not overlap with the snake. Trigger snake growth when food is consumed, adding a new segment to its body. Implement a game-over condition for boundary or body collisions, displaying a message and providing a restart option. Include a scoring system displayed in the user interface, along with clear instructions. }''
\end{tcolorbox}

\begin{figure}[t]
    \centering
    \begin{minipage}{0.3\textwidth}
        \centering
        \includegraphics[width=\textwidth]{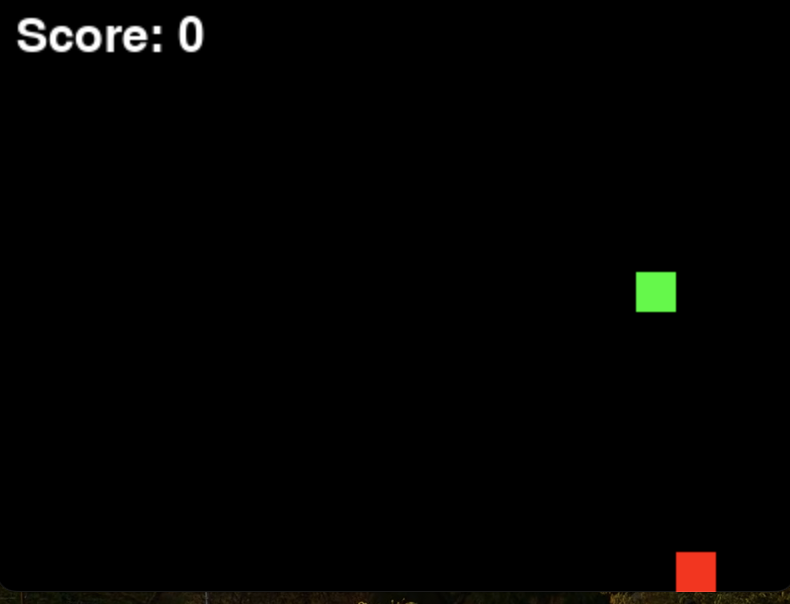}
    \end{minipage}%
    \hfill
    \begin{minipage}{0.3\textwidth}
        \centering
        \includegraphics[width=\textwidth]{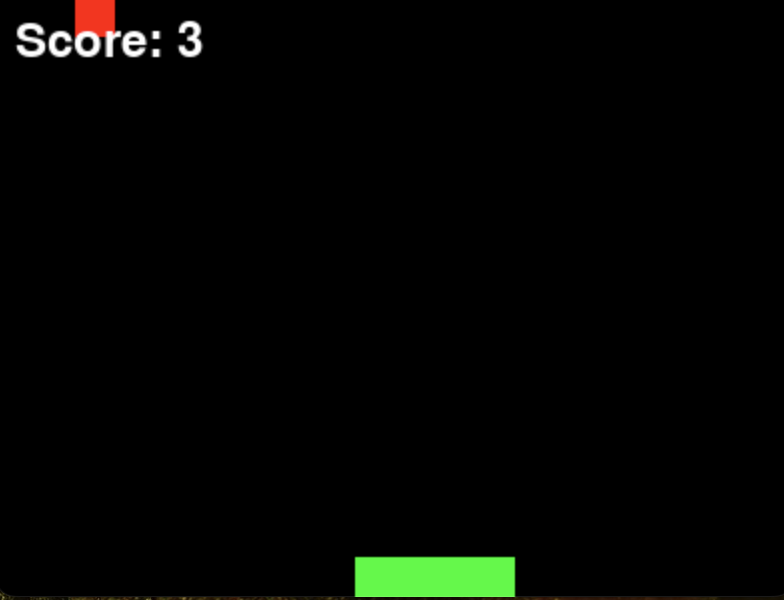}
    \end{minipage}%
    \hfill
    \begin{minipage}{0.3\textwidth}
        \centering
        \includegraphics[width=\textwidth]{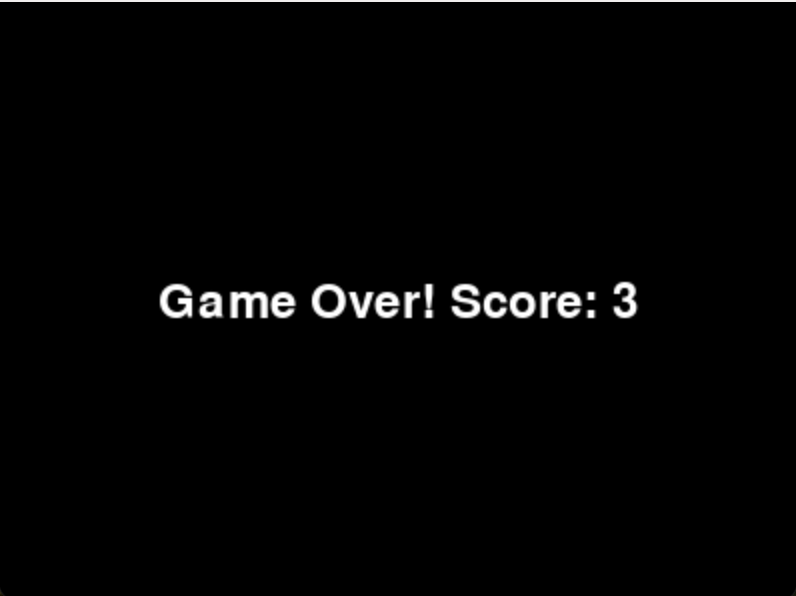}
    \end{minipage}
    \caption{Screen shots of the Snake Game generated by ChatDev.}
    \label{fig:snake}
\end{figure}

While the first attempt to generate the Snake game was unsuccessful, we resubmitted the same prompt to ChatDev, and the second attempt successfully produced a playable version. ChatDev also generated a detailed manual that included information on dependencies, step-by-step instructions for running the game, and an overview of its features. Figure \ref{fig:snake} displays the graphical user interfaces (GUIs) of the generated Snake game, showing the starting state, in-game state, and game-over state. The development process was consistently efficient, taking an average of 76 seconds and costing \$0.019. Upon playing the game, we confirmed that it fulfilled all the requirements outlined in the prompt.

\subsection{Tetris Game}

We present the following prompt to ChatDev to guide the generation of the Tetris game:

\begin{tcolorbox}[colback=green!5!white,colframe=green!75!black,title=Tetris Game Prompt]

``\textit{Design and implement a Tetris game. Start with a randomly chosen piece dropping from the top. Allow players to control the tetromino using arrow keys for movement (left, right, down) and rotation. Enable automatic downward movement with an adjustable speed. Handle collisions with the boundaries and existing pieces, locking the tetromino in place when it cannot move further. Check for complete rows after each placement and remove them. End the game if new tetrominoes cannot spawn due to a full board, displaying a game over message.}''
\end{tcolorbox}

\begin{figure}[t]
    \centering
    \begin{minipage}{0.2\textwidth}
        \centering
        \includegraphics[width=\textwidth]{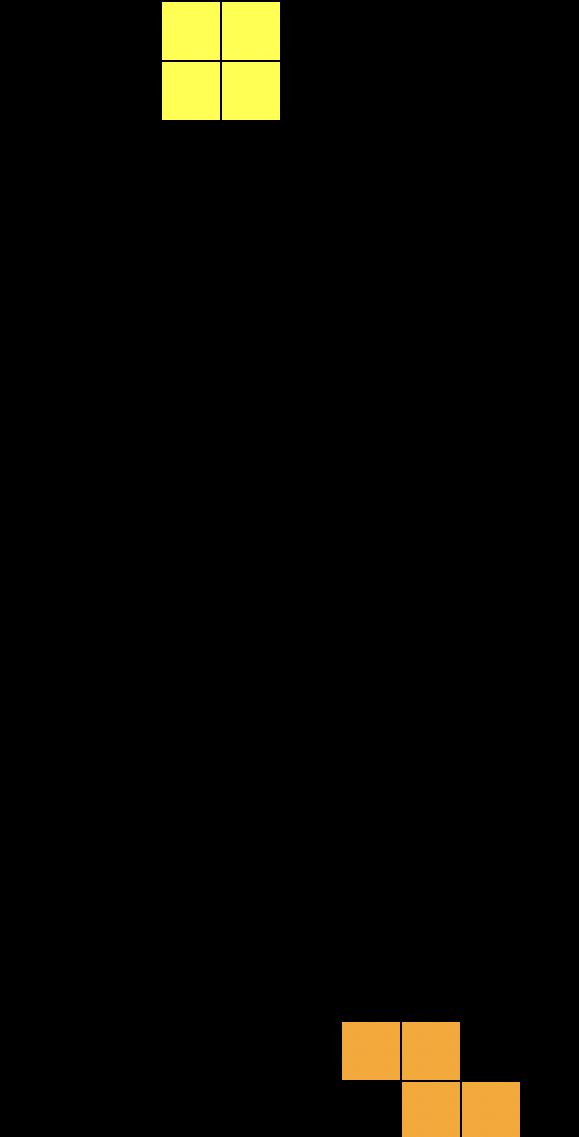}
    \end{minipage}%
    \hfill
    \begin{minipage}{0.2\textwidth}
        \centering
        \includegraphics[width=\textwidth]{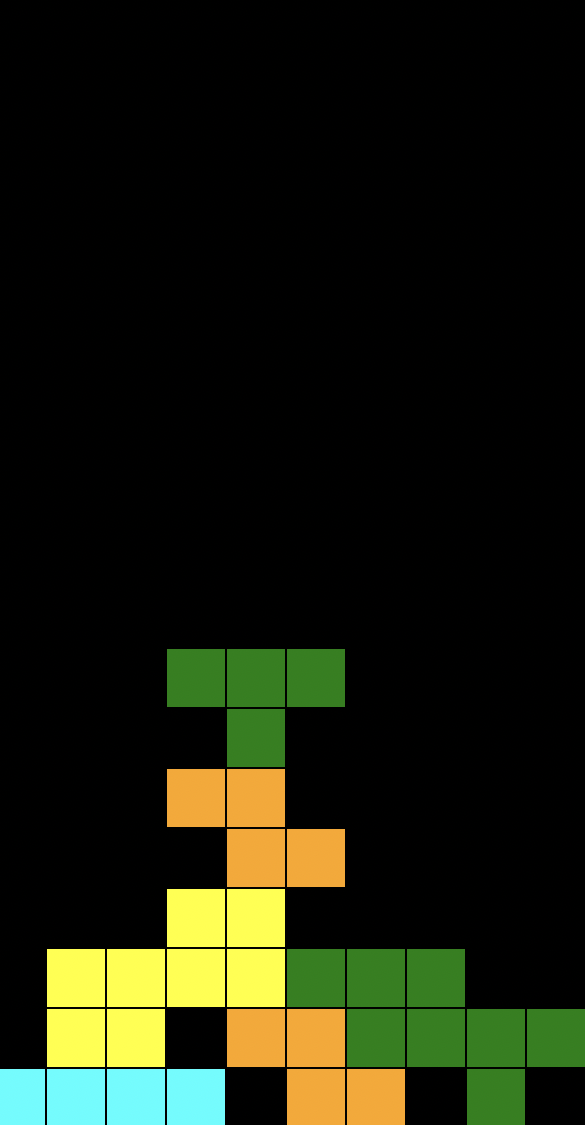}
    \end{minipage}%
    \hfill
    \begin{minipage}{0.2\textwidth}
        \centering
        \includegraphics[width=\textwidth]{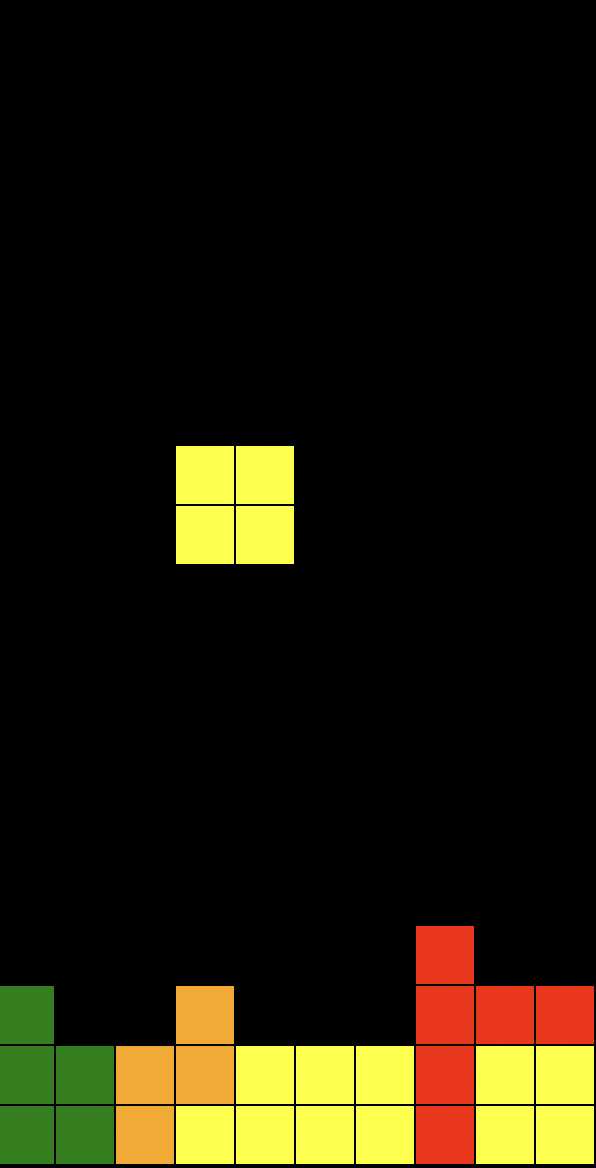}
    \end{minipage}
    \hfill
    \begin{minipage}{0.19\textwidth}
        \centering
        \includegraphics[width=\textwidth]{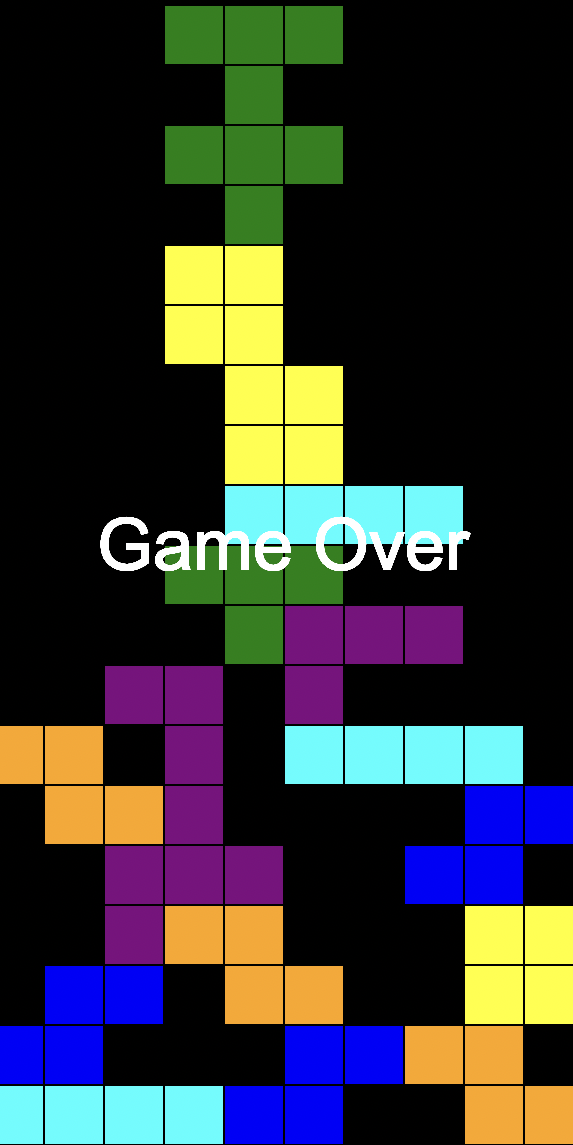}
    \end{minipage}
    \caption{Screen shots of the Tetris Game generated by ChatDev.}
    \label{fig:tetris}
\end{figure}

During development, ChatDev faced challenges in producing functional gameplay across the first nine attempts. Notice that the same prompt was used for each run. 
On the tenth attempt, ChatDev successfully produced a Tetris game that met most of the prompt requirements, as shown in Figure \ref{fig:tetris}. 
The figure illustrates the game's key states: the starting state, in-game states, and game-over state.
However, the game still lacks the core functionality to remove completed rows, as demonstrated in the third subplot of Figure \ref{fig:tetris}. 
Overall, the development process remained efficient, with an average time of 70 seconds and a cost of \$0.020 per attempt.

\noindent \textbf{Summary of Findings}. 
From our case studies, current LMA systems demonstrate strong performance in reasonably complex tasks like developing a Snake game.
The generated Snake game meets all requirements in the prompt within just a few iterations. The process was efficient and cost-effective, with an average completion time of 76 seconds and a cost of \$0.019 per attempt. These results emphasize the suitability of LMA systems for moderately complex software engineering tasks.
However, when tasked with more complex challenges like developing a Tetris game, ChatDev successfully generates a playable Tetris game only by the tenth attempt. The game still lacks the core functionality, i.e., removing completed rows.
This highlights the limitations of current LMA systems in handling more complex tasks that require deeper logical reasoning and abstraction. Nevertheless, development remains efficient and cost-effective, averaging 70 seconds and \$0.020 per run, making the system a promising tool for rapid prototyping.

\section{Research Agenda}
\label{sec:agenta}
Previous research has laid the groundwork for the exploration of LMA systems in software engineering, yet this domain remains in its nascent stages, with many critical challenges awaiting resolution. In this section, we outline our perspective on these challenges and suggest research questions that could advance this burgeoning field. As illustrated in Figure \ref{fig:roadmap}, we envision two phases for the development of LMA systems in software engineering. We discuss each of these phases below and suggest a series of research questions that could form the basis of future research projects.

{\color{red}
\begin{figure}[htbp] 
    \centering       
    \includegraphics[width=\textwidth]{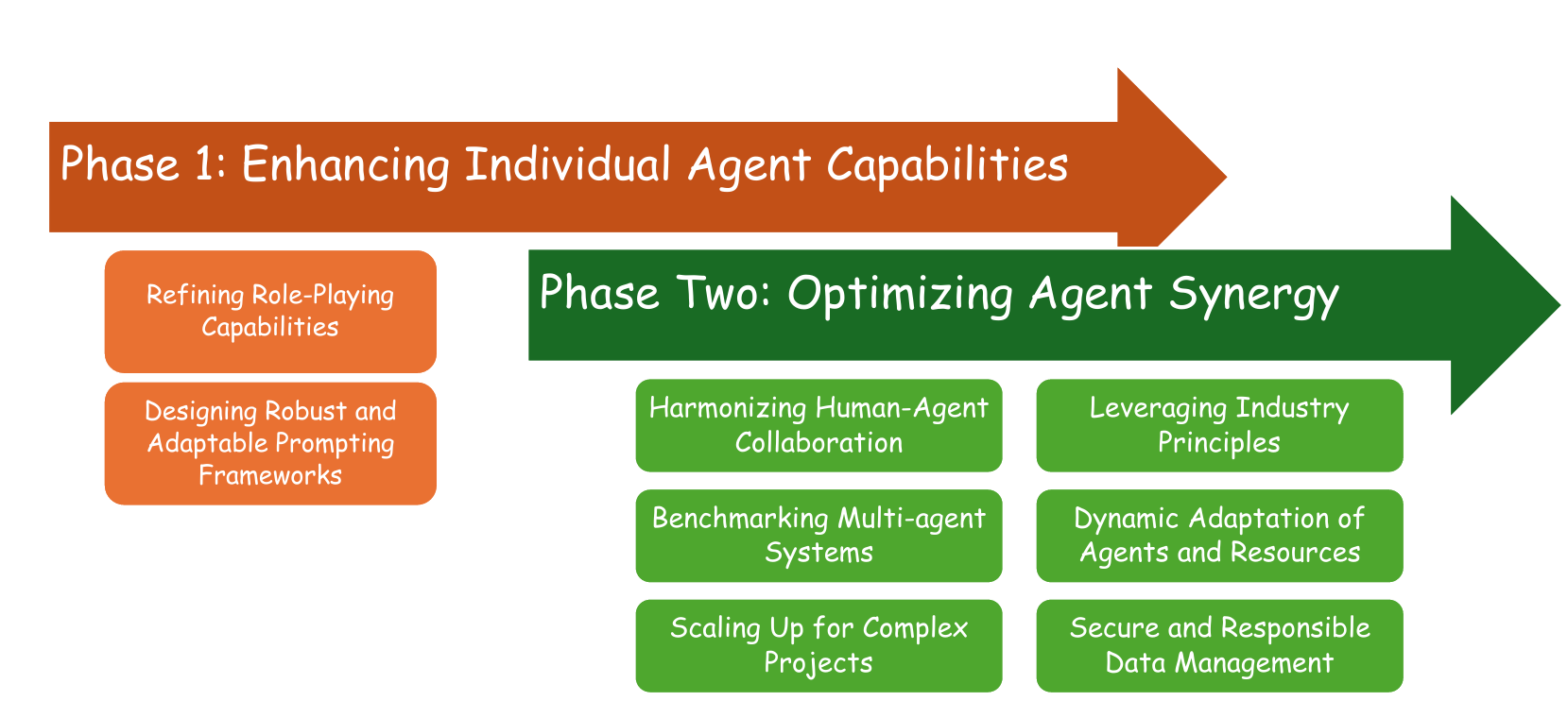} % Adjust width as needed
    \caption{Research Agenda for LLM-Based Multi-Agent Systems in Software Engineering} 
    \label{fig:roadmap}           
\end{figure}
}

\subsection{Phase 1: Enhancing Individual Agent Capabilities}
Indeed, the effectiveness of an LMA system is closely linked to the capabilities of its individual agents. 
This first phase is dedicated to improving these agents' skills, with a particular focus on adaptability and the acquisition of specialized skills in SE. The potential of individual LLM-based agents in SE is further explored through our initial research questions:

\begin{enumerate}[leftmargin=*]
    \item \textit{What SE roles are suitable for LLM-based agents to play and how can their abilities be enhanced to represent these roles?}
    \item \textit{How to design an effective, flexible, and robust prompting language that enhances LLM-based agents’ capabilities?}
\end{enumerate}

\subsubsection{\textbf{Refining Role-Playing Capabilities in Software Engineering}} \hfill

The role-playing capabilities of LLM-based agents are pivotal within LMA systems~\cite{wang2023rolellm}. To address the complexity of software engineering tasks, we need specialized agents capable of adopting diverse roles to tackle intricate challenges throughout the software development lifecycle.

\noindent \textbf{Current State.} Existing LMA systems, such as ChatDev~\cite{qian-etal-2024-chatdev}, MetaGPT~\cite{hong2023metagpt}, and AgileCoder~\cite{nguyen2024agilecoder}, effectively simulate roles like generic software developers and product managers. The agents in these systems rely on general-purpose LLMs such as ChatGPT. Although LLMs like ChatGPT exhibit strong programming skills, they still lack the nuanced expertise required in SE~\cite{hou2023large}. This limitation hampers their ability to simulate other SE-specific roles. For example, roles involving vulnerability detection or security auditing require a deep understanding of security protocols, threat modeling, and the latest vulnerabilities. However, multiple studies have identified deficiencies in ChatGPT's ability to accurately detect and repair vulnerabilities~\cite{fu2023chatgpt, sridhara2023chatgpt, chen2023chatgpt}.
This shortcoming underscores the need to integrate domain-specific expertise into LLMs to better support specialized software engineering roles.

\noindent \textbf{Opportunities.} To address this limitation, we propose a structured and actionable three-step approach encompassing the identification, assessment, and enhancement of role-playing abilities, which are: 

\textsl{Step 1: Identifying and Prioritizing Key SE Roles.} 

\textsl{Step 2: Assessing LLM-Based Agents' Competencies Against Role Requirements.} 

\textsl{Step 3: Enhancing Role-Playing Abilities Through Targeted Training.} 

The first step focuses on identifying key SE roles, prioritizing those with high industry demand and the potential to substantially boost productivity. This involves:
\begin{enumerate} 

\item \textit{Market Analysis}: To begin, we embark on a comprehensive market analysis.
It is crucial to assess not only the current trends and needs within the SE sector but also to anticipate future shifts influenced by the integration of LLM-based agents. 
This analysis involves leveraging various resources such as market reports, job postings, industry forecasts, and technology trend analyses. Platforms like LinkedIn Talent Insights\footnote{\url{https://www.linkedin.com/products/linkedin-talent-insights/}}, Gartner reports\footnote{\url{https://www.gartner.com/en/products/special-reports}}, and Stack Overflow Developer Surveys\footnote{\url{https://survey.stackoverflow.co/2024/}} may also offer valuable data to inform this assessment.
The focus should be on identifying roles that are in high demand and demonstrate rapid growth, especially those requiring specialized skills not typically found among generalist developers. For example, machine learning engineers or cloud architects. Additionally, positions where LLM-based agents could significantly enhance productivity, reduce costs, or accelerate innovation should be evaluated. A key component of this analysis should be determining whether the market has already begun shifting away from recruiting humans for tasks that LLM-based agents can perform, such as routine coding or simple bug fixing. Identifying these trends will help distinguish between roles that are still in demand and those where LLMs have reduced the need for human expertise.

\item \textit{Stakeholder Engagement}:
Engaging comprehensively with a diverse group of stakeholders is essential. This process validates the findings from the market analysis and ensures that the selected roles align with real-world needs.
It involves consulting industry professionals who have hands-on experience in the identified roles. This engagement can provide practical insights and challenges associated with these positions.
Collaboration with HR departments from leading technology companies is also important. It helps gather perspectives on current hiring trends, skill shortages, and the most sought-after competencies. Additionally, academic experts and researchers can offer forward-thinking views on emerging technologies and methodologies. By incorporating feedback from these various sources, the selection of key roles becomes more robust to reflect both current industry demands and future directions.

\item \textit{Value Addition Modeling}: The next crucial step is value addition modeling~\cite{mendes2018towards}, which evaluates the potential advantages that LLM-based agents could bring to each prioritized role. This process involves constructing detailed, data-driven models to analyze key performance indicators such as efficiency improvements, cost reductions, quality enhancements, and the acceleration of innovation resulting from the integration of agents. Pilot projects can be deployed to gather empirical data on these metrics when LLM-based agents are applied to specific tasks. Important factors to consider include the automation of repetitive tasks, the augmentation of human capabilities, and the inclusion of new functionalities that were previously unattainable.
It is important to note that the value added by LLM-based agents can differ significantly across different domains; for example, roles in software development may prioritize automation, whereas domains like systems architecture might see more value in LLMs augmenting complex decision-making around resource allocation or performance optimization, where human expertise and contextual understanding remain essential.
By quantifying these value propositions, organizations can allocate resources more strategically to roles where LLM-based agents are likely to yield the highest return on investment.

\end{enumerate}

The second step involves understanding the limitations of LLM-based agents relative to the demands of the identified SE roles:
\begin{enumerate} 

\item \textit{Competency Mapping}: Competency mapping~\cite{kaur2013competency} entails developing comprehensive competency frameworks for each specialized role. These frameworks define the essential skills, knowledge areas, and competencies required, encompassing both technical and soft skills. For instance, technical skills might encompass proficiency in specific programming languages, tools, methodologies, and domain-specific knowledge. For a machine learning engineer, this would include expertise in algorithms, data preprocessing, model training, and tools such as TensorFlow\footnote{\url{https://www.tensorflow.org/}} or PyTorch\footnote{\url{https://pytorch.org/}}. 
Soft skills include skills like problem-solving, critical thinking, and collaboration. Clearly outlining these competencies creates a benchmark against which the agents' abilities can be measured.

\item \textit{Performance Evaluation}: The next phase is performance evaluation, which involves designing or selecting tasks that closely replicate the real-world challenges associated with each role. These tasks should be practical and scenario-based to accurately gauge the agents' capabilities. They should assess a wide range of competencies, from technical execution to critical thinking. For example, in evaluating a DevOps engineer, the agent might be tasked with automating a deployment pipeline using tools like Jenkins\footnote{\url{https://www.jenkins.io/}} or Docker\footnote{\url{https://www.docker.com/}}, or troubleshooting a continuous integration failure. Such tasks allow for a thorough assessment of both technical and soft skills. 

\item \textit{Gap Analysis}: This step compares the agents' outputs with the expected outcomes for each task. Key areas where the agents underperform--such as misunderstanding domain-specific terminology, neglecting security best practices, or failing to optimize code--are identified and documented. This analysis emphasizes both the agents' strengths and weaknesses, offering valuable insights into recurring patterns of errors or misconceptions.

\item \textit{Expert Consultation and Iterative Refinement}: To further refine the evaluation process, expert consultation and iterative refinement are essential. By engaging with SE professionals who specialize in the assessed roles, qualitative feedback on the agent's performance can be obtained. 
These experts provide insights into subtle nuances that may not be captured through quantitative metrics. For instance, while the agent's code may work, it might not follow best practices or address scalability. This feedback helps refine evaluation methods, update competency frameworks, and uncover deeper issues in the agent's understanding.
\end{enumerate}

The final step involves tailoring the LLM-based agents to effectively represent the identified SE roles through specialized training and prompt engineering. :

\begin{enumerate} 
\item \textit{Curating Specialized Training Data:} At first, this involves creating training datasets that reflect the unique requirements of each specific role. A comprehensive corpus should be built from a variety of sources, including technical documentation such as API guides, technical manuals, and user guides to provide in-depth knowledge of specific technologies. It is also important to incorporate academic and industry research papers, case studies, and whitepapers to capture the latest developments, best practices, and theoretical foundations. 
Additionally, discussions from forums and software Q\&A sites like Stack Overflow\footnote{\url{https://stackoverflow.com/}}, Reddit\footnote{\url{https://www.reddit.com/}}, and specialized industry forums can provide practical problem-solving approaches and real-world challenges faced by professionals. 

\item \textit{Fine-tuning the LLM:} After preparing the data, the curated datasets are used to fine-tune the LLM-based agents. Advanced techniques like parameter-efficient fine-tuning (PEFT)~\cite{liu2022few} are often employed to optimize both efficiency and accuracy.

\item \textit{Designing Customized Prompts:} A key step is designing prompts tailored to improve the agents' role adaptability. These prompts should clearly define the role, tasks, and goals to ensure the agent understands the requirements. For instance, in a cybersecurity analyst role, the prompt should outline specific security protocols, potential vulnerabilities, and compliance standards. Contextual instructions, including relevant background, constraints, and examples, help the agent grasp task nuances. Creating a library of effective prompts for various scenarios can also serve as reusable templates for future tasks.

\item \textit{Continuous Learning and Adaptation:} 
To keep agents aligned with industry developments, continuous adaptation mechanisms are essential. Training data should be regularly updated, and models may be retrained to incorporate new technologies, best practices, and trends in software engineering. Monitoring systems can track agent performance over time, enabling proactive adjustments and continuous improvement. Additionally, agents should be guided to consistently reference the latest documentation and standards to ensure their outputs remain relevant and accurate.

\end{enumerate}

While LMA roles may overlap with traditional software engineering roles, it is important to recognize that they are not necessarily the same, as LMA roles often involve specialized, collaborative tasks suited for agent-based systems. By systematically identifying key roles, assessing agent competencies, and enhancing their capabilities through targeted fine-tuning, we aim to significantly improve the effectiveness of LLM-based agents in specialized SE roles.

\subsubsection{\textbf{Advancing Prompts through Agent-oriented Programming Paradigms}} \hfill

Effective prompts are crucial for the performance of LLM-based agents. However, creating such prompts is challenging due to the need for a framework that is versatile, effective, and robust across diverse scenarios.
Natural language, while flexible, often contains ambiguities and inconsistencies that LLMs may misinterpret. 
Natural language is inherently designed for human communication, where human communication relies on shared context and intuition that LLMs lack. In contrast, LLMs interpret text based on statistical patterns from large datasets, which may lead to different interpretations than those intended for humans~\cite{sun2024ai, zeng2022extensive}.
This highlights the need for a specialized prompting language designed to augment the cognitive functions of LLM-based agents and treats LLMs as the primary audience. Such a language can minimize ambiguities and ensure clear instructions, resulting in more reliable and accurate outputs.

\noindent \textbf{Current State.} Multiple prompting frameworks are released to facilitate the usage of LLMs. For example, DSPy~\cite{khattab2023dspy} and Vieira~\cite{li2024relational} enable fully automated generation of prompts.  AutoGen~\cite{wu2023autogen} and LangChain~\cite{mavroudis2024langchain} support retrieval-augmented generation (RAG)~\cite{gao2023retrieval} and agent-based workflows. 
However, these frameworks are still human-centered. They often prioritize human readability and developer convenience. As a result, there is a lack of research on a language that treats LLMs as the primary audience for prompts.

\noindent \textbf{Opportunities.} Agent-oriented programming (AOP)~\cite{shoham1993agent} offers a promising foundation for this approach. Similar to how Object-Oriented Programming (OOP)~\cite{wegner1990concepts} organize objects, AOP treats agents as fundamental units, focusing on their reasoning, objectives, and interactions. An AOP-based prompting language could enable the precise expression of complex tasks and constraints, allowing LLM-based agents to perform their roles with greater efficiency and accuracy. Extending this concept to Multi-Agent-Oriented Programming (MAOP)~\cite{boissier2020multi, bordini2009multi} allows for the creation of systems where multiple LLM-based agents can collaborate, communicate, and adapt to evolving contexts. By explicitly defining agent behaviors, communication patterns, and task hierarchies, we can reduce ambiguity, mitigate hallucinations, and improve task execution in LMA systems.

Further, such a prompting language must be expressive enough to handle diverse and complex tasks, yet simple enough for users to easily adopt. Conversely, overly simplified languages may lack the expressive power needed to represent complex software engineering workflows.
A complex language that introduce a steep learning curve due to their syntax, hinder adoption, especially for users who require simpler interfaces for prompt creation and modification. Balancing functionality and usability will be another key research question to its success. 

Additionally, this process may involve tailoring prompts specifically for different LLM models and their versions, as variations in model architectures, training data, and capabilities can affect how they interpret and respond to prompts. What works effectively for one model may not perform as well for another, necessitating careful adjustments. Current prompting languages lack mechanisms to easily adapt prompts across models, requiring manual adjustments and experimentation to achieve consistent performance.

While AOP-based prompting may not be the final solution, it represents an important step toward developing an AI-oriented language with grammar tailored specifically for LLMs. This new approach could further refine communication with LLM-based agents, reducing misinterpretation and significantly enhancing overall performance.

\subsection{Phase Two: Optimizing Agent Synergy}
In Phase Two, the spotlight turns towards optimizing agent synergy, underscoring the importance of collaboration and how to leverage the diverse strengths of individual agents.
This phase delves into both the internal dynamics among agents and the role of external human intervention in enhancing the efficacy of the LMA system.  Key research questions guiding this phase include:
\begin{enumerate}[leftmargin=*, start=3]
    \item \textit{How to best allocate tasks between humans and LLM-based agents?}
    \item \textit{How can we quantify the impact of agent collaboration on overall task performance and outcome quality?}
    \item \textit{How to scale LMA systems for large-scale projects? }
    \item \textit{What industrial organization mechanisms can be applied to LMA systems?}
    \item \textit{What strategies allow LMA systems to dynamically adjust their approach?}
    \item \textit{How to ensure security among private data sharing within LMA systems?}
\end{enumerate}

\subsubsection{\textbf{Human-Agent Collaboration}} \hfill

Optimally distributing tasks between humans and LMA agents to leverage their respective strengths is essential. Humans bring unparalleled creativity, critical thinking, ethical judgment, and domain-specific knowledge~\cite{markauskaite2022rethinking}. In contrast, LLM-based agents excel at rapidly processing large datasets, performing repetitive tasks with high accuracy, and detecting patterns that might elude human observers.

\noindent \textbf{Current State.} Several LMA systems incorporate human-in-the-loop designs. For instance, AISD~\cite{zhang2024experimenting} involves human input during requirement analysis and system validation, where users provide feedback on use cases, system designs, and prototypes. Similarly, MARE~\cite{jin2024mare} leverages human assessment to refine generated requirements and specifications.
Although these works demonstrate the feasibility of human contributions, key research questions, including optimizing human roles, enhancing feedback mechanisms, and identifying appropriate intervention points, are still underexplored.

\noindent \textbf{Opportunity.}
Developing role-specific guidelines that outline when and how human intervention should occur is essential. These guidelines should assist in identifying critical decision points where human judgment is indispensable, such as ethical considerations, conflict resolution, ambiguity handling, and creative problem-solving. For example, ethical decisions necessitate human oversight to ensure alignment with societal norms and values, and conflict resolution may require negotiation skills that LLM-based agents lack.

To facilitate seamless collaboration, designing intuitive user-friendly interfaces and interaction protocols is essential~\cite{wang2024mobileagentbench}. Natural language interfaces and adaptive visualization techniques can make interactions more accessible. These interfaces should efficiently present agent outputs in a digestible format and collect user feedback, while also managing the cognitive load on human collaborators.
It is important to note that these interfaces may need to be tailored differently for each human role, as the needs of a project manager, a software developer, and a quality assurance engineer will vary significantly.

Given the complexity of information generated during the agents' workflows~\cite{josifoski2023flows}, designing such interfaces poses challenges.
For instance, presenting modifications suggested by an agent at varying levels of abstraction ensures that each stakeholder can engage with the information at the right depth. A project manager might focus on the broader implications, such as the high-level impact on project timelines or deliverables, whereas a developer or architect might drill down into specific implementation details. Role-specific interfaces will be key to ensuring each stakeholder can effectively collaborate with the agents and extract the necessary information in a manner suited to their specific responsibilities.

Additionally, developing predictive models to determine the optimal human-to-agent ratio across different project types and stages is a fundamental concern. These models must assess factors such as project complexity, time constraints, project priorities, and the specific capabilities and limitations of both human participants and LMA agents. By doing so, tasks can be allocated in a manner that fully harnesses both human ingenuity and agent efficiency throughout the project.  Machine learning techniques could also be leveraged to analyze historical project data to predict effective collaboration strategies.

\subsubsection{\textbf{Evaluating the LMA systems}} \hfill

\noindent \textbf{Current State.} Numerous complex benchmarks have been proposed to challenge the capabilities of LLMs in critical aspects of software development, such as code generation~\cite{zhuo2024bigcodebench, jain2024livecodebench, liu2024your}. While these benchmarks have advanced the field by providing measurable metrics for individual tasks, their focus on isolated problem-solving reveals limitations as software engineering projects become more complex. Software engineering is inherently collaborative, with key activities like joint requirements gathering, code integration, and peer reviews playing essential roles in the process. Current benchmarks often overlook these aspects, failing to assess how well LLMs perform in tasks that require cooperation and collective decision-making.

\noindent \textbf{Opportunities.} There is a growing need for benchmarks that evaluate the cooperative abilities of LLMs in multi-agent settings, particularly for software engineering tasks. These benchmarks should simulate real-world collaborative scenarios where LLM agents work together to achieve common development goals.

Such benchmarks should include tasks where agents must:
\begin{enumerate} 
\item \textit{Participate in collaborative design}: Agents should contribute ideas, propose design solutions, and converge on a unified architecture that balances trade-offs.
\item \textit{Delegate and coordinate tasks}: Effective task division is crucial. Agents should assign responsibilities based on expertise, manage dependencies, and adjust as the project evolves.
\item \textit{Identify conflicts and negotiate}: In collaborative settings, disagreements are inevitable. First, LLMs often struggle to identify conflicts in real time unless explicitly guided to do so~\cite{abdelnabi2023llm}.
Therefore, agents should be evaluated on their ability to recognize these conflicts—whether in logic, goals, or execution. 
Moreover, agents should be tested on their ability to handle conflicts constructively. This includes proposing compromises, engaging in constructive negotiation, and ensuring that the team remains aligned with the overarching objectives. Evaluations should focus on the agents' capacity to balance competing priorities, mitigate misunderstandings, and foster consensus, all while maintaining progress toward shared goals.
\item \textit{Integrate components and perform peer reviews}: Agents should seamlessly integrate their work, review each other's code for quality assurance, and provide constructive feedback.
\item \textit{Proactive Clarification Request
}: Agents should not assume complete understanding when uncertainty arises. Instead, they should preemptively ask for additional information or clarification to avoid potential errors or misunderstandings. Evaluating agents on this ability ensures they are capable of identifying gaps in their knowledge or instructions and can actively seek out the necessary context or data to complete tasks effectively.
\end{enumerate}

To develop such benchmarks, we need to create realistic project scenarios that require multi-agent collaboration over extended periods. These scenarios should reflect common software development challenges, such as evolving requirements and tight deadlines. Additionally, platforms or sandboxes must be built to provide controlled environments where collaborative interactions between agents can be observed and measured. These platforms should establish clear interaction rules, including languages, formats, and communication channels, to facilitate effective information exchange.

Most importantly, comprehensive metrics must be developed to assess not just the final output, but also the collaboration process itself. These metrics could measure communication efficiency, ambiguity resolution, conflict management, adherence to best practices, and overall project success.

\subsubsection{\textbf{Scaling Up for Complex Projects.} } \hfill

 As software projects become more complex, a single LLM-based agent may hit its performance limits. Inspired by the scaling properties of neural models~\cite{bahri2024explaining}, LMA systems can potentially enhance performance by increasing the number of agents within the system. While adding more agents can provide some benefits, handling more complex projects introduces challenges that require more refined solutions.

 \noindent \textbf{Current State.} Existing LMA systems face significant challenges when scaling up to handle complex software projects. Our case studies illustrate these limitations clearly. For instance, ChatDev was unable to autonomously develop a functional Tetris game. In real-world projects with higher complexity, this limitation becomes even more pronounced.

\noindent \textbf{Opportunities.} First, as software projects grow in size and complexity, breaking down high-level requirements into manageable sub-tasks becomes more difficult. It is not just about handling more tasks but also managing the intricate interdependencies between them. A hierarchical task decomposition approach can help, where higher-level agents oversee broader objectives and delegate specific tasks to lower-level agents. This structure streamlines planning and makes global task allocation more efficient.

Second, as the number of agents increases, so does the complexity of communication. Coordinating multiple agents can lead to communication bottlenecks and information overload. Additionally, large-scale software projects challenge the memory capacity of individual agents, making it harder to store and process the extensive information required. Efficient communication protocols and message prioritization are crucial to mitigating these issues. For instance, agents can use summarized updates instead of detailed reports, reducing communication overhead and memory usage. From the outset, the system should be designed with scalability in mind, ensuring that both software and hardware resources can expand efficiently as the number of agents increases.

Moreover, with more agents comes the risk of inconsistencies and conflicts in the shared information. A centralized knowledge repository or shared blackboard system can ensure that all agents have access to consistent, up-to-date information, acting as a single source of truth and minimizing the spread of misinformation. Robust error handling mechanisms should also be implemented to detect and correct issues autonomously before they escalate into significant failures.

Finally, as the number of agents grows, so do the rounds of discussion and decision-making, which can slow down progress. To avoid this, decision-making hierarchies or consensus algorithms can streamline the process. For example, only a subset of agents responsible for a specific module may need to reach a consensus, rather than involving the entire agent network.

% To address these challenges, innovative solutions in architectural design are required. 
% These solutions should focus on facilitating efficient and reliable information exchange within the expanded LMA systems, ensuring the timely and effective completion of tasks. Streamlining communication and coordination among a growing number of agents, while managing the increased computational demands, are key to scaling LMA systems for complex software projects effectively.

\subsubsection{\textbf{Leveraging Industry Principles}} \hfill

As LLM-based agents can closely mimic human developers in SE tasks, they can greatly benefit from adopting established industry principles and management strategies. By emulating organizational frameworks used by successful companies, LMA systems can improve their design and optimization processes. These industrial mechanisms enable LMA systems to remain agile, efficient, and effective, even as project complexities grow. 

\noindent \textbf{Current State.} As we described in Section \ref{sec:review}, numerous works~\cite{qian-etal-2024-chatdev, hong2023metagpt, nguyen2024agilecoder, al2020agile} are designed using popular process models like the Waterfall and Agile. 
For example, ChatDev~\cite{qian-etal-2024-chatdev} emulates a traditional Waterfall approach, breaking tasks into distinct phases (e.g., requirement analysis, design, implementation, testing), with agents dedicated to each phase. AgileCoder~\cite{nguyen2024agilecoder} incorporates the Agile methodologies, leveraging iterative development, continuous feedback loops, and collaborative sprints.

\noindent \textbf{Opportunities.} However, current LMA systems often do not leverage more specialized and modern industry practices, such as Value Stream Mapping, Design Thinking, or Model-Based Systems Engineering (MBSE). Additionally, frameworks like Domain-Driven Design (DDD), Behavior-Driven Development (BDD), and Team Topologies remain underutilized. These methodologies emphasize aligning development with business goals, improving user-centric design, and optimizing team structures—key components that could further enhance the efficiency, adaptability, and effectiveness of LMA systems.

Leadership and governance structures from industrial organizations provide valuable insights for designing LMA systems. Project management tools and practices, essential for coordinating large development teams, can be applied to LMA systems to enhance their operational efficiency. Using established project management frameworks, LMA systems can monitor progress, allocate resources, and manage timelines effectively. Agents can dynamically update task boards, report milestones, and adjust workloads in real-time based on project data. This not only improves transparency but also allows for early detection of bottlenecks or delays, ensuring projects stay on track.

Incorporating design patterns and software architecture best practices further strengthens LMA systems~\cite{lee2024unified}. By adhering to these principles, agents can produce well-structured, maintainable code that is scalable and reusable. This reduces technical debt and ensures that the solutions developed by LMA systems are easier to integrate, maintain, and expand in the future.

\subsubsection{\textbf{Dynamic Adaptation. }} \hfill

In the context of software development, predicting the optimal configuration for LMA systems at the outset is unrealistic due to the inherent complexity and variability of tasks~\cite{leffingwell2000managing}. The dynamic nature of software requirements and the unpredictable challenges that arise during development necessitate systems that can adapt on the fly~\cite{liu2023dynamic}. For example, a sudden shift in project requirements or unexpected delays caused by dependencies on external components. Therefore, LMA systems must be capable of dynamically adjusting their scale, strategies, and structures throughout the development process.

\noindent \textbf{Current State.} Most existing LMA systems~\cite{qian-etal-2024-chatdev, hong2023metagpt} operate with static architectures characterized by fixed agent roles and predefined communication patterns. Recent research efforts~\cite{liu2024dynamic,zhang2024g} have introduced mechanisms for adaptive agent team selection and task-specific collaboration strategies. These methods enable the selection of suitable agent team configurations for specific tasks, however, they still fall short of true dynamic adaptation and lack the capability to adjust to real-time changes. To the best of our knowledge, no previous work addresses the need for on-the-fly adjustments in response to evolving project demands.

\noindent \textbf{Opportunities.} To minimize redundant work, LMA systems should continuously evaluate existing solutions~\cite{wang2024benchmark}, identifying reusable elements for new requirements. By learning from each development cycle, the system can recognize patterns of efficiency and inefficiency, enabling it to make informed decisions when handling similar tasks in the future or adapting existing solutions to new requirements.

A key element of dynamic adaptation is the ability to automatically adjust the number of agents involved in a project~\cite{guo2024large}. This includes not only scaling the number of agents up or down as needed but also generating new agents with new specialized roles to meet emerging task requirements, ensuring both efficiency and responsiveness. Additionally, the system can replicate agents in existing roles to manage increased workloads. Furthermore, LMA systems can generate new agents that come equipped with contextual knowledge of the project—such as its history, current state, and objectives—by accessing shared knowledge bases, project documentation, and recent communications. This allows new agents to integrate smoothly and contribute effectively right from the start, reducing onboarding time and minimizing disruptions.

Another key component is the dynamic redefinition of agent roles~\cite{jouvin2002role}. As the project evolves, certain roles may become obsolete while new ones emerge. LMA systems should be capable of reassigning roles to agents or modifying their responsibilities to better align with current project needs. This flexibility enhances the system's ability to adapt to changing requirements and priorities.

Dynamic adaptation also involves the reallocation of memory and computational resources. As agents are added or removed and tasks shift in complexity, the system must efficiently distribute resources to where they are most needed. This may include scaling computational power for agents handling intensive tasks or increasing memory allocation for agents processing large datasets. Effective resource management ensures that the system operates optimally without unnecessary strain on infrastructure.

Finally, the uncertainty of the software development process makes it challenging to define effective termination conditions~\cite{smith2020proactive}. Relying solely on predefined criteria may result in infinite loops or premature task completion. To address this, LMA systems must incorporate real-time monitoring and feedback loops to continuously evaluate progress. Machine learning techniques can help predict optimal stopping points by analyzing historical data and current performance metrics, allowing for informed adjustments to task completion criteria as the project evolves.

\subsubsection{\textbf{Privacy and Partial Information.} } \hfill

In multi-organizational software development projects, data often resides in silos due to privacy concerns, proprietary restrictions, and regulatory compliance requirements~\cite{paasivaara2008distributed}. Each entity may have its own data governance policies and competitive considerations that limit data sharing. This fragmentation poses significant challenges in enabling agents to access necessary information while ensuring that privacy is maintained. Moreover, a lack of transparency in data sources and processes can exacerbate the risk of privacy violations, which may go unnoticed if data handling activities are not fully visible to all parties~\cite{crawford2014big}.

\noindent \textbf{Current State.} The challenge of ensuring privacy while managing partial information has been extensively studied in the field of computer security. ~\cite{dinur2003revealing,bennett1988privacy}. To the best of our knowledge, existing research has yet to provide a solution to these challenges for LMA systems in SE.

\noindent \textbf{Opportunities.} To address these challenges, robust and fine-grained access control mechanisms must be implemented across organizational boundaries. It is essential to prevent unauthorized access while still accommodating the varied data access needs of the system. Traditional models like Role-Based Access Control (RBAC)~\cite{sandhu1998role} and Attribute-Based Access Control (ABAC)~\cite{hu2015attribute} may need to be extended to handle the dynamic nature of multi-agent systems effectively. Establishing protocols that allow agents to share insights derived from sensitive data, without exposing the data itself, is critical. Advanced privacy-preserving techniques like Differential Privacy~\cite{dwork2006differential}, Secure Multi-Party Computation (SMPC)~\cite{goldreich1998secure}, Federated Learning~\cite{kairouz2021advances}, or Homomorphic Encryption~\cite{yi2014homomorphic} can be leveraged to ensure that agents collaborate without compromising data privacy.

Moreover, compliance with data protection laws such as the General Data Protection Regulation (GDPR)~\cite{voigt2017eu} in the EU and the California Consumer Privacy Act (CCPA)~\cite{pardau2018california} in the U.S. is crucial. LMA systems should follow privacy-by-design principles, ensuring that data subjects' rights are upheld, and that data processing activities remain transparent and lawful. This includes implementing mechanisms for data minimization, consent management, and honoring the right to be forgotten.

For non-sensitive data, integrated data storage solutions can reduce redundancy, improve data consistency, and increase efficiency. This can be achieved through distributed databases accessible to authorized agents, along with data synchronization mechanisms to ensure agents have up-to-date information in real time. Additionally, using technologies like blockchain~\cite{zheng2018blockchain} and distributed ledgers~\cite{sunyaev2020distributed} can enhance transparency, traceability, and tamper-resistance in recording agent transactions and data access events, fostering greater trust among collaborating entities.

\section{Discussion}
\label{sec:discuss}

\subsection{A Comparison with the Mixture of Experts Paradigm}
Another paradigm that has recently attracted much attention from both academia and industry is the Mixture of Experts (MoE) paradigm~\cite{zhu2024llama, cai2024survey}. MoE organizes an LLM into multiple specialized components known as ``experts." Each expert is designed to focus on specialized tasks. Further, a gating mechanism is employed to dynamically activate the most relevant subset of experts based on the input. While MoE is promising, LMA systems offer several distinct advantages:

One limitation of MoE is its high resource consumption. MoE models contain multiple experts within a single architecture, which makes the total number of parameters rather huge.
Furthermore, training MoE is more resource-intensive and time-consuming than standard LLMs.
This is mainly due to the complex training process for the gating mechanism. Training the gating mechanism involves optimizing the selection process for the most relevant experts, which adds considerable overhead.

Since specific experts are dynamically activated based on input, MoE can be viewed as a method to learn the internal routing of LLMs. However, there is no interaction and communication between experts in MoE. On the other hand, LMA systems usually are designed to resemble real-world collaborative workflows. Agents in LMA systems can actively communicate with each other, exchange information, and iteratively refine the output based on feedback from other agents. More importantly, LMA systems can also integrate external feedback from tools such as compilers, static analyzers, or testing frameworks. 
LMA systems also facilitate seamless and continuous human-in-the-loop collaboration, enabling human experts to intervene, validate outputs, and provide guidance at any stage of the process. As a result, we consider LMA systems to be a more appropriate approach to MoE to address the multifaceted challenges of software engineering.

\subsection{Threat to Validity}
One potential threat to validity lies in the possibility of inadvertently excluding relevant studies during the literature search and selection process. To mitigate this risk, we conducted a comprehensive search on the DBLP database, ensuring coverage of a broad spectrum of studies, including preprints. Additionally, we enhanced the search process by combining automated querying with forward and backward snowballing, aiming to identify and include all pertinent studies.

\section{Conclusion and Future Work}
This paper explores the evolving role of LMA systems in shaping the future of Software Engineering 2.0~\cite{lo2023trustworthy}. To support this vision, we first present a systematic review of recent applications of LMA systems across different stages of the software development lifecycle. Our review highlights key advancements in areas such as requirements engineering, code generation, software quality assurance, and maintenance.
To further understand the current landscape, we conduct two case studies that illustrate the practical uses and challenges of LMA systems. Based on these insights, we propose a structured research agenda aimed at advancing LMA integration in software engineering. Future work will focus on addressing critical research questions to enhance LMA capabilities and optimize their synergy with software development processes.

In the immediate term, efforts will be dedicated to enhancing the capabilities of LLM-based agents in representing specialized SE roles. This will involve creating specialized datasets and pre-training tasks that mirror the complex realities of SE tasks. Additionally, there will be a focus on formulating advanced prompting strategies, which can refine the agents' cognitive functions and decision-making skills. Looking towards the longer-term objectives, the emphasis will transition towards optimizing the synergy between agents. The initial step involves examining optimal strategies for task allocation between humans and LLM-based agents, capitalizing on the unique strengths of both entities. 
Further, we need to develop scalable methodologies for LMA systems, which would enable them to orchestrate and complete large-scale, multifaceted SE projects efficiently. 
Moreover, ensuring the privacy and confidentiality of data within LMA systems is also critical. This entails exploring data management and access control mechanisms to protect sensitive information while still enabling the essential exchange of insights among project stakeholders.
By systematically exploring these research questions, we aim to drive innovation in LMA systems for software engineering and create a more cohesive, effective, and flexible LMA-driven development process.

\balance
\bibliographystyle{ACM-Reference-Format}
\bibliography{reference}


\begin{thebibliography}{171}

%%% the Web address.

\ifx \showCODEN    \undefined \def \showCODEN     #1{\unskip}     \fi
\ifx \showDOI      \undefined \def \showDOI       #1{#1}\fi
\ifx \showISBNx    \undefined \def \showISBNx     #1{\unskip}     \fi
\ifx \showISBNxiii \undefined \def \showISBNxiii  #1{\unskip}     \fi
\ifx \showISSN     \undefined \def \showISSN      #1{\unskip}     \fi
\ifx \showLCCN     \undefined \def \showLCCN      #1{\unskip}     \fi
\ifx \shownote     \undefined \def \shownote      #1{#1}          \fi
\ifx \showarticletitle \undefined \def \showarticletitle #1{#1}   \fi
\ifx \showURL      \undefined \def \showURL       {\relax}        \fi
% invisible to TeX
\providecommand\bibfield[2]{#2}
\providecommand\bibinfo[2]{#2}
\providecommand\natexlab[1]{#1}
\providecommand\showeprint[2][]{arXiv:#2}

\bibitem[Abdelnabi et~al\mbox{.}(2023)]%
        {abdelnabi2023llm}
\bibfield{author}{\bibinfo{person}{Sahar Abdelnabi}, \bibinfo{person}{Amr Gomaa}, \bibinfo{person}{Sarath Sivaprasad}, \bibinfo{person}{Lea Sch{\"o}nherr}, {and} \bibinfo{person}{Mario Fritz}.} \bibinfo{year}{2023}\natexlab{}.
\newblock \showarticletitle{Llm-deliberation: Evaluating llms with interactive multi-agent negotiation games}.
\newblock \bibinfo{journal}{\emph{arXiv preprint arXiv:2309.17234}} (\bibinfo{year}{2023}).
\newblock


\bibitem[Abrahamsson et~al\mbox{.}(2017)]%
        {abrahamsson2017agile}
\bibfield{author}{\bibinfo{person}{Pekka Abrahamsson}, \bibinfo{person}{Outi Salo}, \bibinfo{person}{Jussi Ronkainen}, {and} \bibinfo{person}{Juhani Warsta}.} \bibinfo{year}{2017}\natexlab{}.
\newblock \showarticletitle{Agile software development methods: Review and analysis}.
\newblock \bibinfo{journal}{\emph{arXiv preprint arXiv:1709.08439}} (\bibinfo{year}{2017}).
\newblock


\bibitem[Achiam et~al\mbox{.}(2023)]%
        {gpt4}
\bibfield{author}{\bibinfo{person}{Josh Achiam}, \bibinfo{person}{Steven Adler}, \bibinfo{person}{Sandhini Agarwal}, \bibinfo{person}{Lama Ahmad}, \bibinfo{person}{Ilge Akkaya}, \bibinfo{person}{Florencia~Leoni Aleman}, \bibinfo{person}{Diogo Almeida}, \bibinfo{person}{Janko Altenschmidt}, \bibinfo{person}{Sam Altman}, \bibinfo{person}{Shyamal Anadkat}, {et~al\mbox{.}}} \bibinfo{year}{2023}\natexlab{}.
\newblock \showarticletitle{Gpt-4 technical report}.
\newblock \bibinfo{journal}{\emph{arXiv preprint arXiv:2303.08774}} (\bibinfo{year}{2023}).
\newblock


\bibitem[Agashe(2023)]%
        {agashe2023llm}
\bibfield{author}{\bibinfo{person}{Saaket Agashe}.} \bibinfo{year}{2023}\natexlab{}.
\newblock \bibinfo{booktitle}{\emph{LLM-Coordination: Developing Coordinating Agents with Large Language Models}}.
\newblock \bibinfo{publisher}{University of California, Santa Cruz}.
\newblock


\bibitem[Al-Saqqa et~al\mbox{.}(2020)]%
        {al2020agile}
\bibfield{author}{\bibinfo{person}{Samar Al-Saqqa}, \bibinfo{person}{Samer Sawalha}, {and} \bibinfo{person}{Hiba AbdelNabi}.} \bibinfo{year}{2020}\natexlab{}.
\newblock \showarticletitle{Agile software development: Methodologies and trends.}
\newblock \bibinfo{journal}{\emph{International Journal of Interactive Mobile Technologies}} \bibinfo{volume}{14}, \bibinfo{number}{11} (\bibinfo{year}{2020}).
\newblock


\bibitem[Albrecht and Stone(2018)]%
        {albrecht2018autonomous}
\bibfield{author}{\bibinfo{person}{Stefano~V Albrecht} {and} \bibinfo{person}{Peter Stone}.} \bibinfo{year}{2018}\natexlab{}.
\newblock \showarticletitle{Autonomous agents modelling other agents: A comprehensive survey and open problems}.
\newblock \bibinfo{journal}{\emph{Artificial Intelligence}}  \bibinfo{volume}{258} (\bibinfo{year}{2018}), \bibinfo{pages}{66--95}.
\newblock


\bibitem[Antoniades et~al\mbox{.}(2024)]%
        {antoniades2024swe}
\bibfield{author}{\bibinfo{person}{Antonis Antoniades}, \bibinfo{person}{Albert {\"O}rwall}, \bibinfo{person}{Kexun Zhang}, \bibinfo{person}{Yuxi Xie}, \bibinfo{person}{Anirudh Goyal}, {and} \bibinfo{person}{William Wang}.} \bibinfo{year}{2024}\natexlab{}.
\newblock \showarticletitle{SWE-Search: Enhancing Software Agents with Monte Carlo Tree Search and Iterative Refinement}.
\newblock \bibinfo{journal}{\emph{arXiv preprint arXiv:2410.20285}} (\bibinfo{year}{2024}).
\newblock


\bibitem[Arora et~al\mbox{.}(2024)]%
        {arora2024masai}
\bibfield{author}{\bibinfo{person}{Daman Arora}, \bibinfo{person}{Atharv Sonwane}, \bibinfo{person}{Nalin Wadhwa}, \bibinfo{person}{Abhav Mehrotra}, \bibinfo{person}{Saiteja Utpala}, \bibinfo{person}{Ramakrishna Bairi}, \bibinfo{person}{Aditya Kanade}, {and} \bibinfo{person}{Nagarajan Natarajan}.} \bibinfo{year}{2024}\natexlab{}.
\newblock \showarticletitle{MASAI: Modular Architecture for Software-engineering AI Agents}.
\newblock \bibinfo{journal}{\emph{arXiv preprint arXiv:2406.11638}} (\bibinfo{year}{2024}).
\newblock


\bibitem[Ataei et~al\mbox{.}(2024)]%
        {ataei2024elicitron}
\bibfield{author}{\bibinfo{person}{Mohammadmehdi Ataei}, \bibinfo{person}{Hyunmin Cheong}, \bibinfo{person}{Daniele Grandi}, \bibinfo{person}{Ye Wang}, \bibinfo{person}{Nigel Morris}, {and} \bibinfo{person}{Alexander Tessier}.} \bibinfo{year}{2024}\natexlab{}.
\newblock \showarticletitle{Elicitron: An LLM Agent-Based Simulation Framework for Design Requirements Elicitation}.
\newblock \bibinfo{journal}{\emph{arXiv preprint arXiv:2404.16045}} (\bibinfo{year}{2024}).
\newblock


\bibitem[Bahri et~al\mbox{.}(2024)]%
        {bahri2024explaining}
\bibfield{author}{\bibinfo{person}{Yasaman Bahri}, \bibinfo{person}{Ethan Dyer}, \bibinfo{person}{Jared Kaplan}, \bibinfo{person}{Jaehoon Lee}, {and} \bibinfo{person}{Utkarsh Sharma}.} \bibinfo{year}{2024}\natexlab{}.
\newblock \showarticletitle{Explaining neural scaling laws}.
\newblock \bibinfo{journal}{\emph{Proceedings of the National Academy of Sciences}} \bibinfo{volume}{121}, \bibinfo{number}{27} (\bibinfo{year}{2024}), \bibinfo{pages}{e2311878121}.
\newblock


\bibitem[Bennett et~al\mbox{.}(1988)]%
        {bennett1988privacy}
\bibfield{author}{\bibinfo{person}{Charles~H Bennett}, \bibinfo{person}{Gilles Brassard}, {and} \bibinfo{person}{Jean-Marc Robert}.} \bibinfo{year}{1988}\natexlab{}.
\newblock \showarticletitle{Privacy amplification by public discussion}.
\newblock \bibinfo{journal}{\emph{SIAM journal on Computing}} \bibinfo{volume}{17}, \bibinfo{number}{2} (\bibinfo{year}{1988}), \bibinfo{pages}{210--229}.
\newblock


\bibitem[Bettenburg et~al\mbox{.}(2008)]%
        {bettenburg2008makes}
\bibfield{author}{\bibinfo{person}{Nicolas Bettenburg}, \bibinfo{person}{Sascha Just}, \bibinfo{person}{Adrian Schr{\"o}ter}, \bibinfo{person}{Cathrin Weiss}, \bibinfo{person}{Rahul Premraj}, {and} \bibinfo{person}{Thomas Zimmermann}.} \bibinfo{year}{2008}\natexlab{}.
\newblock \showarticletitle{What makes a good bug report?}. In \bibinfo{booktitle}{\emph{Proceedings of the 16th ACM SIGSOFT International Symposium on Foundations of software engineering}}. \bibinfo{pages}{308--318}.
\newblock


\bibitem[Boissier et~al\mbox{.}(2020)]%
        {boissier2020multi}
\bibfield{author}{\bibinfo{person}{Olivier Boissier}, \bibinfo{person}{Rafael~H Bordini}, \bibinfo{person}{Jomi Hubner}, {and} \bibinfo{person}{Alessandro Ricci}.} \bibinfo{year}{2020}\natexlab{}.
\newblock \bibinfo{booktitle}{\emph{Multi-agent oriented programming: programming multi-agent systems using JaCaMo}}.
\newblock \bibinfo{publisher}{Mit Press}.
\newblock


\bibitem[Bordini et~al\mbox{.}(2009)]%
        {bordini2009multi}
\bibfield{author}{\bibinfo{person}{Rafael~H Bordini}, \bibinfo{person}{Mehdi Dastani}, \bibinfo{person}{J{\"u}rgen Dix}, {and} \bibinfo{person}{Amal El~Fallah Seghrouchni}.} \bibinfo{year}{2009}\natexlab{}.
\newblock \bibinfo{booktitle}{\emph{Multi-agent programming}}.
\newblock \bibinfo{publisher}{Springer}.
\newblock


\bibitem[Budinsky et~al\mbox{.}(1996)]%
        {budinsky1996automatic}
\bibfield{author}{\bibinfo{person}{Frank~J. Budinsky}, \bibinfo{person}{Marilyn~A. Finnie}, \bibinfo{person}{John~M. Vlissides}, {and} \bibinfo{person}{Patsy~S. Yu}.} \bibinfo{year}{1996}\natexlab{}.
\newblock \showarticletitle{Automatic code generation from design patterns}.
\newblock \bibinfo{journal}{\emph{IBM systems Journal}} \bibinfo{volume}{35}, \bibinfo{number}{2} (\bibinfo{year}{1996}), \bibinfo{pages}{151--171}.
\newblock


\bibitem[Cai et~al\mbox{.}(2024)]%
        {cai2024survey}
\bibfield{author}{\bibinfo{person}{Weilin Cai}, \bibinfo{person}{Juyong Jiang}, \bibinfo{person}{Fan Wang}, \bibinfo{person}{Jing Tang}, \bibinfo{person}{Sunghun Kim}, {and} \bibinfo{person}{Jiayi Huang}.} \bibinfo{year}{2024}\natexlab{}.
\newblock \showarticletitle{A survey on mixture of experts}.
\newblock \bibinfo{journal}{\emph{Authorea Preprints}} (\bibinfo{year}{2024}).
\newblock


\bibitem[Cai et~al\mbox{.}(2023)]%
        {cai2023low}
\bibfield{author}{\bibinfo{person}{Yuzhe Cai}, \bibinfo{person}{Shaoguang Mao}, \bibinfo{person}{Wenshan Wu}, \bibinfo{person}{Zehua Wang}, \bibinfo{person}{Yaobo Liang}, \bibinfo{person}{Tao Ge}, \bibinfo{person}{Chenfei Wu}, \bibinfo{person}{Wang You}, \bibinfo{person}{Ting Song}, \bibinfo{person}{Yan Xia}, {et~al\mbox{.}}} \bibinfo{year}{2023}\natexlab{}.
\newblock \showarticletitle{Low-code llm: Visual programming over llms}.
\newblock \bibinfo{journal}{\emph{arXiv preprint arXiv:2304.08103}}  \bibinfo{volume}{2} (\bibinfo{year}{2023}).
\newblock


\bibitem[Chen et~al\mbox{.}(2024b)]%
        {chen2023chatgpt}
\bibfield{author}{\bibinfo{person}{Chong Chen}, \bibinfo{person}{Jianzhong Su}, \bibinfo{person}{Jiachi Chen}, \bibinfo{person}{Yanlin Wang}, \bibinfo{person}{Tingting Bi}, \bibinfo{person}{Jianxing Yu}, \bibinfo{person}{Yanli Wang}, \bibinfo{person}{Xingwei Lin}, \bibinfo{person}{Ting Chen}, {and} \bibinfo{person}{Zibin Zheng}.} \bibinfo{year}{2024}\natexlab{b}.
\newblock \showarticletitle{When ChatGPT Meets Smart Contract Vulnerability Detection: How Far Are We?}
\newblock \bibinfo{journal}{\emph{ACM Trans. Softw. Eng. Methodol.}} (\bibinfo{date}{Nov.} \bibinfo{year}{2024}).
\newblock
\showISSN{1049-331X}
\urldef\tempurl%
\url{https://doi.org/10.1145/3702973}
\showDOI{\tempurl}
\newblock
\shownote{Just Accepted}.


\bibitem[Chen et~al\mbox{.}(2024a)]%
        {chen2024coder}
\bibfield{author}{\bibinfo{person}{Dong Chen}, \bibinfo{person}{Shaoxin Lin}, \bibinfo{person}{Muhan Zeng}, \bibinfo{person}{Daoguang Zan}, \bibinfo{person}{Jian-Gang Wang}, \bibinfo{person}{Anton Cheshkov}, \bibinfo{person}{Jun Sun}, \bibinfo{person}{Hao Yu}, \bibinfo{person}{Guoliang Dong}, \bibinfo{person}{Artem Aliev}, {et~al\mbox{.}}} \bibinfo{year}{2024}\natexlab{a}.
\newblock \showarticletitle{CodeR: Issue Resolving with Multi-Agent and Task Graphs}.
\newblock \bibinfo{journal}{\emph{arXiv preprint arXiv:2406.01304}} (\bibinfo{year}{2024}).
\newblock


\bibitem[Chen et~al\mbox{.}(2020)]%
        {chen2020survey}
\bibfield{author}{\bibinfo{person}{Junjie Chen}, \bibinfo{person}{Jibesh Patra}, \bibinfo{person}{Michael Pradel}, \bibinfo{person}{Yingfei Xiong}, \bibinfo{person}{Hongyu Zhang}, \bibinfo{person}{Dan Hao}, {and} \bibinfo{person}{Lu Zhang}.} \bibinfo{year}{2020}\natexlab{}.
\newblock \showarticletitle{A survey of compiler testing}.
\newblock \bibinfo{journal}{\emph{ACM Computing Surveys (CSUR)}} \bibinfo{volume}{53}, \bibinfo{number}{1} (\bibinfo{year}{2020}), \bibinfo{pages}{1--36}.
\newblock


\bibitem[Chen et~al\mbox{.}(2023b)]%
        {chen2023agentverse}
\bibfield{author}{\bibinfo{person}{Weize Chen}, \bibinfo{person}{Yusheng Su}, \bibinfo{person}{Jingwei Zuo}, \bibinfo{person}{Cheng Yang}, \bibinfo{person}{Chenfei Yuan}, \bibinfo{person}{Chi-Min Chan}, \bibinfo{person}{Heyang Yu}, \bibinfo{person}{Yaxi Lu}, \bibinfo{person}{Yi-Hsin Hung}, \bibinfo{person}{Chen Qian}, {et~al\mbox{.}}} \bibinfo{year}{2023}\natexlab{b}.
\newblock \showarticletitle{Agentverse: Facilitating multi-agent collaboration and exploring emergent behaviors}. In \bibinfo{booktitle}{\emph{The Twelfth International Conference on Learning Representations}}.
\newblock


\bibitem[Chen et~al\mbox{.}(2023a)]%
        {chen2023scalable}
\bibfield{author}{\bibinfo{person}{Yongchao Chen}, \bibinfo{person}{Jacob Arkin}, \bibinfo{person}{Yang Zhang}, \bibinfo{person}{Nicholas Roy}, {and} \bibinfo{person}{Chuchu Fan}.} \bibinfo{year}{2023}\natexlab{a}.
\newblock \showarticletitle{Scalable multi-robot collaboration with large language models: Centralized or decentralized systems?}
\newblock \bibinfo{journal}{\emph{arXiv preprint arXiv:2309.15943}} (\bibinfo{year}{2023}).
\newblock


\bibitem[Chen et~al\mbox{.}(2024c)]%
        {chen2024fairness}
\bibfield{author}{\bibinfo{person}{Zhenpeng Chen}, \bibinfo{person}{Jie~M Zhang}, \bibinfo{person}{Max Hort}, \bibinfo{person}{Mark Harman}, {and} \bibinfo{person}{Federica Sarro}.} \bibinfo{year}{2024}\natexlab{c}.
\newblock \showarticletitle{Fairness testing: A comprehensive survey and analysis of trends}.
\newblock \bibinfo{journal}{\emph{ACM Transactions on Software Engineering and Methodology}} \bibinfo{volume}{33}, \bibinfo{number}{5} (\bibinfo{year}{2024}), \bibinfo{pages}{1--59}.
\newblock


\bibitem[Cheng et~al\mbox{.}(2024)]%
        {cheng2024exploring}
\bibfield{author}{\bibinfo{person}{Yuheng Cheng}, \bibinfo{person}{Ceyao Zhang}, \bibinfo{person}{Zhengwen Zhang}, \bibinfo{person}{Xiangrui Meng}, \bibinfo{person}{Sirui Hong}, \bibinfo{person}{Wenhao Li}, \bibinfo{person}{Zihao Wang}, \bibinfo{person}{Zekai Wang}, \bibinfo{person}{Feng Yin}, \bibinfo{person}{Junhua Zhao}, {et~al\mbox{.}}} \bibinfo{year}{2024}\natexlab{}.
\newblock \showarticletitle{Exploring Large Language Model based Intelligent Agents: Definitions, Methods, and Prospects}.
\newblock \bibinfo{journal}{\emph{arXiv preprint arXiv:2401.03428}} (\bibinfo{year}{2024}).
\newblock


\bibitem[Christel and Kang(1992)]%
        {christel1992issues}
\bibfield{author}{\bibinfo{person}{Michael~G Christel} {and} \bibinfo{person}{Kyo~C Kang}.} \bibinfo{year}{1992}\natexlab{}.
\newblock \bibinfo{title}{Issues in requirements elicitation}.
\newblock
\newblock


\bibitem[Cohen et~al\mbox{.}(2004)]%
        {cohen2004introduction}
\bibfield{author}{\bibinfo{person}{David Cohen}, \bibinfo{person}{Mikael Lindvall}, {and} \bibinfo{person}{Patricia Costa}.} \bibinfo{year}{2004}\natexlab{}.
\newblock \showarticletitle{An introduction to agile methods.}
\newblock \bibinfo{journal}{\emph{Adv. Comput.}} \bibinfo{volume}{62}, \bibinfo{number}{03} (\bibinfo{year}{2004}), \bibinfo{pages}{1--66}.
\newblock


\bibitem[Crawford and Schultz(2014)]%
        {crawford2014big}
\bibfield{author}{\bibinfo{person}{Kate Crawford} {and} \bibinfo{person}{Jason Schultz}.} \bibinfo{year}{2014}\natexlab{}.
\newblock \showarticletitle{Big data and due process: Toward a framework to redress predictive privacy harms}.
\newblock \bibinfo{journal}{\emph{BCL Rev.}}  \bibinfo{volume}{55} (\bibinfo{year}{2014}), \bibinfo{pages}{93}.
\newblock


\bibitem[{DBLP Computer Science Bibliography}(2024)]%
        {dblp}
\bibfield{author}{\bibinfo{person}{{DBLP Computer Science Bibliography}}.} \bibinfo{year}{2024}\natexlab{}.
\newblock \bibinfo{title}{DBLP: Computer Science Bibliography}.
\newblock \bibinfo{howpublished}{\url{https://dblp.org}}.
\newblock
\newblock
\shownote{Accessed: 2024-11-13}.


\bibitem[Deng et~al\mbox{.}(2023)]%
        {deng2023pentestgpt}
\bibfield{author}{\bibinfo{person}{Gelei Deng}, \bibinfo{person}{Yi Liu}, \bibinfo{person}{V{\'\i}ctor Mayoral-Vilches}, \bibinfo{person}{Peng Liu}, \bibinfo{person}{Yuekang Li}, \bibinfo{person}{Yuan Xu}, \bibinfo{person}{Tianwei Zhang}, \bibinfo{person}{Yang Liu}, \bibinfo{person}{Martin Pinzger}, {and} \bibinfo{person}{Stefan Rass}.} \bibinfo{year}{2023}\natexlab{}.
\newblock \showarticletitle{Pentestgpt: An llm-empowered automatic penetration testing tool}.
\newblock \bibinfo{journal}{\emph{arXiv preprint arXiv:2308.06782}} (\bibinfo{year}{2023}).
\newblock


\bibitem[Dinur and Nissim(2003)]%
        {dinur2003revealing}
\bibfield{author}{\bibinfo{person}{Irit Dinur} {and} \bibinfo{person}{Kobbi Nissim}.} \bibinfo{year}{2003}\natexlab{}.
\newblock \showarticletitle{Revealing information while preserving privacy}. In \bibinfo{booktitle}{\emph{Proceedings of the twenty-second ACM SIGMOD-SIGACT-SIGART symposium on Principles of database systems}}. \bibinfo{pages}{202--210}.
\newblock


\bibitem[Dong et~al\mbox{.}(2023)]%
        {dong2023self}
\bibfield{author}{\bibinfo{person}{Yihong Dong}, \bibinfo{person}{Xue Jiang}, \bibinfo{person}{Zhi Jin}, {and} \bibinfo{person}{Ge Li}.} \bibinfo{year}{2023}\natexlab{}.
\newblock \showarticletitle{Self-collaboration Code Generation via ChatGPT}.
\newblock \bibinfo{journal}{\emph{arXiv preprint arXiv:2304.07590}} (\bibinfo{year}{2023}).
\newblock


\bibitem[Du et~al\mbox{.}(2023)]%
        {du2023improving}
\bibfield{author}{\bibinfo{person}{Yilun Du}, \bibinfo{person}{Shuang Li}, \bibinfo{person}{Antonio Torralba}, \bibinfo{person}{Joshua~B Tenenbaum}, {and} \bibinfo{person}{Igor Mordatch}.} \bibinfo{year}{2023}\natexlab{}.
\newblock \showarticletitle{Improving factuality and reasoning in language models through multiagent debate}.
\newblock \bibinfo{journal}{\emph{arXiv preprint arXiv:2305.14325}} (\bibinfo{year}{2023}).
\newblock


\bibitem[Du et~al\mbox{.}(2024)]%
        {du2024multi}
\bibfield{author}{\bibinfo{person}{Zhuoyun Du}, \bibinfo{person}{Chen Qian}, \bibinfo{person}{Wei Liu}, \bibinfo{person}{Zihao Xie}, \bibinfo{person}{Yifei Wang}, \bibinfo{person}{Yufan Dang}, \bibinfo{person}{Weize Chen}, {and} \bibinfo{person}{Cheng Yang}.} \bibinfo{year}{2024}\natexlab{}.
\newblock \showarticletitle{Multi-Agent Software Development through Cross-Team Collaboration}.
\newblock \bibinfo{journal}{\emph{arXiv preprint arXiv:2406.08979}} (\bibinfo{year}{2024}).
\newblock


\bibitem[Dwork(2006)]%
        {dwork2006differential}
\bibfield{author}{\bibinfo{person}{Cynthia Dwork}.} \bibinfo{year}{2006}\natexlab{}.
\newblock \showarticletitle{Differential privacy}. In \bibinfo{booktitle}{\emph{International colloquium on automata, languages, and programming}}. Springer, \bibinfo{pages}{1--12}.
\newblock


\bibitem[Fan et~al\mbox{.}(2023)]%
        {fan2023static}
\bibfield{author}{\bibinfo{person}{Gang Fan}, \bibinfo{person}{Xiaoheng Xie}, \bibinfo{person}{Xunjin Zheng}, \bibinfo{person}{Yinan Liang}, {and} \bibinfo{person}{Peng Di}.} \bibinfo{year}{2023}\natexlab{}.
\newblock \showarticletitle{Static Code Analysis in the AI Era: An In-depth Exploration of the Concept, Function, and Potential of Intelligent Code Analysis Agents}.
\newblock \bibinfo{journal}{\emph{arXiv preprint arXiv:2310.08837}} (\bibinfo{year}{2023}).
\newblock


\bibitem[Franklin and Graesser(1996)]%
        {franklin1996agent}
\bibfield{author}{\bibinfo{person}{Stan Franklin} {and} \bibinfo{person}{Art Graesser}.} \bibinfo{year}{1996}\natexlab{}.
\newblock \showarticletitle{Is it an Agent, or just a Program?: A Taxonomy for Autonomous Agents}. In \bibinfo{booktitle}{\emph{International workshop on agent theories, architectures, and languages}}. Springer, \bibinfo{pages}{21--35}.
\newblock


\bibitem[Fu et~al\mbox{.}(2023)]%
        {fu2023chatgpt}
\bibfield{author}{\bibinfo{person}{Michael Fu}, \bibinfo{person}{Chakkrit~Kla Tantithamthavorn}, \bibinfo{person}{Van Nguyen}, {and} \bibinfo{person}{Trung Le}.} \bibinfo{year}{2023}\natexlab{}.
\newblock \showarticletitle{Chatgpt for vulnerability detection, classification, and repair: How far are we?}. In \bibinfo{booktitle}{\emph{2023 30th Asia-Pacific Software Engineering Conference (APSEC)}}. IEEE, \bibinfo{pages}{632--636}.
\newblock


\bibitem[Gao et~al\mbox{.}(2023)]%
        {gao2023retrieval}
\bibfield{author}{\bibinfo{person}{Yunfan Gao}, \bibinfo{person}{Yun Xiong}, \bibinfo{person}{Xinyu Gao}, \bibinfo{person}{Kangxiang Jia}, \bibinfo{person}{Jinliu Pan}, \bibinfo{person}{Yuxi Bi}, \bibinfo{person}{Yi Dai}, \bibinfo{person}{Jiawei Sun}, {and} \bibinfo{person}{Haofen Wang}.} \bibinfo{year}{2023}\natexlab{}.
\newblock \showarticletitle{Retrieval-augmented generation for large language models: A survey}.
\newblock \bibinfo{journal}{\emph{arXiv preprint arXiv:2312.10997}} (\bibinfo{year}{2023}).
\newblock


\bibitem[Goguen and Linde(1993)]%
        {goguen1993techniques}
\bibfield{author}{\bibinfo{person}{Joseph~A Goguen} {and} \bibinfo{person}{Charlotte Linde}.} \bibinfo{year}{1993}\natexlab{}.
\newblock \showarticletitle{Techniques for requirements elicitation}. In \bibinfo{booktitle}{\emph{[1993] Proceedings of the IEEE International Symposium on Requirements Engineering}}. IEEE, \bibinfo{pages}{152--164}.
\newblock


\bibitem[Goldreich(1998)]%
        {goldreich1998secure}
\bibfield{author}{\bibinfo{person}{Oded Goldreich}.} \bibinfo{year}{1998}\natexlab{}.
\newblock \showarticletitle{Secure multi-party computation}.
\newblock \bibinfo{journal}{\emph{Manuscript. Preliminary version}} \bibinfo{volume}{78}, \bibinfo{number}{110} (\bibinfo{year}{1998}), \bibinfo{pages}{1--108}.
\newblock


\bibitem[Guo et~al\mbox{.}(2024)]%
        {guo2024large}
\bibfield{author}{\bibinfo{person}{Taicheng Guo}, \bibinfo{person}{Xiuying Chen}, \bibinfo{person}{Yaqi Wang}, \bibinfo{person}{Ruidi Chang}, \bibinfo{person}{Shichao Pei}, \bibinfo{person}{Nitesh~V Chawla}, \bibinfo{person}{Olaf Wiest}, {and} \bibinfo{person}{Xiangliang Zhang}.} \bibinfo{year}{2024}\natexlab{}.
\newblock \showarticletitle{Large language model based multi-agents: A survey of progress and challenges}.
\newblock \bibinfo{journal}{\emph{arXiv preprint arXiv:2402.01680}} (\bibinfo{year}{2024}).
\newblock


\bibitem[He et~al\mbox{.}(2024a)]%
        {he2024ptm4tag+}
\bibfield{author}{\bibinfo{person}{Junda He}, \bibinfo{person}{Bowen Xu}, \bibinfo{person}{Zhou Yang}, \bibinfo{person}{DongGyun Han}, \bibinfo{person}{Chengran Yang}, \bibinfo{person}{Jiakun Liu}, \bibinfo{person}{Zhipeng Zhao}, {and} \bibinfo{person}{David Lo}.} \bibinfo{year}{2024}\natexlab{a}.
\newblock \showarticletitle{PTM4Tag+: Tag Recommendation of Stack Overflow Posts with Pre-trained Models}.
\newblock \bibinfo{journal}{\emph{arXiv preprint arXiv:2408.02311}} (\bibinfo{year}{2024}).
\newblock


\bibitem[He et~al\mbox{.}(2022)]%
        {he2022ptm4tag}
\bibfield{author}{\bibinfo{person}{Junda He}, \bibinfo{person}{Bowen Xu}, \bibinfo{person}{Zhou Yang}, \bibinfo{person}{DongGyun Han}, \bibinfo{person}{Chengran Yang}, {and} \bibinfo{person}{David Lo}.} \bibinfo{year}{2022}\natexlab{}.
\newblock \showarticletitle{Ptm4tag: sharpening tag recommendation of stack overflow posts with pre-trained models}. In \bibinfo{booktitle}{\emph{Proceedings of the 30th IEEE/ACM International Conference on Program Comprehension}}. \bibinfo{pages}{1--11}.
\newblock


\bibitem[He et~al\mbox{.}(2024b)]%
        {he2024representation}
\bibfield{author}{\bibinfo{person}{Junda He}, \bibinfo{person}{Xin Zhou}, \bibinfo{person}{Bowen Xu}, \bibinfo{person}{Ting Zhang}, \bibinfo{person}{Kisub Kim}, \bibinfo{person}{Zhou Yang}, \bibinfo{person}{Ferdian Thung}, \bibinfo{person}{Ivana~Clairine Irsan}, {and} \bibinfo{person}{David Lo}.} \bibinfo{year}{2024}\natexlab{b}.
\newblock \showarticletitle{Representation learning for stack overflow posts: How far are we?}
\newblock \bibinfo{journal}{\emph{ACM Transactions on Software Engineering and Methodology}} \bibinfo{volume}{33}, \bibinfo{number}{3} (\bibinfo{year}{2024}), \bibinfo{pages}{1--24}.
\newblock


\bibitem[Herrington(2003)]%
        {herrington2003code}
\bibfield{author}{\bibinfo{person}{Jack Herrington}.} \bibinfo{year}{2003}\natexlab{}.
\newblock \bibinfo{booktitle}{\emph{Code generation in action}}.
\newblock \bibinfo{publisher}{Manning Publications Co.}
\newblock


\bibitem[Hickey and Davis(2004)]%
        {hickey2004unified}
\bibfield{author}{\bibinfo{person}{Ann~M Hickey} {and} \bibinfo{person}{Alan~M Davis}.} \bibinfo{year}{2004}\natexlab{}.
\newblock \showarticletitle{A unified model of requirements elicitation}.
\newblock \bibinfo{journal}{\emph{Journal of management information systems}} \bibinfo{volume}{20}, \bibinfo{number}{4} (\bibinfo{year}{2004}), \bibinfo{pages}{65--84}.
\newblock


\bibitem[Holt et~al\mbox{.}(2023)]%
        {holt2023l2mac}
\bibfield{author}{\bibinfo{person}{Samuel Holt}, \bibinfo{person}{Max~Ruiz Luyten}, {and} \bibinfo{person}{Mihaela van~der Schaar}.} \bibinfo{year}{2023}\natexlab{}.
\newblock \showarticletitle{L2mac: Large language model automatic computer for unbounded code generation}.
\newblock \bibinfo{journal}{\emph{arXiv preprint arXiv:2310.02003}} (\bibinfo{year}{2023}).
\newblock


\bibitem[Hong et~al\mbox{.}(2023)]%
        {hong2023metagpt}
\bibfield{author}{\bibinfo{person}{Sirui Hong}, \bibinfo{person}{Xiawu Zheng}, \bibinfo{person}{Jonathan Chen}, \bibinfo{person}{Yuheng Cheng}, \bibinfo{person}{Jinlin Wang}, \bibinfo{person}{Ceyao Zhang}, \bibinfo{person}{Zili Wang}, \bibinfo{person}{Steven Ka~Shing Yau}, \bibinfo{person}{Zijuan Lin}, \bibinfo{person}{Liyang Zhou}, {et~al\mbox{.}}} \bibinfo{year}{2023}\natexlab{}.
\newblock \showarticletitle{Metagpt: Meta programming for multi-agent collaborative framework}.
\newblock \bibinfo{journal}{\emph{arXiv preprint arXiv:2308.00352}} (\bibinfo{year}{2023}).
\newblock


\bibitem[Horton(2023)]%
        {horton2023large}
\bibfield{author}{\bibinfo{person}{John~J Horton}.} \bibinfo{year}{2023}\natexlab{}.
\newblock \bibinfo{booktitle}{\emph{Large language models as simulated economic agents: What can we learn from homo silicus?}}
\newblock \bibinfo{type}{{T}echnical {R}eport}. \bibinfo{institution}{National Bureau of Economic Research}.
\newblock


\bibitem[Hou et~al\mbox{.}(2024)]%
        {hou2023large}
\bibfield{author}{\bibinfo{person}{Xinyi Hou}, \bibinfo{person}{Yanjie Zhao}, \bibinfo{person}{Yue Liu}, \bibinfo{person}{Zhou Yang}, \bibinfo{person}{Kailong Wang}, \bibinfo{person}{Li Li}, \bibinfo{person}{Xiapu Luo}, \bibinfo{person}{David Lo}, \bibinfo{person}{John Grundy}, {and} \bibinfo{person}{Haoyu Wang}.} \bibinfo{year}{2024}\natexlab{}.
\newblock \showarticletitle{Large Language Models for Software Engineering: A Systematic Literature Review}.
\newblock \bibinfo{journal}{\emph{ACM Trans. Softw. Eng. Methodol.}} \bibinfo{volume}{33}, \bibinfo{number}{8}, Article \bibinfo{articleno}{220} (\bibinfo{date}{Dec.} \bibinfo{year}{2024}), \bibinfo{numpages}{79}~pages.
\newblock
\showISSN{1049-331X}
\urldef\tempurl%
\url{https://doi.org/10.1145/3695988}
\showDOI{\tempurl}


\bibitem[Hu et~al\mbox{.}(2023)]%
        {hu2023large}
\bibfield{author}{\bibinfo{person}{Sihao Hu}, \bibinfo{person}{Tiansheng Huang}, \bibinfo{person}{Fatih {\.I}lhan}, \bibinfo{person}{Selim~Furkan Tekin}, {and} \bibinfo{person}{Ling Liu}.} \bibinfo{year}{2023}\natexlab{}.
\newblock \showarticletitle{Large language model-powered smart contract vulnerability detection: New perspectives}. In \bibinfo{booktitle}{\emph{2023 5th IEEE International Conference on Trust, Privacy and Security in Intelligent Systems and Applications (TPS-ISA)}}. IEEE, \bibinfo{pages}{297--306}.
\newblock


\bibitem[Hu et~al\mbox{.}(2015)]%
        {hu2015attribute}
\bibfield{author}{\bibinfo{person}{Vincent~C Hu}, \bibinfo{person}{D~Richard Kuhn}, \bibinfo{person}{David~F Ferraiolo}, {and} \bibinfo{person}{Jeffrey Voas}.} \bibinfo{year}{2015}\natexlab{}.
\newblock \showarticletitle{Attribute-based access control}.
\newblock \bibinfo{journal}{\emph{Computer}} \bibinfo{volume}{48}, \bibinfo{number}{2} (\bibinfo{year}{2015}), \bibinfo{pages}{85--88}.
\newblock


\bibitem[Hu et~al\mbox{.}(2024)]%
        {hu2024self}
\bibfield{author}{\bibinfo{person}{Yue Hu}, \bibinfo{person}{Yuzhu Cai}, \bibinfo{person}{Yaxin Du}, \bibinfo{person}{Xinyu Zhu}, \bibinfo{person}{Xiangrui Liu}, \bibinfo{person}{Zijie Yu}, \bibinfo{person}{Yuchen Hou}, \bibinfo{person}{Shuo Tang}, {and} \bibinfo{person}{Siheng Chen}.} \bibinfo{year}{2024}\natexlab{}.
\newblock \showarticletitle{Self-Evolving Multi-Agent Collaboration Networks for Software Development}.
\newblock \bibinfo{journal}{\emph{arXiv preprint arXiv:2410.16946}} (\bibinfo{year}{2024}).
\newblock


\bibitem[Huang et~al\mbox{.}(2023)]%
        {huang2023agentcoder}
\bibfield{author}{\bibinfo{person}{Dong Huang}, \bibinfo{person}{Qingwen Bu}, \bibinfo{person}{Jie~M Zhang}, \bibinfo{person}{Michael Luck}, {and} \bibinfo{person}{Heming Cui}.} \bibinfo{year}{2023}\natexlab{}.
\newblock \showarticletitle{AgentCoder: Multi-Agent-based Code Generation with Iterative Testing and Optimisation}.
\newblock \bibinfo{journal}{\emph{arXiv preprint arXiv:2312.13010}} (\bibinfo{year}{2023}).
\newblock


\bibitem[Imtiaz et~al\mbox{.}(2021)]%
        {imtiaz2021comparative}
\bibfield{author}{\bibinfo{person}{Nasif Imtiaz}, \bibinfo{person}{Seaver Thorn}, {and} \bibinfo{person}{Laurie Williams}.} \bibinfo{year}{2021}\natexlab{}.
\newblock \showarticletitle{A comparative study of vulnerability reporting by software composition analysis tools}. In \bibinfo{booktitle}{\emph{Proceedings of the 15th ACM/IEEE International Symposium on Empirical Software Engineering and Measurement (ESEM)}}. \bibinfo{pages}{1--11}.
\newblock


\bibitem[Ishibashi and Nishimura(2024)]%
        {ishibashi2024self}
\bibfield{author}{\bibinfo{person}{Yoichi Ishibashi} {and} \bibinfo{person}{Yoshimasa Nishimura}.} \bibinfo{year}{2024}\natexlab{}.
\newblock \showarticletitle{Self-organized agents: A llm multi-agent framework toward ultra large-scale code generation and optimization}.
\newblock \bibinfo{journal}{\emph{arXiv preprint arXiv:2404.02183}} (\bibinfo{year}{2024}).
\newblock


\bibitem[Islam et~al\mbox{.}(2024)]%
        {islam-etal-2024-mapcoder}
\bibfield{author}{\bibinfo{person}{Md.~Ashraful Islam}, \bibinfo{person}{Mohammed~Eunus Ali}, {and} \bibinfo{person}{Md~Rizwan Parvez}.} \bibinfo{year}{2024}\natexlab{}.
\newblock \showarticletitle{{M}ap{C}oder: Multi-Agent Code Generation for Competitive Problem Solving}. In \bibinfo{booktitle}{\emph{Proceedings of the 62nd Annual Meeting of the Association for Computational Linguistics (Volume 1: Long Papers)}}, \bibfield{editor}{\bibinfo{person}{Lun-Wei Ku}, \bibinfo{person}{Andre Martins}, {and} \bibinfo{person}{Vivek Srikumar}} (Eds.). \bibinfo{publisher}{Association for Computational Linguistics}, \bibinfo{address}{Bangkok, Thailand}, \bibinfo{pages}{4912--4944}.
\newblock
\urldef\tempurl%
\url{https://aclanthology.org/2024.acl-long.269}
\showURL{%
\tempurl}


\bibitem[Jain et~al\mbox{.}(2024)]%
        {jain2024livecodebench}
\bibfield{author}{\bibinfo{person}{Naman Jain}, \bibinfo{person}{King Han}, \bibinfo{person}{Alex Gu}, \bibinfo{person}{Wen-Ding Li}, \bibinfo{person}{Fanjia Yan}, \bibinfo{person}{Tianjun Zhang}, \bibinfo{person}{Sida Wang}, \bibinfo{person}{Armando Solar-Lezama}, \bibinfo{person}{Koushik Sen}, {and} \bibinfo{person}{Ion Stoica}.} \bibinfo{year}{2024}\natexlab{}.
\newblock \showarticletitle{Livecodebench: Holistic and contamination free evaluation of large language models for code}.
\newblock \bibinfo{journal}{\emph{arXiv preprint arXiv:2403.07974}} (\bibinfo{year}{2024}).
\newblock


\bibitem[Jin et~al\mbox{.}(2024)]%
        {jin2024mare}
\bibfield{author}{\bibinfo{person}{Dongming Jin}, \bibinfo{person}{Zhi Jin}, \bibinfo{person}{Xiaohong Chen}, {and} \bibinfo{person}{Chunhui Wang}.} \bibinfo{year}{2024}\natexlab{}.
\newblock \showarticletitle{MARE: Multi-Agents Collaboration Framework for Requirements Engineering}.
\newblock \bibinfo{journal}{\emph{arXiv preprint arXiv:2405.03256}} (\bibinfo{year}{2024}).
\newblock


\bibitem[Josifoski et~al\mbox{.}(2023)]%
        {josifoski2023flows}
\bibfield{author}{\bibinfo{person}{Martin Josifoski}, \bibinfo{person}{Lars Klein}, \bibinfo{person}{Maxime Peyrard}, \bibinfo{person}{Nicolas Baldwin}, \bibinfo{person}{Yifei Li}, \bibinfo{person}{Saibo Geng}, \bibinfo{person}{Julian~Paul Schnitzler}, \bibinfo{person}{Yuxing Yao}, \bibinfo{person}{Jiheng Wei}, \bibinfo{person}{Debjit Paul}, {et~al\mbox{.}}} \bibinfo{year}{2023}\natexlab{}.
\newblock \showarticletitle{Flows: Building blocks of reasoning and collaborating ai}.
\newblock \bibinfo{journal}{\emph{arXiv preprint arXiv:2308.01285}} (\bibinfo{year}{2023}).
\newblock


\bibitem[Jouvin and Hassas(2002)]%
        {jouvin2002role}
\bibfield{author}{\bibinfo{person}{Denis Jouvin} {and} \bibinfo{person}{Salima Hassas}.} \bibinfo{year}{2002}\natexlab{}.
\newblock \showarticletitle{Role delegation as multi-agent oriented dynamic composition}. In \bibinfo{booktitle}{\emph{Proceedings of Net Object Days (NOD), AgeS workshop, Erfurt, Germany}}.
\newblock


\bibitem[Kairouz et~al\mbox{.}(2021)]%
        {kairouz2021advances}
\bibfield{author}{\bibinfo{person}{Peter Kairouz}, \bibinfo{person}{H~Brendan McMahan}, \bibinfo{person}{Brendan Avent}, \bibinfo{person}{Aur{\'e}lien Bellet}, \bibinfo{person}{Mehdi Bennis}, \bibinfo{person}{Arjun~Nitin Bhagoji}, \bibinfo{person}{Kallista Bonawitz}, \bibinfo{person}{Zachary Charles}, \bibinfo{person}{Graham Cormode}, \bibinfo{person}{Rachel Cummings}, {et~al\mbox{.}}} \bibinfo{year}{2021}\natexlab{}.
\newblock \showarticletitle{Advances and open problems in federated learning}.
\newblock \bibinfo{journal}{\emph{Foundations and trends{\textregistered} in machine learning}} \bibinfo{volume}{14}, \bibinfo{number}{1--2} (\bibinfo{year}{2021}), \bibinfo{pages}{1--210}.
\newblock


\bibitem[Kan(2003)]%
        {kan2003metrics}
\bibfield{author}{\bibinfo{person}{Stephen~H Kan}.} \bibinfo{year}{2003}\natexlab{}.
\newblock \bibinfo{booktitle}{\emph{Metrics and models in software quality engineering}}.
\newblock \bibinfo{publisher}{Addison-Wesley Professional}.
\newblock


\bibitem[Kang et~al\mbox{.}(2023)]%
        {kang2023explainable}
\bibfield{author}{\bibinfo{person}{Sungmin Kang}, \bibinfo{person}{Bei Chen}, \bibinfo{person}{Shin Yoo}, {and} \bibinfo{person}{Jian-Guang Lou}.} \bibinfo{year}{2023}\natexlab{}.
\newblock \showarticletitle{Explainable automated debugging via large language model-driven scientific debugging}.
\newblock \bibinfo{journal}{\emph{arXiv preprint arXiv:2304.02195}} (\bibinfo{year}{2023}).
\newblock


\bibitem[Kasneci et~al\mbox{.}(2023)]%
        {kasneci2023chatgpt}
\bibfield{author}{\bibinfo{person}{Enkelejda Kasneci}, \bibinfo{person}{Kathrin Se{\ss}ler}, \bibinfo{person}{Stefan K{\"u}chemann}, \bibinfo{person}{Maria Bannert}, \bibinfo{person}{Daryna Dementieva}, \bibinfo{person}{Frank Fischer}, \bibinfo{person}{Urs Gasser}, \bibinfo{person}{Georg Groh}, \bibinfo{person}{Stephan G{\"u}nnemann}, \bibinfo{person}{Eyke H{\"u}llermeier}, {et~al\mbox{.}}} \bibinfo{year}{2023}\natexlab{}.
\newblock \showarticletitle{ChatGPT for good? On opportunities and challenges of large language models for education}.
\newblock \bibinfo{journal}{\emph{Learning and individual differences}}  \bibinfo{volume}{103} (\bibinfo{year}{2023}), \bibinfo{pages}{102274}.
\newblock


\bibitem[Kaur and Kumar(2013)]%
        {kaur2013competency}
\bibfield{author}{\bibinfo{person}{Jaideep Kaur} {and} \bibinfo{person}{Vikas Kumar}.} \bibinfo{year}{2013}\natexlab{}.
\newblock \showarticletitle{Competency mapping: A gap Analysis}.
\newblock \bibinfo{journal}{\emph{International Journal of Education and Research}} \bibinfo{volume}{1}, \bibinfo{number}{1} (\bibinfo{year}{2013}), \bibinfo{pages}{1--9}.
\newblock


\bibitem[Khattab et~al\mbox{.}(2023)]%
        {khattab2023dspy}
\bibfield{author}{\bibinfo{person}{Omar Khattab}, \bibinfo{person}{Arnav Singhvi}, \bibinfo{person}{Paridhi Maheshwari}, \bibinfo{person}{Zhiyuan Zhang}, \bibinfo{person}{Keshav Santhanam}, \bibinfo{person}{Sri Vardhamanan}, \bibinfo{person}{Saiful Haq}, \bibinfo{person}{Ashutosh Sharma}, \bibinfo{person}{Thomas~T Joshi}, \bibinfo{person}{Hanna Moazam}, {et~al\mbox{.}}} \bibinfo{year}{2023}\natexlab{}.
\newblock \showarticletitle{Dspy: Compiling declarative language model calls into self-improving pipelines}.
\newblock \bibinfo{journal}{\emph{arXiv preprint arXiv:2310.03714}} (\bibinfo{year}{2023}).
\newblock


\bibitem[Larman(2004)]%
        {larman2004agile}
\bibfield{author}{\bibinfo{person}{Craig Larman}.} \bibinfo{year}{2004}\natexlab{}.
\newblock \bibinfo{booktitle}{\emph{Agile and iterative development: a manager's guide}}.
\newblock \bibinfo{publisher}{Addison-Wesley Professional}.
\newblock


\bibitem[Le et~al\mbox{.}(2024)]%
        {le2024indict}
\bibfield{author}{\bibinfo{person}{Hung Le}, \bibinfo{person}{Yingbo Zhou}, \bibinfo{person}{Caiming Xiong}, \bibinfo{person}{Silvio Savarese}, {and} \bibinfo{person}{Doyen Sahoo}.} \bibinfo{year}{2024}\natexlab{}.
\newblock \showarticletitle{INDICT: Code Generation with Internal Dialogues of Critiques for Both Security and Helpfulness}.
\newblock \bibinfo{journal}{\emph{arXiv preprint arXiv:2407.02518}} (\bibinfo{year}{2024}).
\newblock


\bibitem[Lee et~al\mbox{.}(2024)]%
        {lee2024unified}
\bibfield{author}{\bibinfo{person}{Cheryl Lee}, \bibinfo{person}{Chunqiu~Steven Xia}, \bibinfo{person}{Jen-tse Huang}, \bibinfo{person}{Zhouruixin Zhu}, \bibinfo{person}{Lingming Zhang}, {and} \bibinfo{person}{Michael~R Lyu}.} \bibinfo{year}{2024}\natexlab{}.
\newblock \showarticletitle{A Unified Debugging Approach via LLM-Based Multi-Agent Synergy}.
\newblock \bibinfo{journal}{\emph{arXiv preprint arXiv:2404.17153}} (\bibinfo{year}{2024}).
\newblock


\bibitem[Leffingwell and Widrig(2000)]%
        {leffingwell2000managing}
\bibfield{author}{\bibinfo{person}{Dean Leffingwell} {and} \bibinfo{person}{Don Widrig}.} \bibinfo{year}{2000}\natexlab{}.
\newblock \bibinfo{booktitle}{\emph{Managing software requirements: a unified approach}}.
\newblock \bibinfo{publisher}{Addison-Wesley Professional}.
\newblock


\bibitem[Lei et~al\mbox{.}(2024a)]%
        {lei2024autocoder}
\bibfield{author}{\bibinfo{person}{Bin Lei}, \bibinfo{person}{Yuchen Li}, {and} \bibinfo{person}{Qiuwu Chen}.} \bibinfo{year}{2024}\natexlab{a}.
\newblock \showarticletitle{AutoCoder: Enhancing Code Large Language Model with$\backslash$textsc $\{$AIEV-Instruct$\}$}.
\newblock \bibinfo{journal}{\emph{arXiv preprint arXiv:2405.14906}} (\bibinfo{year}{2024}).
\newblock


\bibitem[Lei et~al\mbox{.}(2024b)]%
        {lei2024infant}
\bibfield{author}{\bibinfo{person}{Bin Lei}, \bibinfo{person}{Yuchen Li}, \bibinfo{person}{Yiming Zeng}, \bibinfo{person}{Tao Ren}, \bibinfo{person}{Yi Luo}, \bibinfo{person}{Tianyu Shi}, \bibinfo{person}{Zitian Gao}, \bibinfo{person}{Zeyu Hu}, \bibinfo{person}{Weitai Kang}, {and} \bibinfo{person}{Qiuwu Chen}.} \bibinfo{year}{2024}\natexlab{b}.
\newblock \showarticletitle{Infant Agent: A Tool-Integrated, Logic-Driven Agent with Cost-Effective API Usage}.
\newblock \bibinfo{journal}{\emph{arXiv preprint arXiv:2411.01114}} (\bibinfo{year}{2024}).
\newblock


\bibitem[Lemner et~al\mbox{.}(2024)]%
        {lemner2024exploring}
\bibfield{author}{\bibinfo{person}{Ludvig Lemner}, \bibinfo{person}{Linnea Wahlgren}, \bibinfo{person}{Gregory Gay}, \bibinfo{person}{Nasser Mohammadiha}, \bibinfo{person}{Jingxiong Liu}, {and} \bibinfo{person}{Joakim Wennerberg}.} \bibinfo{year}{2024}\natexlab{}.
\newblock \showarticletitle{Exploring the Integration of Large Language Models in Industrial Test Maintenance Processes}.
\newblock \bibinfo{journal}{\emph{arXiv preprint arXiv:2409.06416}} (\bibinfo{year}{2024}).
\newblock


\bibitem[Li et~al\mbox{.}(2024a)]%
        {li2024camel}
\bibfield{author}{\bibinfo{person}{Guohao Li}, \bibinfo{person}{Hasan Hammoud}, \bibinfo{person}{Hani Itani}, \bibinfo{person}{Dmitrii Khizbullin}, {and} \bibinfo{person}{Bernard Ghanem}.} \bibinfo{year}{2024}\natexlab{a}.
\newblock \showarticletitle{Camel: Communicative agents for" mind" exploration of large language model society}.
\newblock \bibinfo{journal}{\emph{Advances in Neural Information Processing Systems}}  \bibinfo{volume}{36} (\bibinfo{year}{2024}).
\newblock


\bibitem[Li et~al\mbox{.}(2024c)]%
        {li2024codetree}
\bibfield{author}{\bibinfo{person}{Jierui Li}, \bibinfo{person}{Hung Le}, \bibinfo{person}{Yinbo Zhou}, \bibinfo{person}{Caiming Xiong}, \bibinfo{person}{Silvio Savarese}, {and} \bibinfo{person}{Doyen Sahoo}.} \bibinfo{year}{2024}\natexlab{c}.
\newblock \showarticletitle{CodeTree: Agent-guided Tree Search for Code Generation with Large Language Models}.
\newblock \bibinfo{journal}{\emph{arXiv preprint arXiv:2411.04329}} (\bibinfo{year}{2024}).
\newblock


\bibitem[Li et~al\mbox{.}(2024d)]%
        {li2024more}
\bibfield{author}{\bibinfo{person}{Junyou Li}, \bibinfo{person}{Qin Zhang}, \bibinfo{person}{Yangbin Yu}, \bibinfo{person}{Qiang Fu}, {and} \bibinfo{person}{Deheng Ye}.} \bibinfo{year}{2024}\natexlab{d}.
\newblock \showarticletitle{More agents is all you need}.
\newblock \bibinfo{journal}{\emph{arXiv preprint arXiv:2402.05120}} (\bibinfo{year}{2024}).
\newblock


\bibitem[Li et~al\mbox{.}(2023)]%
        {li2023metaagents}
\bibfield{author}{\bibinfo{person}{Yuan Li}, \bibinfo{person}{Yixuan Zhang}, {and} \bibinfo{person}{Lichao Sun}.} \bibinfo{year}{2023}\natexlab{}.
\newblock \bibinfo{title}{MetaAgents: Simulating Interactions of Human Behaviors for LLM-based Task-oriented Coordination via Collaborative Generative Agents (arXiv: 2310.06500). arXiv}.
\newblock
\newblock


\bibitem[Li et~al\mbox{.}(2024b)]%
        {li2024relational}
\bibfield{author}{\bibinfo{person}{Ziyang Li}, \bibinfo{person}{Jiani Huang}, \bibinfo{person}{Jason Liu}, \bibinfo{person}{Felix Zhu}, \bibinfo{person}{Eric Zhao}, \bibinfo{person}{William Dodds}, \bibinfo{person}{Neelay Velingker}, \bibinfo{person}{Rajeev Alur}, {and} \bibinfo{person}{Mayur Naik}.} \bibinfo{year}{2024}\natexlab{b}.
\newblock \showarticletitle{Relational Programming with Foundational Models}. In \bibinfo{booktitle}{\emph{Proceedings of the AAAI Conference on Artificial Intelligence}}, Vol.~\bibinfo{volume}{38}. \bibinfo{pages}{10635--10644}.
\newblock


\bibitem[Liang et~al\mbox{.}(2023)]%
        {liang2023encouraging}
\bibfield{author}{\bibinfo{person}{Tian Liang}, \bibinfo{person}{Zhiwei He}, \bibinfo{person}{Wenxiang Jiao}, \bibinfo{person}{Xing Wang}, \bibinfo{person}{Yan Wang}, \bibinfo{person}{Rui Wang}, \bibinfo{person}{Yujiu Yang}, \bibinfo{person}{Zhaopeng Tu}, {and} \bibinfo{person}{Shuming Shi}.} \bibinfo{year}{2023}\natexlab{}.
\newblock \showarticletitle{Encouraging divergent thinking in large language models through multi-agent debate}.
\newblock \bibinfo{journal}{\emph{arXiv preprint arXiv:2305.19118}} (\bibinfo{year}{2023}).
\newblock


\bibitem[Lin et~al\mbox{.}(2024a)]%
        {lin2024llm}
\bibfield{author}{\bibinfo{person}{Feng Lin}, \bibinfo{person}{Dong~Jae Kim}, {et~al\mbox{.}}} \bibinfo{year}{2024}\natexlab{a}.
\newblock \showarticletitle{SOEN-101: Code Generation by Emulating Software Process Models Using Large Language Model Agents}.
\newblock \bibinfo{journal}{\emph{arXiv preprint arXiv:2403.15852}} (\bibinfo{year}{2024}).
\newblock


\bibitem[Lin et~al\mbox{.}(2024b)]%
        {lin2024think}
\bibfield{author}{\bibinfo{person}{Leilei Lin}, \bibinfo{person}{Yingming Zhou}, \bibinfo{person}{Wenlong Chen}, {and} \bibinfo{person}{Chen Qian}.} \bibinfo{year}{2024}\natexlab{b}.
\newblock \showarticletitle{Think-on-Process: Dynamic Process Generation for Collaborative Development of Multi-Agent System}.
\newblock \bibinfo{journal}{\emph{arXiv preprint arXiv:2409.06568}} (\bibinfo{year}{2024}).
\newblock


\bibitem[Liu et~al\mbox{.}(2022)]%
        {liu2022few}
\bibfield{author}{\bibinfo{person}{Haokun Liu}, \bibinfo{person}{Derek Tam}, \bibinfo{person}{Mohammed Muqeeth}, \bibinfo{person}{Jay Mohta}, \bibinfo{person}{Tenghao Huang}, \bibinfo{person}{Mohit Bansal}, {and} \bibinfo{person}{Colin~A Raffel}.} \bibinfo{year}{2022}\natexlab{}.
\newblock \showarticletitle{Few-shot parameter-efficient fine-tuning is better and cheaper than in-context learning}.
\newblock \bibinfo{journal}{\emph{Advances in Neural Information Processing Systems}}  \bibinfo{volume}{35} (\bibinfo{year}{2022}), \bibinfo{pages}{1950--1965}.
\newblock


\bibitem[Liu et~al\mbox{.}(2024c)]%
        {liu2024your}
\bibfield{author}{\bibinfo{person}{Jiawei Liu}, \bibinfo{person}{Chunqiu~Steven Xia}, \bibinfo{person}{Yuyao Wang}, {and} \bibinfo{person}{Lingming Zhang}.} \bibinfo{year}{2024}\natexlab{c}.
\newblock \showarticletitle{Is your code generated by chatgpt really correct? rigorous evaluation of large language models for code generation}.
\newblock \bibinfo{journal}{\emph{Advances in Neural Information Processing Systems}}  \bibinfo{volume}{36} (\bibinfo{year}{2024}).
\newblock


\bibitem[Liu et~al\mbox{.}(2024b)]%
        {liu2024codexgraph}
\bibfield{author}{\bibinfo{person}{Xiangyan Liu}, \bibinfo{person}{Bo Lan}, \bibinfo{person}{Zhiyuan Hu}, \bibinfo{person}{Yang Liu}, \bibinfo{person}{Zhicheng Zhang}, \bibinfo{person}{Wenmeng Zhou}, \bibinfo{person}{Fei Wang}, {and} \bibinfo{person}{Michael Shieh}.} \bibinfo{year}{2024}\natexlab{b}.
\newblock \showarticletitle{CodexGraph: Bridging Large Language Models and Code Repositories via Code Graph Databases}.
\newblock \bibinfo{journal}{\emph{arXiv preprint arXiv:2408.03910}} (\bibinfo{year}{2024}).
\newblock


\bibitem[Liu et~al\mbox{.}(2024a)]%
        {liu2024marscode}
\bibfield{author}{\bibinfo{person}{Yizhou Liu}, \bibinfo{person}{Pengfei Gao}, \bibinfo{person}{Xinchen Wang}, \bibinfo{person}{Chao Peng}, {and} \bibinfo{person}{Zhao Zhang}.} \bibinfo{year}{2024}\natexlab{a}.
\newblock \showarticletitle{MarsCode Agent: AI-native Automated Bug Fixing}.
\newblock \bibinfo{journal}{\emph{arXiv preprint arXiv:2409.00899}} (\bibinfo{year}{2024}).
\newblock


\bibitem[Liu et~al\mbox{.}(2024d)]%
        {liu2024agents4plc}
\bibfield{author}{\bibinfo{person}{Zihan Liu}, \bibinfo{person}{Ruinan Zeng}, \bibinfo{person}{Dongxia Wang}, \bibinfo{person}{Gengyun Peng}, \bibinfo{person}{Jingyi Wang}, \bibinfo{person}{Qiang Liu}, \bibinfo{person}{Peiyu Liu}, {and} \bibinfo{person}{Wenhai Wang}.} \bibinfo{year}{2024}\natexlab{d}.
\newblock \showarticletitle{Agents4PLC: Automating Closed-loop PLC Code Generation and Verification in Industrial Control Systems using LLM-based Agents}.
\newblock \bibinfo{journal}{\emph{arXiv preprint arXiv:2410.14209}} (\bibinfo{year}{2024}).
\newblock


\bibitem[Liu et~al\mbox{.}(2023)]%
        {liu2023dynamic}
\bibfield{author}{\bibinfo{person}{Zijun Liu}, \bibinfo{person}{Yanzhe Zhang}, \bibinfo{person}{Peng Li}, \bibinfo{person}{Yang Liu}, {and} \bibinfo{person}{Diyi Yang}.} \bibinfo{year}{2023}\natexlab{}.
\newblock \showarticletitle{Dynamic llm-agent network: An llm-agent collaboration framework with agent team optimization}.
\newblock \bibinfo{journal}{\emph{arXiv preprint arXiv:2310.02170}} (\bibinfo{year}{2023}).
\newblock


\bibitem[Liu et~al\mbox{.}(2024e)]%
        {liu2024dynamic}
\bibfield{author}{\bibinfo{person}{Zijun Liu}, \bibinfo{person}{Yanzhe Zhang}, \bibinfo{person}{Peng Li}, \bibinfo{person}{Yang Liu}, {and} \bibinfo{person}{Diyi Yang}.} \bibinfo{year}{2024}\natexlab{e}.
\newblock \showarticletitle{A dynamic LLM-powered agent network for task-oriented agent collaboration}. In \bibinfo{booktitle}{\emph{First Conference on Language Modeling}}.
\newblock


\bibitem[Lo(2023)]%
        {lo2023trustworthy}
\bibfield{author}{\bibinfo{person}{David Lo}.} \bibinfo{year}{2023}\natexlab{}.
\newblock \showarticletitle{Trustworthy and Synergistic Artificial Intelligence for Software Engineering: Vision and Roadmaps}.
\newblock \bibinfo{journal}{\emph{arXiv preprint arXiv:2309.04142}} (\bibinfo{year}{2023}).
\newblock


\bibitem[Lo(2024)]%
        {lo2024requirements}
\bibfield{author}{\bibinfo{person}{David Lo}.} \bibinfo{year}{2024}\natexlab{}.
\newblock \showarticletitle{Requirements Engineering for Trustworthy Human-AI Synergy in Software Engineering 2.0}. In \bibinfo{booktitle}{\emph{2024 IEEE 32nd International Requirements Engineering Conference (RE)}}. IEEE, \bibinfo{pages}{3--4}.
\newblock


\bibitem[Ma et~al\mbox{.}(2024)]%
        {ma2024understand}
\bibfield{author}{\bibinfo{person}{Yingwei Ma}, \bibinfo{person}{Qingping Yang}, \bibinfo{person}{Rongyu Cao}, \bibinfo{person}{Binhua Li}, \bibinfo{person}{Fei Huang}, {and} \bibinfo{person}{Yongbin Li}.} \bibinfo{year}{2024}\natexlab{}.
\newblock \showarticletitle{How to Understand Whole Software Repository?}
\newblock \bibinfo{journal}{\emph{arXiv preprint arXiv:2406.01422}} (\bibinfo{year}{2024}).
\newblock


\bibitem[Maes(1993)]%
        {maes1993modeling}
\bibfield{author}{\bibinfo{person}{Pattie Maes}.} \bibinfo{year}{1993}\natexlab{}.
\newblock \showarticletitle{Modeling adaptive autonomous agents}.
\newblock \bibinfo{journal}{\emph{Artificial life}} \bibinfo{volume}{1}, \bibinfo{number}{1\_2} (\bibinfo{year}{1993}), \bibinfo{pages}{135--162}.
\newblock


\bibitem[Mao et~al\mbox{.}(2024)]%
        {mao2024multi}
\bibfield{author}{\bibinfo{person}{Zhenyu Mao}, \bibinfo{person}{Jialong Li}, \bibinfo{person}{Munan Li}, {and} \bibinfo{person}{Kenji Tei}.} \bibinfo{year}{2024}\natexlab{}.
\newblock \showarticletitle{Multi-role Consensus through LLMs Discussions for Vulnerability Detection}.
\newblock \bibinfo{journal}{\emph{arXiv preprint arXiv:2403.14274}} (\bibinfo{year}{2024}).
\newblock


\bibitem[Markauskaite et~al\mbox{.}(2022)]%
        {markauskaite2022rethinking}
\bibfield{author}{\bibinfo{person}{Lina Markauskaite}, \bibinfo{person}{Rebecca Marrone}, \bibinfo{person}{Oleksandra Poquet}, \bibinfo{person}{Simon Knight}, \bibinfo{person}{Roberto Martinez-Maldonado}, \bibinfo{person}{Sarah Howard}, \bibinfo{person}{Jo Tondeur}, \bibinfo{person}{Maarten De~Laat}, \bibinfo{person}{Simon~Buckingham Shum}, \bibinfo{person}{Dragan Ga{\v{s}}evi{\'c}}, {et~al\mbox{.}}} \bibinfo{year}{2022}\natexlab{}.
\newblock \showarticletitle{Rethinking the entwinement between artificial intelligence and human learning: What capabilities do learners need for a world with AI?}
\newblock \bibinfo{journal}{\emph{Computers and Education: Artificial Intelligence}}  \bibinfo{volume}{3} (\bibinfo{year}{2022}), \bibinfo{pages}{100056}.
\newblock


\bibitem[Mathews and Nagappan(2024)]%
        {mathews2024test}
\bibfield{author}{\bibinfo{person}{Noble~Saji Mathews} {and} \bibinfo{person}{Meiyappan Nagappan}.} \bibinfo{year}{2024}\natexlab{}.
\newblock \showarticletitle{Test-Driven Development for Code Generation}.
\newblock \bibinfo{journal}{\emph{arXiv preprint arXiv:2402.13521}} (\bibinfo{year}{2024}).
\newblock


\bibitem[Mavroudis(2024)]%
        {mavroudis2024langchain}
\bibfield{author}{\bibinfo{person}{Vasilios Mavroudis}.} \bibinfo{year}{2024}\natexlab{}.
\newblock \showarticletitle{LangChain}.
\newblock  (\bibinfo{year}{2024}).
\newblock


\bibitem[Mele(2001)]%
        {mele2001autonomous}
\bibfield{author}{\bibinfo{person}{Alfred~R Mele}.} \bibinfo{year}{2001}\natexlab{}.
\newblock \bibinfo{booktitle}{\emph{Autonomous agents: From self-control to autonomy}}.
\newblock \bibinfo{publisher}{Oxford University Press, USA}.
\newblock


\bibitem[Meline(2006)]%
        {meline2006selecting}
\bibfield{author}{\bibinfo{person}{Timothy Meline}.} \bibinfo{year}{2006}\natexlab{}.
\newblock \showarticletitle{Selecting studies for systemic review: Inclusion and exclusion criteria}.
\newblock \bibinfo{journal}{\emph{Contemporary issues in communication science and disorders}} \bibinfo{volume}{33}, \bibinfo{number}{Spring} (\bibinfo{year}{2006}), \bibinfo{pages}{21--27}.
\newblock


\bibitem[Mendes et~al\mbox{.}(2018)]%
        {mendes2018towards}
\bibfield{author}{\bibinfo{person}{Emilia Mendes}, \bibinfo{person}{Pilar Rodriguez}, \bibinfo{person}{Vitor Freitas}, \bibinfo{person}{Simon Baker}, {and} \bibinfo{person}{Mohamed~Amine Atoui}.} \bibinfo{year}{2018}\natexlab{}.
\newblock \showarticletitle{Towards improving decision making and estimating the value of decisions in value-based software engineering: the VALUE framework}.
\newblock \bibinfo{journal}{\emph{Software Quality Journal}}  \bibinfo{volume}{26} (\bibinfo{year}{2018}), \bibinfo{pages}{607--656}.
\newblock


\bibitem[Nguyen et~al\mbox{.}(2024)]%
        {nguyen2024agilecoder}
\bibfield{author}{\bibinfo{person}{Minh~Huynh Nguyen}, \bibinfo{person}{Thang~Phan Chau}, \bibinfo{person}{Phong~X Nguyen}, {and} \bibinfo{person}{Nghi~DQ Bui}.} \bibinfo{year}{2024}\natexlab{}.
\newblock \showarticletitle{AgileCoder: Dynamic Collaborative Agents for Software Development based on Agile Methodology}.
\newblock \bibinfo{journal}{\emph{arXiv preprint arXiv:2406.11912}} (\bibinfo{year}{2024}).
\newblock


\bibitem[Olausson et~al\mbox{.}(2023)]%
        {olausson2023self}
\bibfield{author}{\bibinfo{person}{Theo~X Olausson}, \bibinfo{person}{Jeevana~Priya Inala}, \bibinfo{person}{Chenglong Wang}, \bibinfo{person}{Jianfeng Gao}, {and} \bibinfo{person}{Armando Solar-Lezama}.} \bibinfo{year}{2023}\natexlab{}.
\newblock \showarticletitle{Is Self-Repair a Silver Bullet for Code Generation?}. In \bibinfo{booktitle}{\emph{The Twelfth International Conference on Learning Representations}}.
\newblock


\bibitem[Paasivaara et~al\mbox{.}(2008)]%
        {paasivaara2008distributed}
\bibfield{author}{\bibinfo{person}{Maria Paasivaara}, \bibinfo{person}{Sandra Durasiewicz}, {and} \bibinfo{person}{Casper Lassenius}.} \bibinfo{year}{2008}\natexlab{}.
\newblock \showarticletitle{Distributed agile development: Using scrum in a large project}. In \bibinfo{booktitle}{\emph{2008 IEEE International Conference on Global Software Engineering}}. IEEE, \bibinfo{pages}{87--95}.
\newblock


\bibitem[Pardau(2018)]%
        {pardau2018california}
\bibfield{author}{\bibinfo{person}{Stuart~L Pardau}.} \bibinfo{year}{2018}\natexlab{}.
\newblock \showarticletitle{The california consumer privacy act: Towards a european-style privacy regime in the united states}.
\newblock \bibinfo{journal}{\emph{J. Tech. L. \& Pol'y}}  \bibinfo{volume}{23} (\bibinfo{year}{2018}), \bibinfo{pages}{68}.
\newblock


\bibitem[Petersen et~al\mbox{.}(2009)]%
        {petersen2009waterfall}
\bibfield{author}{\bibinfo{person}{Kai Petersen}, \bibinfo{person}{Claes Wohlin}, {and} \bibinfo{person}{Dejan Baca}.} \bibinfo{year}{2009}\natexlab{}.
\newblock \showarticletitle{The waterfall model in large-scale development}. In \bibinfo{booktitle}{\emph{Product-Focused Software Process Improvement: 10th International Conference, PROFES 2009, Oulu, Finland, June 15-17, 2009. Proceedings 10}}. Springer, \bibinfo{pages}{386--400}.
\newblock


\bibitem[Phan et~al\mbox{.}(2024)]%
        {phan2024hyperagent}
\bibfield{author}{\bibinfo{person}{Huy~Nhat Phan}, \bibinfo{person}{Tien~N Nguyen}, \bibinfo{person}{Phong~X Nguyen}, {and} \bibinfo{person}{Nghi~DQ Bui}.} \bibinfo{year}{2024}\natexlab{}.
\newblock \showarticletitle{Hyperagent: Generalist software engineering agents to solve coding tasks at scale}.
\newblock \bibinfo{journal}{\emph{arXiv preprint arXiv:2409.16299}} (\bibinfo{year}{2024}).
\newblock


\bibitem[Qian et~al\mbox{.}(2024a)]%
        {qian-etal-2024-experiential}
\bibfield{author}{\bibinfo{person}{Chen Qian}, \bibinfo{person}{Yufan Dang}, \bibinfo{person}{Jiahao Li}, \bibinfo{person}{Wei Liu}, \bibinfo{person}{Zihao Xie}, \bibinfo{person}{YiFei Wang}, \bibinfo{person}{Weize Chen}, \bibinfo{person}{Cheng Yang}, \bibinfo{person}{Xin Cong}, \bibinfo{person}{Xiaoyin Che}, \bibinfo{person}{Zhiyuan Liu}, {and} \bibinfo{person}{Maosong Sun}.} \bibinfo{year}{2024}\natexlab{a}.
\newblock \showarticletitle{Experiential Co-Learning of Software-Developing Agents}. In \bibinfo{booktitle}{\emph{Proceedings of the 62nd Annual Meeting of the Association for Computational Linguistics (Volume 1: Long Papers)}}, \bibfield{editor}{\bibinfo{person}{Lun-Wei Ku}, \bibinfo{person}{Andre Martins}, {and} \bibinfo{person}{Vivek Srikumar}} (Eds.). \bibinfo{publisher}{Association for Computational Linguistics}, \bibinfo{address}{Bangkok, Thailand}, \bibinfo{pages}{5628--5640}.
\newblock
\urldef\tempurl%
\url{https://aclanthology.org/2024.acl-long.305}
\showURL{%
\tempurl}


\bibitem[Qian et~al\mbox{.}(2024b)]%
        {qian2024iterative}
\bibfield{author}{\bibinfo{person}{Chen Qian}, \bibinfo{person}{Jiahao Li}, \bibinfo{person}{Yufan Dang}, \bibinfo{person}{Wei Liu}, \bibinfo{person}{YiFei Wang}, \bibinfo{person}{Zihao Xie}, \bibinfo{person}{Weize Chen}, \bibinfo{person}{Cheng Yang}, \bibinfo{person}{Yingli Zhang}, \bibinfo{person}{Zhiyuan Liu}, {et~al\mbox{.}}} \bibinfo{year}{2024}\natexlab{b}.
\newblock \showarticletitle{Iterative Experience Refinement of Software-Developing Agents}.
\newblock \bibinfo{journal}{\emph{arXiv preprint arXiv:2405.04219}} (\bibinfo{year}{2024}).
\newblock


\bibitem[Qian et~al\mbox{.}(2024c)]%
        {qian-etal-2024-chatdev}
\bibfield{author}{\bibinfo{person}{Chen Qian}, \bibinfo{person}{Wei Liu}, \bibinfo{person}{Hongzhang Liu}, \bibinfo{person}{Nuo Chen}, \bibinfo{person}{Yufan Dang}, \bibinfo{person}{Jiahao Li}, \bibinfo{person}{Cheng Yang}, \bibinfo{person}{Weize Chen}, \bibinfo{person}{Yusheng Su}, \bibinfo{person}{Xin Cong}, \bibinfo{person}{Juyuan Xu}, \bibinfo{person}{Dahai Li}, \bibinfo{person}{Zhiyuan Liu}, {and} \bibinfo{person}{Maosong Sun}.} \bibinfo{year}{2024}\natexlab{c}.
\newblock \showarticletitle{{C}hat{D}ev: Communicative Agents for Software Development}. In \bibinfo{booktitle}{\emph{Proceedings of the 62nd Annual Meeting of the Association for Computational Linguistics (Volume 1: Long Papers)}}, \bibfield{editor}{\bibinfo{person}{Lun-Wei Ku}, \bibinfo{person}{Andre Martins}, {and} \bibinfo{person}{Vivek Srikumar}} (Eds.). \bibinfo{publisher}{Association for Computational Linguistics}, \bibinfo{address}{Bangkok, Thailand}, \bibinfo{pages}{15174--15186}.
\newblock
\urldef\tempurl%
\url{https://aclanthology.org/2024.acl-long.810}
\showURL{%
\tempurl}


\bibitem[Qin et~al\mbox{.}(2024)]%
        {qin2024agentfl}
\bibfield{author}{\bibinfo{person}{Yihao Qin}, \bibinfo{person}{Shangwen Wang}, \bibinfo{person}{Yiling Lou}, \bibinfo{person}{Jinhao Dong}, \bibinfo{person}{Kaixin Wang}, \bibinfo{person}{Xiaoling Li}, {and} \bibinfo{person}{Xiaoguang Mao}.} \bibinfo{year}{2024}\natexlab{}.
\newblock \showarticletitle{AgentFL: Scaling LLM-based Fault Localization to Project-Level Context}.
\newblock \bibinfo{journal}{\emph{arXiv preprint arXiv:2403.16362}} (\bibinfo{year}{2024}).
\newblock


\bibitem[Rasheed et~al\mbox{.}(2024a)]%
        {rasheed2024ai}
\bibfield{author}{\bibinfo{person}{Zeeshan Rasheed}, \bibinfo{person}{Malik~Abdul Sami}, \bibinfo{person}{Muhammad Waseem}, \bibinfo{person}{Kai-Kristian Kemell}, \bibinfo{person}{Xiaofeng Wang}, \bibinfo{person}{Anh Nguyen}, \bibinfo{person}{Kari Syst{\"a}}, {and} \bibinfo{person}{Pekka Abrahamsson}.} \bibinfo{year}{2024}\natexlab{a}.
\newblock \showarticletitle{AI-powered Code Review with LLMs: Early Results}.
\newblock \bibinfo{journal}{\emph{arXiv preprint arXiv:2404.18496}} (\bibinfo{year}{2024}).
\newblock


\bibitem[Rasheed et~al\mbox{.}(2024b)]%
        {rasheed2024codepori}
\bibfield{author}{\bibinfo{person}{Zeeshan Rasheed}, \bibinfo{person}{Muhammad Waseem}, \bibinfo{person}{Mika Saari}, \bibinfo{person}{Kari Syst{\"a}}, {and} \bibinfo{person}{Pekka Abrahamsson}.} \bibinfo{year}{2024}\natexlab{b}.
\newblock \showarticletitle{Codepori: Large scale model for autonomous software development by using multi-agents}.
\newblock \bibinfo{journal}{\emph{arXiv preprint arXiv:2402.01411}} (\bibinfo{year}{2024}).
\newblock


\bibitem[Ruan et~al\mbox{.}(2024)]%
        {ruan2024specrover}
\bibfield{author}{\bibinfo{person}{Haifeng Ruan}, \bibinfo{person}{Yuntong Zhang}, {and} \bibinfo{person}{Abhik Roychoudhury}.} \bibinfo{year}{2024}\natexlab{}.
\newblock \showarticletitle{SpecRover: Code Intent Extraction via LLMs}.
\newblock \bibinfo{journal}{\emph{arXiv preprint arXiv:2408.02232}} (\bibinfo{year}{2024}).
\newblock


\bibitem[Sami et~al\mbox{.}(2024a)]%
        {sami2024experimenting}
\bibfield{author}{\bibinfo{person}{Malik~Abdul Sami}, \bibinfo{person}{Muhammad Waseem}, \bibinfo{person}{Zeeshan Rasheed}, \bibinfo{person}{Mika Saari}, \bibinfo{person}{Kari Syst{\"a}}, {and} \bibinfo{person}{Pekka Abrahamsson}.} \bibinfo{year}{2024}\natexlab{a}.
\newblock \showarticletitle{Experimenting with Multi-Agent Software Development: Towards a Unified Platform}.
\newblock \bibinfo{journal}{\emph{arXiv preprint arXiv:2406.05381}} (\bibinfo{year}{2024}).
\newblock


\bibitem[Sami et~al\mbox{.}(2024b)]%
        {sami2024ai}
\bibfield{author}{\bibinfo{person}{Malik~Abdul Sami}, \bibinfo{person}{Muhammad Waseem}, \bibinfo{person}{Zheying Zhang}, \bibinfo{person}{Zeeshan Rasheed}, \bibinfo{person}{Kari Syst{\"a}}, {and} \bibinfo{person}{Pekka Abrahamsson}.} \bibinfo{year}{2024}\natexlab{b}.
\newblock \showarticletitle{AI based Multiagent Approach for Requirements Elicitation and Analysis}.
\newblock \bibinfo{journal}{\emph{arXiv preprint arXiv:2409.00038}} (\bibinfo{year}{2024}).
\newblock


\bibitem[Sandhu(1998)]%
        {sandhu1998role}
\bibfield{author}{\bibinfo{person}{Ravi~S Sandhu}.} \bibinfo{year}{1998}\natexlab{}.
\newblock \showarticletitle{Role-based access control}.
\newblock In \bibinfo{booktitle}{\emph{Advances in computers}}. Vol.~\bibinfo{volume}{46}. \bibinfo{publisher}{Elsevier}, \bibinfo{pages}{237--286}.
\newblock


\bibitem[Shinn et~al\mbox{.}(2024)]%
        {shinn2024reflexion}
\bibfield{author}{\bibinfo{person}{Noah Shinn}, \bibinfo{person}{Federico Cassano}, \bibinfo{person}{Ashwin Gopinath}, \bibinfo{person}{Karthik Narasimhan}, {and} \bibinfo{person}{Shunyu Yao}.} \bibinfo{year}{2024}\natexlab{}.
\newblock \showarticletitle{Reflexion: Language agents with verbal reinforcement learning}.
\newblock \bibinfo{journal}{\emph{Advances in Neural Information Processing Systems}}  \bibinfo{volume}{36} (\bibinfo{year}{2024}).
\newblock


\bibitem[Shoham(1993)]%
        {shoham1993agent}
\bibfield{author}{\bibinfo{person}{Yoav Shoham}.} \bibinfo{year}{1993}\natexlab{}.
\newblock \showarticletitle{Agent-oriented programming}.
\newblock \bibinfo{journal}{\emph{Artificial intelligence}} \bibinfo{volume}{60}, \bibinfo{number}{1} (\bibinfo{year}{1993}), \bibinfo{pages}{51--92}.
\newblock


\bibitem[Smith and Merritt(2020)]%
        {smith2020proactive}
\bibfield{author}{\bibinfo{person}{Preston~G Smith} {and} \bibinfo{person}{Guy~M Merritt}.} \bibinfo{year}{2020}\natexlab{}.
\newblock \bibinfo{booktitle}{\emph{Proactive risk management: Controlling uncertainty in product development}}.
\newblock \bibinfo{publisher}{productivity press}.
\newblock


\bibitem[Sridhara et~al\mbox{.}(2023)]%
        {sridhara2023chatgpt}
\bibfield{author}{\bibinfo{person}{Giriprasad Sridhara}, \bibinfo{person}{Sourav Mazumdar}, {et~al\mbox{.}}} \bibinfo{year}{2023}\natexlab{}.
\newblock \showarticletitle{Chatgpt: A study on its utility for ubiquitous software engineering tasks}.
\newblock \bibinfo{journal}{\emph{arXiv preprint arXiv:2305.16837}} (\bibinfo{year}{2023}).
\newblock


\bibitem[Sun et~al\mbox{.}(2024)]%
        {sun2024ai}
\bibfield{author}{\bibinfo{person}{Zhensu Sun}, \bibinfo{person}{Xiaoning Du}, \bibinfo{person}{Zhou Yang}, \bibinfo{person}{Li Li}, {and} \bibinfo{person}{David Lo}.} \bibinfo{year}{2024}\natexlab{}.
\newblock \showarticletitle{AI Coders Are Among Us: Rethinking Programming Language Grammar Towards Efficient Code Generation}. In \bibinfo{booktitle}{\emph{Proceedings of the 33rd ACM SIGSOFT International Symposium on Software Testing and Analysis}}. \bibinfo{pages}{1124--1136}.
\newblock


\bibitem[Sunyaev and Sunyaev(2020)]%
        {sunyaev2020distributed}
\bibfield{author}{\bibinfo{person}{Ali Sunyaev} {and} \bibinfo{person}{Ali Sunyaev}.} \bibinfo{year}{2020}\natexlab{}.
\newblock \showarticletitle{Distributed ledger technology}.
\newblock \bibinfo{journal}{\emph{Internet computing: Principles of distributed systems and emerging internet-based technologies}} (\bibinfo{year}{2020}), \bibinfo{pages}{265--299}.
\newblock


\bibitem[Taeb et~al\mbox{.}(2024)]%
        {taeb2024axnav}
\bibfield{author}{\bibinfo{person}{Maryam Taeb}, \bibinfo{person}{Amanda Swearngin}, \bibinfo{person}{Eldon Schoop}, \bibinfo{person}{Ruijia Cheng}, \bibinfo{person}{Yue Jiang}, {and} \bibinfo{person}{Jeffrey Nichols}.} \bibinfo{year}{2024}\natexlab{}.
\newblock \showarticletitle{Axnav: Replaying accessibility tests from natural language}. In \bibinfo{booktitle}{\emph{Proceedings of the CHI Conference on Human Factors in Computing Systems}}. \bibinfo{pages}{1--16}.
\newblock


\bibitem[Tang et~al\mbox{.}(2024)]%
        {tang2024collaborative}
\bibfield{author}{\bibinfo{person}{Daniel Tang}, \bibinfo{person}{Zhenghan Chen}, \bibinfo{person}{Kisub Kim}, \bibinfo{person}{Yewei Song}, \bibinfo{person}{Haoye Tian}, \bibinfo{person}{Saad Ezzini}, \bibinfo{person}{Yongfeng Huang}, {and} \bibinfo{person}{Jacques Klein Tegawende~F Bissyande}.} \bibinfo{year}{2024}\natexlab{}.
\newblock \showarticletitle{Collaborative agents for software engineering}.
\newblock \bibinfo{journal}{\emph{arXiv preprint arXiv:2402.02172}} (\bibinfo{year}{2024}).
\newblock


\bibitem[Tao et~al\mbox{.}(2024)]%
        {tao2024magis}
\bibfield{author}{\bibinfo{person}{Wei Tao}, \bibinfo{person}{Yucheng Zhou}, \bibinfo{person}{Wenqiang Zhang}, {and} \bibinfo{person}{Yu Cheng}.} \bibinfo{year}{2024}\natexlab{}.
\newblock \showarticletitle{MAGIS: LLM-Based Multi-Agent Framework for GitHub Issue Resolution}.
\newblock \bibinfo{journal}{\emph{arXiv preprint arXiv:2403.17927}} (\bibinfo{year}{2024}).
\newblock


\bibitem[Tian(2005)]%
        {tian2005software}
\bibfield{author}{\bibinfo{person}{Jeff Tian}.} \bibinfo{year}{2005}\natexlab{}.
\newblock \bibinfo{booktitle}{\emph{Software quality engineering: testing, quality assurance, and quantifiable improvement}}.
\newblock \bibinfo{publisher}{John Wiley \& Sons}.
\newblock


\bibitem[Unland(2015)]%
        {unland2015software}
\bibfield{author}{\bibinfo{person}{Rainer Unland}.} \bibinfo{year}{2015}\natexlab{}.
\newblock \showarticletitle{Software agent systems}.
\newblock In \bibinfo{booktitle}{\emph{Industrial Agents}}. \bibinfo{publisher}{Elsevier}, \bibinfo{pages}{3--22}.
\newblock


\bibitem[Van~Dinter et~al\mbox{.}(2021)]%
        {van2021automation}
\bibfield{author}{\bibinfo{person}{Raymon Van~Dinter}, \bibinfo{person}{Bedir Tekinerdogan}, {and} \bibinfo{person}{Cagatay Catal}.} \bibinfo{year}{2021}\natexlab{}.
\newblock \showarticletitle{Automation of systematic literature reviews: A systematic literature review}.
\newblock \bibinfo{journal}{\emph{Information and Software Technology}}  \bibinfo{volume}{136} (\bibinfo{year}{2021}), \bibinfo{pages}{106589}.
\newblock


\bibitem[Van~Lamsweerde(2000)]%
        {van2000requirements}
\bibfield{author}{\bibinfo{person}{Axel Van~Lamsweerde}.} \bibinfo{year}{2000}\natexlab{}.
\newblock \showarticletitle{Requirements engineering in the year 00: A research perspective}. In \bibinfo{booktitle}{\emph{Proceedings of the 22nd international conference on Software engineering}}. \bibinfo{pages}{5--19}.
\newblock


\bibitem[Voigt and Von~dem Bussche(2017)]%
        {voigt2017eu}
\bibfield{author}{\bibinfo{person}{Paul Voigt} {and} \bibinfo{person}{Axel Von~dem Bussche}.} \bibinfo{year}{2017}\natexlab{}.
\newblock \showarticletitle{The eu general data protection regulation (gdpr)}.
\newblock \bibinfo{journal}{\emph{A Practical Guide, 1st Ed., Cham: Springer International Publishing}} \bibinfo{volume}{10}, \bibinfo{number}{3152676} (\bibinfo{year}{2017}), \bibinfo{pages}{10--5555}.
\newblock


\bibitem[Wang et~al\mbox{.}(2023e)]%
        {wang2023voyager}
\bibfield{author}{\bibinfo{person}{Guanzhi Wang}, \bibinfo{person}{Yuqi Xie}, \bibinfo{person}{Yunfan Jiang}, \bibinfo{person}{Ajay Mandlekar}, \bibinfo{person}{Chaowei Xiao}, \bibinfo{person}{Yuke Zhu}, \bibinfo{person}{Linxi Fan}, {and} \bibinfo{person}{Anima Anandkumar}.} \bibinfo{year}{2023}\natexlab{e}.
\newblock \showarticletitle{Voyager: An open-ended embodied agent with large language models}.
\newblock \bibinfo{journal}{\emph{arXiv preprint arXiv:2305.16291}} (\bibinfo{year}{2023}).
\newblock


\bibitem[Wang et~al\mbox{.}(2024b)]%
        {wang2024intervenor}
\bibfield{author}{\bibinfo{person}{Hanbin Wang}, \bibinfo{person}{Zhenghao Liu}, \bibinfo{person}{Shuo Wang}, \bibinfo{person}{Ganqu Cui}, \bibinfo{person}{Ning Ding}, \bibinfo{person}{Zhiyuan Liu}, {and} \bibinfo{person}{Ge Yu}.} \bibinfo{year}{2024}\natexlab{b}.
\newblock \showarticletitle{INTERVENOR: Prompting the Coding Ability of Large Language Models with the Interactive Chain of Repair}. In \bibinfo{booktitle}{\emph{Findings of the Association for Computational Linguistics ACL 2024}}. \bibinfo{pages}{2081--2107}.
\newblock


\bibitem[Wang et~al\mbox{.}(2024a)]%
        {wang2024mobileagentbench}
\bibfield{author}{\bibinfo{person}{Luyuan Wang}, \bibinfo{person}{Yongyu Deng}, \bibinfo{person}{Yiwei Zha}, \bibinfo{person}{Guodong Mao}, \bibinfo{person}{Qinmin Wang}, \bibinfo{person}{Tianchen Min}, \bibinfo{person}{Wei Chen}, {and} \bibinfo{person}{Shoufa Chen}.} \bibinfo{year}{2024}\natexlab{a}.
\newblock \showarticletitle{MobileAgentBench: An Efficient and User-Friendly Benchmark for Mobile LLM Agents}.
\newblock \bibinfo{journal}{\emph{arXiv preprint arXiv:2406.08184}} (\bibinfo{year}{2024}).
\newblock


\bibitem[Wang et~al\mbox{.}(2023c)]%
        {wang2023survey}
\bibfield{author}{\bibinfo{person}{Lei Wang}, \bibinfo{person}{Chen Ma}, \bibinfo{person}{Xueyang Feng}, \bibinfo{person}{Zeyu Zhang}, \bibinfo{person}{Hao Yang}, \bibinfo{person}{Jingsen Zhang}, \bibinfo{person}{Zhiyuan Chen}, \bibinfo{person}{Jiakai Tang}, \bibinfo{person}{Xu Chen}, \bibinfo{person}{Yankai Lin}, {et~al\mbox{.}}} \bibinfo{year}{2023}\natexlab{c}.
\newblock \bibinfo{title}{A Survey on Large Language Model based Autonomous Agents. CoRR abs/2308.11432 (2023)}.
\newblock
\newblock


\bibitem[Wang et~al\mbox{.}(2024d)]%
        {wang2024megaagent}
\bibfield{author}{\bibinfo{person}{Qian Wang}, \bibinfo{person}{Tianyu Wang}, \bibinfo{person}{Qinbin Li}, \bibinfo{person}{Jingsheng Liang}, {and} \bibinfo{person}{Bingsheng He}.} \bibinfo{year}{2024}\natexlab{d}.
\newblock \showarticletitle{MegaAgent: A Practical Framework for Autonomous Cooperation in Large-Scale LLM Agent Systems}.
\newblock \bibinfo{journal}{\emph{arXiv preprint arXiv:2408.09955}} (\bibinfo{year}{2024}).
\newblock


\bibitem[Wang et~al\mbox{.}(2024c)]%
        {wang2024benchmark}
\bibfield{author}{\bibinfo{person}{Siyuan Wang}, \bibinfo{person}{Zhuohan Long}, \bibinfo{person}{Zhihao Fan}, \bibinfo{person}{Zhongyu Wei}, {and} \bibinfo{person}{Xuanjing Huang}.} \bibinfo{year}{2024}\natexlab{c}.
\newblock \showarticletitle{Benchmark Self-Evolving: A Multi-Agent Framework for Dynamic LLM Evaluation}.
\newblock \bibinfo{journal}{\emph{arXiv preprint arXiv:2402.11443}} (\bibinfo{year}{2024}).
\newblock


\bibitem[Wang et~al\mbox{.}(2023a)]%
        {wang2023describe}
\bibfield{author}{\bibinfo{person}{Zihao Wang}, \bibinfo{person}{Shaofei Cai}, \bibinfo{person}{Guanzhou Chen}, \bibinfo{person}{Anji Liu}, \bibinfo{person}{Xiaojian Ma}, {and} \bibinfo{person}{Yitao Liang}.} \bibinfo{year}{2023}\natexlab{a}.
\newblock \showarticletitle{Describe, explain, plan and select: Interactive planning with large language models enables open-world multi-task agents}.
\newblock \bibinfo{journal}{\emph{arXiv preprint arXiv:2302.01560}} (\bibinfo{year}{2023}).
\newblock


\bibitem[Wang et~al\mbox{.}(2023b)]%
        {wang2023rcagent}
\bibfield{author}{\bibinfo{person}{Zefan Wang}, \bibinfo{person}{Zichuan Liu}, \bibinfo{person}{Yingying Zhang}, \bibinfo{person}{Aoxiao Zhong}, \bibinfo{person}{Lunting Fan}, \bibinfo{person}{Lingfei Wu}, {and} \bibinfo{person}{Qingsong Wen}.} \bibinfo{year}{2023}\natexlab{b}.
\newblock \showarticletitle{RCAgent: Cloud Root Cause Analysis by Autonomous Agents with Tool-Augmented Large Language Models}.
\newblock \bibinfo{journal}{\emph{arXiv preprint arXiv:2310.16340}} (\bibinfo{year}{2023}).
\newblock


\bibitem[Wang et~al\mbox{.}(2024e)]%
        {wang2024xuat}
\bibfield{author}{\bibinfo{person}{Zhitao Wang}, \bibinfo{person}{Wei Wang}, \bibinfo{person}{Zirao Li}, \bibinfo{person}{Long Wang}, \bibinfo{person}{Can Yi}, \bibinfo{person}{Xinjie Xu}, \bibinfo{person}{Luyang Cao}, \bibinfo{person}{Hanjing Su}, \bibinfo{person}{Shouzhi Chen}, {and} \bibinfo{person}{Jun Zhou}.} \bibinfo{year}{2024}\natexlab{e}.
\newblock \showarticletitle{XUAT-Copilot: Multi-Agent Collaborative System for Automated User Acceptance Testing with Large Language Model}.
\newblock \bibinfo{journal}{\emph{arXiv preprint arXiv:2401.02705}} (\bibinfo{year}{2024}).
\newblock


\bibitem[Wang et~al\mbox{.}(2023d)]%
        {wang2023rolellm}
\bibfield{author}{\bibinfo{person}{Zekun~Moore Wang}, \bibinfo{person}{Zhongyuan Peng}, \bibinfo{person}{Haoran Que}, \bibinfo{person}{Jiaheng Liu}, \bibinfo{person}{Wangchunshu Zhou}, \bibinfo{person}{Yuhan Wu}, \bibinfo{person}{Hongcheng Guo}, \bibinfo{person}{Ruitong Gan}, \bibinfo{person}{Zehao Ni}, \bibinfo{person}{Man Zhang}, {et~al\mbox{.}}} \bibinfo{year}{2023}\natexlab{d}.
\newblock \showarticletitle{Rolellm: Benchmarking, eliciting, and enhancing role-playing abilities of large language models}.
\newblock \bibinfo{journal}{\emph{arXiv preprint arXiv:2310.00746}} (\bibinfo{year}{2023}).
\newblock


\bibitem[Wegner(1990)]%
        {wegner1990concepts}
\bibfield{author}{\bibinfo{person}{Peter Wegner}.} \bibinfo{year}{1990}\natexlab{}.
\newblock \showarticletitle{Concepts and paradigms of object-oriented programming}.
\newblock \bibinfo{journal}{\emph{ACM Sigplan Oops Messenger}} \bibinfo{volume}{1}, \bibinfo{number}{1} (\bibinfo{year}{1990}), \bibinfo{pages}{7--87}.
\newblock


\bibitem[Widyasari et~al\mbox{.}(2024)]%
        {widyasari2024beyond}
\bibfield{author}{\bibinfo{person}{Ratnadira Widyasari}, \bibinfo{person}{David Lo}, {and} \bibinfo{person}{Lizi Liao}.} \bibinfo{year}{2024}\natexlab{}.
\newblock \showarticletitle{Beyond ChatGPT: Enhancing Software Quality Assurance Tasks with Diverse LLMs and Validation Techniques}.
\newblock \bibinfo{journal}{\emph{arXiv preprint arXiv:2409.01001}} (\bibinfo{year}{2024}).
\newblock


\bibitem[Wohlin(2014)]%
        {wohlin2014guidelines}
\bibfield{author}{\bibinfo{person}{Claes Wohlin}.} \bibinfo{year}{2014}\natexlab{}.
\newblock \showarticletitle{Guidelines for snowballing in systematic literature studies and a replication in software engineering}. In \bibinfo{booktitle}{\emph{Proceedings of the 18th international conference on evaluation and assessment in software engineering}}. \bibinfo{pages}{1--10}.
\newblock


\bibitem[Wooldridge(2009)]%
        {wooldridge2009introduction}
\bibfield{author}{\bibinfo{person}{Michael Wooldridge}.} \bibinfo{year}{2009}\natexlab{}.
\newblock \bibinfo{booktitle}{\emph{An introduction to multiagent systems}}.
\newblock \bibinfo{publisher}{John wiley \& sons}.
\newblock


\bibitem[Wu et~al\mbox{.}(2023a)]%
        {wu2023autogen}
\bibfield{author}{\bibinfo{person}{Qingyun Wu}, \bibinfo{person}{Gagan Bansal}, \bibinfo{person}{Jieyu Zhang}, \bibinfo{person}{Yiran Wu}, \bibinfo{person}{Shaokun Zhang}, \bibinfo{person}{Erkang Zhu}, \bibinfo{person}{Beibin Li}, \bibinfo{person}{Li Jiang}, \bibinfo{person}{Xiaoyun Zhang}, {and} \bibinfo{person}{Chi Wang}.} \bibinfo{year}{2023}\natexlab{a}.
\newblock \showarticletitle{Autogen: Enabling next-gen llm applications via multi-agent conversation framework}.
\newblock \bibinfo{journal}{\emph{arXiv preprint arXiv:2308.08155}} (\bibinfo{year}{2023}).
\newblock


\bibitem[Wu et~al\mbox{.}(2023b)]%
        {wu2023empirical}
\bibfield{author}{\bibinfo{person}{Yiran Wu}, \bibinfo{person}{Feiran Jia}, \bibinfo{person}{Shaokun Zhang}, \bibinfo{person}{Qingyun Wu}, \bibinfo{person}{Hangyu Li}, \bibinfo{person}{Erkang Zhu}, \bibinfo{person}{Yue Wang}, \bibinfo{person}{Yin~Tat Lee}, \bibinfo{person}{Richard Peng}, {and} \bibinfo{person}{Chi Wang}.} \bibinfo{year}{2023}\natexlab{b}.
\newblock \showarticletitle{An empirical study on challenging math problem solving with gpt-4}.
\newblock \bibinfo{journal}{\emph{arXiv preprint arXiv:2306.01337}} (\bibinfo{year}{2023}).
\newblock


\bibitem[Wu et~al\mbox{.}(2024)]%
        {wu2024shall}
\bibfield{author}{\bibinfo{person}{Zengqing Wu}, \bibinfo{person}{Shuyuan Zheng}, \bibinfo{person}{Qianying Liu}, \bibinfo{person}{Xu Han}, \bibinfo{person}{Brian~Inhyuk Kwon}, \bibinfo{person}{Makoto Onizuka}, \bibinfo{person}{Shaojie Tang}, \bibinfo{person}{Run Peng}, {and} \bibinfo{person}{Chuan Xiao}.} \bibinfo{year}{2024}\natexlab{}.
\newblock \showarticletitle{Shall We Talk: Exploring Spontaneous Collaborations of Competing LLM Agents}.
\newblock \bibinfo{journal}{\emph{arXiv preprint arXiv:2402.12327}} (\bibinfo{year}{2024}).
\newblock


\bibitem[Xi et~al\mbox{.}(2023)]%
        {xi2023rise}
\bibfield{author}{\bibinfo{person}{Zhiheng Xi}, \bibinfo{person}{Wenxiang Chen}, \bibinfo{person}{Xin Guo}, \bibinfo{person}{Wei He}, \bibinfo{person}{Yiwen Ding}, \bibinfo{person}{Boyang Hong}, \bibinfo{person}{Ming Zhang}, \bibinfo{person}{Junzhe Wang}, \bibinfo{person}{Senjie Jin}, \bibinfo{person}{Enyu Zhou}, {et~al\mbox{.}}} \bibinfo{year}{2023}\natexlab{}.
\newblock \showarticletitle{The rise and potential of large language model based agents: A survey}.
\newblock \bibinfo{journal}{\emph{arXiv preprint arXiv:2309.07864}} (\bibinfo{year}{2023}).
\newblock


\bibitem[Xia et~al\mbox{.}(2024)]%
        {xia2024fuzz4all}
\bibfield{author}{\bibinfo{person}{Chunqiu~Steven Xia}, \bibinfo{person}{Matteo Paltenghi}, \bibinfo{person}{Jia Le~Tian}, \bibinfo{person}{Michael Pradel}, {and} \bibinfo{person}{Lingming Zhang}.} \bibinfo{year}{2024}\natexlab{}.
\newblock \showarticletitle{Fuzz4all: Universal fuzzing with large language models}. In \bibinfo{booktitle}{\emph{Proceedings of the IEEE/ACM 46th International Conference on Software Engineering}}. \bibinfo{pages}{1--13}.
\newblock


\bibitem[Yang et~al\mbox{.}(2023a)]%
        {yang2023white}
\bibfield{author}{\bibinfo{person}{Chenyuan Yang}, \bibinfo{person}{Yinlin Deng}, \bibinfo{person}{Runyu Lu}, \bibinfo{person}{Jiayi Yao}, \bibinfo{person}{Jiawei Liu}, \bibinfo{person}{Reyhaneh Jabbarvand}, {and} \bibinfo{person}{Lingming Zhang}.} \bibinfo{year}{2023}\natexlab{a}.
\newblock \showarticletitle{White-box compiler fuzzing empowered by large language models}.
\newblock \bibinfo{journal}{\emph{arXiv preprint arXiv:2310.15991}} (\bibinfo{year}{2023}).
\newblock


\bibitem[Yang et~al\mbox{.}(2023b)]%
        {yang2023apidocbooster}
\bibfield{author}{\bibinfo{person}{Chengran Yang}, \bibinfo{person}{Jiakun Liu}, \bibinfo{person}{Bowen Xu}, \bibinfo{person}{Christoph Treude}, \bibinfo{person}{Yunbo Lyu}, \bibinfo{person}{Ming Li}, {and} \bibinfo{person}{David Lo}.} \bibinfo{year}{2023}\natexlab{b}.
\newblock \showarticletitle{APIDocBooster: An Extract-Then-Abstract Framework Leveraging Large Language Models for Augmenting API Documentation}.
\newblock \bibinfo{journal}{\emph{arXiv preprint arXiv:2312.10934}} (\bibinfo{year}{2023}).
\newblock


\bibitem[Yang et~al\mbox{.}(2024)]%
        {yang2024enhancing}
\bibfield{author}{\bibinfo{person}{Weiqing Yang}, \bibinfo{person}{Hanbin Wang}, \bibinfo{person}{Zhenghao Liu}, \bibinfo{person}{Xinze Li}, \bibinfo{person}{Yukun Yan}, \bibinfo{person}{Shuo Wang}, \bibinfo{person}{Yu Gu}, \bibinfo{person}{Minghe Yu}, \bibinfo{person}{Zhiyuan Liu}, {and} \bibinfo{person}{Ge Yu}.} \bibinfo{year}{2024}\natexlab{}.
\newblock \showarticletitle{Enhancing the Code Debugging Ability of LLMs via Communicative Agent Based Data Refinement}.
\newblock \bibinfo{journal}{\emph{arXiv preprint arXiv:2408.05006}} (\bibinfo{year}{2024}).
\newblock


\bibitem[Yi et~al\mbox{.}(2014)]%
        {yi2014homomorphic}
\bibfield{author}{\bibinfo{person}{Xun Yi}, \bibinfo{person}{Russell Paulet}, \bibinfo{person}{Elisa Bertino}, \bibinfo{person}{Xun Yi}, \bibinfo{person}{Russell Paulet}, {and} \bibinfo{person}{Elisa Bertino}.} \bibinfo{year}{2014}\natexlab{}.
\newblock \bibinfo{booktitle}{\emph{Homomorphic encryption}}.
\newblock \bibinfo{publisher}{Springer}.
\newblock


\bibitem[Yoon et~al\mbox{.}(2024)]%
        {yoon2024intent}
\bibfield{author}{\bibinfo{person}{Juyeon Yoon}, \bibinfo{person}{Robert Feldt}, {and} \bibinfo{person}{Shin Yoo}.} \bibinfo{year}{2024}\natexlab{}.
\newblock \showarticletitle{Intent-Driven Mobile GUI Testing with Autonomous Large Language Model Agents}. In \bibinfo{booktitle}{\emph{2024 IEEE Conference on Software Testing, Verification and Validation (ICST)}}. IEEE, \bibinfo{pages}{129--139}.
\newblock


\bibitem[Zan et~al\mbox{.}(2024)]%
        {zan2024codes}
\bibfield{author}{\bibinfo{person}{Daoguang Zan}, \bibinfo{person}{Ailun Yu}, \bibinfo{person}{Wei Liu}, \bibinfo{person}{Dong Chen}, \bibinfo{person}{Bo Shen}, \bibinfo{person}{Wei Li}, \bibinfo{person}{Yafen Yao}, \bibinfo{person}{Yongshun Gong}, \bibinfo{person}{Xiaolin Chen}, \bibinfo{person}{Bei Guan}, {et~al\mbox{.}}} \bibinfo{year}{2024}\natexlab{}.
\newblock \showarticletitle{CodeS: Natural Language to Code Repository via Multi-Layer Sketch}.
\newblock \bibinfo{journal}{\emph{arXiv preprint arXiv:2403.16443}} (\bibinfo{year}{2024}).
\newblock


\bibitem[Zeng et~al\mbox{.}(2022)]%
        {zeng2022extensive}
\bibfield{author}{\bibinfo{person}{Zhengran Zeng}, \bibinfo{person}{Hanzhuo Tan}, \bibinfo{person}{Haotian Zhang}, \bibinfo{person}{Jing Li}, \bibinfo{person}{Yuqun Zhang}, {and} \bibinfo{person}{Lingming Zhang}.} \bibinfo{year}{2022}\natexlab{}.
\newblock \showarticletitle{An extensive study on pre-trained models for program understanding and generation}. In \bibinfo{booktitle}{\emph{Proceedings of the 31st ACM SIGSOFT international symposium on software testing and analysis}}. \bibinfo{pages}{39--51}.
\newblock


\bibitem[Zhang et~al\mbox{.}(2024g)]%
        {zhang2024g}
\bibfield{author}{\bibinfo{person}{Guibin Zhang}, \bibinfo{person}{Yanwei Yue}, \bibinfo{person}{Xiangguo Sun}, \bibinfo{person}{Guancheng Wan}, \bibinfo{person}{Miao Yu}, \bibinfo{person}{Junfeng Fang}, \bibinfo{person}{Kun Wang}, {and} \bibinfo{person}{Dawei Cheng}.} \bibinfo{year}{2024}\natexlab{g}.
\newblock \showarticletitle{G-designer: Architecting multi-agent communication topologies via graph neural networks}.
\newblock \bibinfo{journal}{\emph{arXiv preprint arXiv:2410.11782}} (\bibinfo{year}{2024}).
\newblock


\bibitem[Zhang et~al\mbox{.}(2024a)]%
        {zhang2024pair}
\bibfield{author}{\bibinfo{person}{Huan Zhang}, \bibinfo{person}{Wei Cheng}, \bibinfo{person}{Yuhan Wu}, {and} \bibinfo{person}{Wei Hu}.} \bibinfo{year}{2024}\natexlab{a}.
\newblock \showarticletitle{A Pair Programming Framework for Code Generation via Multi-Plan Exploration and Feedback-Driven Refinement}.
\newblock \bibinfo{journal}{\emph{arXiv preprint arXiv:2409.05001}} (\bibinfo{year}{2024}).
\newblock


\bibitem[Zhang et~al\mbox{.}(2024f)]%
        {zhang2024diversity}
\bibfield{author}{\bibinfo{person}{Kexun Zhang}, \bibinfo{person}{Weiran Yao}, \bibinfo{person}{Zuxin Liu}, \bibinfo{person}{Yihao Feng}, \bibinfo{person}{Zhiwei Liu}, \bibinfo{person}{Rithesh Murthy}, \bibinfo{person}{Tian Lan}, \bibinfo{person}{Lei Li}, \bibinfo{person}{Renze Lou}, \bibinfo{person}{Jiacheng Xu}, {et~al\mbox{.}}} \bibinfo{year}{2024}\natexlab{f}.
\newblock \showarticletitle{Diversity Empowers Intelligence: Integrating Expertise of Software Engineering Agents}.
\newblock \bibinfo{journal}{\emph{arXiv preprint arXiv:2408.07060}} (\bibinfo{year}{2024}).
\newblock


\bibitem[Zhang et~al\mbox{.}(2024b)]%
        {zhang2024acfix}
\bibfield{author}{\bibinfo{person}{Lyuye Zhang}, \bibinfo{person}{Kaixuan Li}, \bibinfo{person}{Kairan Sun}, \bibinfo{person}{Daoyuan Wu}, \bibinfo{person}{Ye Liu}, \bibinfo{person}{Haoye Tian}, {and} \bibinfo{person}{Yang Liu}.} \bibinfo{year}{2024}\natexlab{b}.
\newblock \showarticletitle{Acfix: Guiding llms with mined common rbac practices for context-aware repair of access control vulnerabilities in smart contracts}.
\newblock \bibinfo{journal}{\emph{arXiv preprint arXiv:2403.06838}} (\bibinfo{year}{2024}).
\newblock


\bibitem[Zhang et~al\mbox{.}(2018)]%
        {zhang2018empirical}
\bibfield{author}{\bibinfo{person}{Li Zhang}, \bibinfo{person}{Jia-Hao Tian}, \bibinfo{person}{Jing Jiang}, \bibinfo{person}{Yi-Jun Liu}, \bibinfo{person}{Meng-Yuan Pu}, {and} \bibinfo{person}{Tao Yue}.} \bibinfo{year}{2018}\natexlab{}.
\newblock \showarticletitle{Empirical research in software engineering—a literature survey}.
\newblock \bibinfo{journal}{\emph{Journal of Computer Science and Technology}}  \bibinfo{volume}{33} (\bibinfo{year}{2018}), \bibinfo{pages}{876--899}.
\newblock


\bibitem[Zhang et~al\mbox{.}(2024d)]%
        {zhang2024experimenting}
\bibfield{author}{\bibinfo{person}{Simiao Zhang}, \bibinfo{person}{Jiaping Wang}, \bibinfo{person}{Guoliang Dong}, \bibinfo{person}{Jun Sun}, \bibinfo{person}{Yueling Zhang}, {and} \bibinfo{person}{Geguang Pu}.} \bibinfo{year}{2024}\natexlab{d}.
\newblock \showarticletitle{Experimenting a New Programming Practice with LLMs}.
\newblock \bibinfo{journal}{\emph{arXiv preprint arXiv:2401.01062}} (\bibinfo{year}{2024}).
\newblock


\bibitem[Zhang et~al\mbox{.}(2024e)]%
        {zhang2024empowering}
\bibfield{author}{\bibinfo{person}{Sai Zhang}, \bibinfo{person}{Zhenchang Xing}, \bibinfo{person}{Ronghui Guo}, \bibinfo{person}{Fangzhou Xu}, \bibinfo{person}{Lei Chen}, \bibinfo{person}{Zhaoyuan Zhang}, \bibinfo{person}{Xiaowang Zhang}, \bibinfo{person}{Zhiyong Feng}, {and} \bibinfo{person}{Zhiqiang Zhuang}.} \bibinfo{year}{2024}\natexlab{e}.
\newblock \showarticletitle{Empowering Agile-Based Generative Software Development through Human-AI Teamwork}.
\newblock \bibinfo{journal}{\emph{arXiv preprint arXiv:2407.15568}} (\bibinfo{year}{2024}).
\newblock


\bibitem[Zhang et~al\mbox{.}(2023)]%
        {zhang2023siren}
\bibfield{author}{\bibinfo{person}{Yue Zhang}, \bibinfo{person}{Yafu Li}, \bibinfo{person}{Leyang Cui}, \bibinfo{person}{Deng Cai}, \bibinfo{person}{Lemao Liu}, \bibinfo{person}{Tingchen Fu}, \bibinfo{person}{Xinting Huang}, \bibinfo{person}{Enbo Zhao}, \bibinfo{person}{Yu Zhang}, \bibinfo{person}{Yulong Chen}, {et~al\mbox{.}}} \bibinfo{year}{2023}\natexlab{}.
\newblock \showarticletitle{Siren's song in the AI ocean: a survey on hallucination in large language models}.
\newblock \bibinfo{journal}{\emph{arXiv preprint arXiv:2309.01219}} (\bibinfo{year}{2023}).
\newblock


\bibitem[Zhang et~al\mbox{.}(2024c)]%
        {zhang2024autocoderover}
\bibfield{author}{\bibinfo{person}{Yuntong Zhang}, \bibinfo{person}{Haifeng Ruan}, \bibinfo{person}{Zhiyu Fan}, {and} \bibinfo{person}{Abhik Roychoudhury}.} \bibinfo{year}{2024}\natexlab{c}.
\newblock \showarticletitle{Autocoderover: Autonomous program improvement}. In \bibinfo{booktitle}{\emph{Proceedings of the 33rd ACM SIGSOFT International Symposium on Software Testing and Analysis}}. \bibinfo{pages}{1592--1604}.
\newblock


\bibitem[Zhao et~al\mbox{.}(2024)]%
        {zhao2024hierarchical}
\bibfield{author}{\bibinfo{person}{Zhonghan Zhao}, \bibinfo{person}{Kewei Chen}, \bibinfo{person}{Dongxu Guo}, \bibinfo{person}{Wenhao Chai}, \bibinfo{person}{Tian Ye}, \bibinfo{person}{Yanting Zhang}, {and} \bibinfo{person}{Gaoang Wang}.} \bibinfo{year}{2024}\natexlab{}.
\newblock \showarticletitle{Hierarchical Auto-Organizing System for Open-Ended Multi-Agent Navigation}.
\newblock \bibinfo{journal}{\emph{arXiv preprint arXiv:2403.08282}} (\bibinfo{year}{2024}).
\newblock


\bibitem[Zheng et~al\mbox{.}(2018)]%
        {zheng2018blockchain}
\bibfield{author}{\bibinfo{person}{Zibin Zheng}, \bibinfo{person}{Shaoan Xie}, \bibinfo{person}{Hong-Ning Dai}, \bibinfo{person}{Xiangping Chen}, {and} \bibinfo{person}{Huaimin Wang}.} \bibinfo{year}{2018}\natexlab{}.
\newblock \showarticletitle{Blockchain challenges and opportunities: A survey}.
\newblock \bibinfo{journal}{\emph{International journal of web and grid services}} \bibinfo{volume}{14}, \bibinfo{number}{4} (\bibinfo{year}{2018}), \bibinfo{pages}{352--375}.
\newblock


\bibitem[Zhou et~al\mbox{.}(2024)]%
        {zhou2024large}
\bibfield{author}{\bibinfo{person}{Xin Zhou}, \bibinfo{person}{Sicong Cao}, \bibinfo{person}{Xiaobing Sun}, {and} \bibinfo{person}{David Lo}.} \bibinfo{year}{2024}\natexlab{}.
\newblock \showarticletitle{Large Language Model for Vulnerability Detection and Repair: Literature Review and the Road Ahead.}
\newblock \bibinfo{journal}{\emph{arXiv preprint arXiv:2404.02525}} (\bibinfo{year}{2024}).
\newblock


\bibitem[Zhou et~al\mbox{.}(2023)]%
        {zhou2023ccbert}
\bibfield{author}{\bibinfo{person}{Xin Zhou}, \bibinfo{person}{Bowen Xu}, \bibinfo{person}{DongGyun Han}, \bibinfo{person}{Zhou Yang}, \bibinfo{person}{Junda He}, {and} \bibinfo{person}{David Lo}.} \bibinfo{year}{2023}\natexlab{}.
\newblock \showarticletitle{CCBERT: Self-Supervised Code Change Representation Learning}. In \bibinfo{booktitle}{\emph{2023 IEEE International Conference on Software Maintenance and Evolution (ICSME)}}. IEEE, \bibinfo{pages}{182--193}.
\newblock


\bibitem[Zhu et~al\mbox{.}(2024)]%
        {zhu2024llama}
\bibfield{author}{\bibinfo{person}{Tong Zhu}, \bibinfo{person}{Xiaoye Qu}, \bibinfo{person}{Daize Dong}, \bibinfo{person}{Jiacheng Ruan}, \bibinfo{person}{Jingqi Tong}, \bibinfo{person}{Conghui He}, {and} \bibinfo{person}{Yu Cheng}.} \bibinfo{year}{2024}\natexlab{}.
\newblock \showarticletitle{Llama-moe: Building mixture-of-experts from llama with continual pre-training}. In \bibinfo{booktitle}{\emph{Proceedings of the 2024 Conference on Empirical Methods in Natural Language Processing}}. \bibinfo{pages}{15913--15923}.
\newblock


\bibitem[Zhuo et~al\mbox{.}(2024)]%
        {zhuo2024bigcodebench}
\bibfield{author}{\bibinfo{person}{Terry~Yue Zhuo}, \bibinfo{person}{Minh~Chien Vu}, \bibinfo{person}{Jenny Chim}, \bibinfo{person}{Han Hu}, \bibinfo{person}{Wenhao Yu}, \bibinfo{person}{Ratnadira Widyasari}, \bibinfo{person}{Imam Nur~Bani Yusuf}, \bibinfo{person}{Haolan Zhan}, \bibinfo{person}{Junda He}, \bibinfo{person}{Indraneil Paul}, {et~al\mbox{.}}} \bibinfo{year}{2024}\natexlab{}.
\newblock \showarticletitle{Bigcodebench: Benchmarking code generation with diverse function calls and complex instructions}.
\newblock \bibinfo{journal}{\emph{arXiv preprint arXiv:2406.15877}} (\bibinfo{year}{2024}).
\newblock


\end{thebibliography}

\end{document}